# An Introduction to the Interacting Boson Model of the Atomic Nucleus

**Walter Pfeifer**

----------------------------------------------------------------

Dr. Walter Pfeifer : Street address: Stapfenackerweg 9
CH 5034 Suhr
Switzerland

email: mailbox@walterpfeifer.ch

# Contents

# Preface

The interacting boson model (IBM) is suitable for describing intermediate and heavy atomic nuclei. Adjusting a small number of parameters, it reproduces the majority of the low-lying states of such nuclei. Figure 0.1 gives a survey of nuclei which have been handled with the model variant IBM2. Figures 10.7 and 14.3 show the nuclei for which IBM1-calculations have been performed.

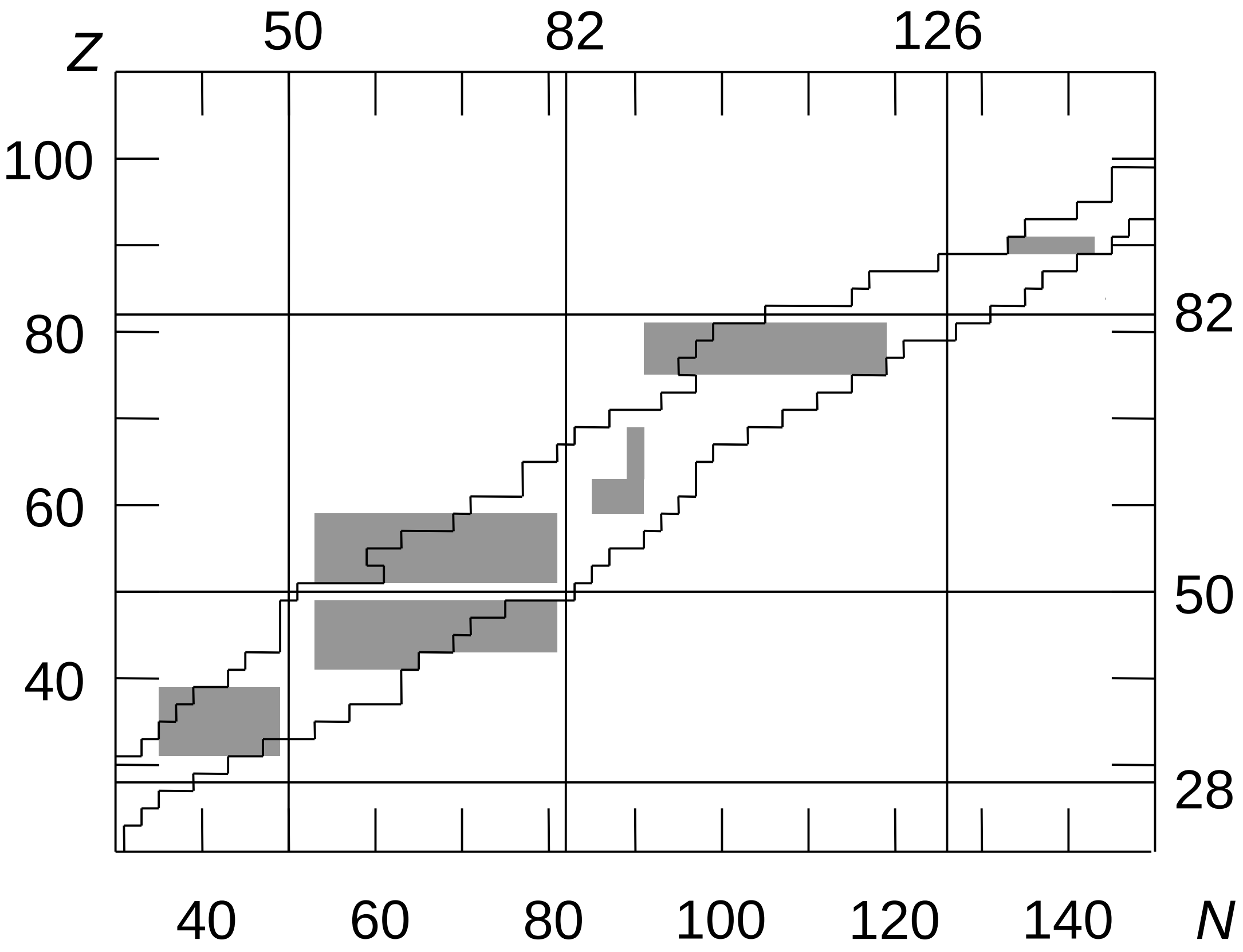

Figure 0.1. Card of even-even nuclei. $Z$ = number of protons, $N$ = number of neutrons. The dark areas denote nuclei which have been calculated using the IBM2 approximation (Iachello, 1988, p. 110).

The IBM is based on the well-known shell model and on geometrical collective models of the atomic nucleus. Despite its relatively simple structure, it has proved to be a powerful tool. In addition, it is of considerable theoretical interest since it shows the dynamical symmetries of several nuclei, which are made visible using Lie algebras.

The IBM was created in 1974 by F. Iachello and A. Arima ( Arima and Iachello, 1975 ). Subsequently in numerous papers it has been checked, extended, and discussed. In 1990 in Santa Fe, New Mexico, Akito Arima was awarded the Weatherhill medal by the Franklin Institute for his many contributions to the field of nuclear physics. In the same year, Francesco Iachello received the Wigner medal given by the Group Theory and Fundamental Physics Foundation, which cited him for " developing powerful algebraic tools and models in nuclear physics ". In 1993, A. Arima and F. Iachello were awarded the T. W. Bonner Prize in Nuclear Physics by the American Physical Society.

The international symposia of Erice ( Italy, 1978 ), Granada ( Spain, 1981 ), Drexel ( USA, 1983 ), Gull Lake ( USA, 1984 ), La Rábida ( Spain, 1985 ), Dubrovnik ( Yugoslavia, 1986 ) as well as other events focusing on the IBM have clearly demonstrated the wide interest in this theory and its further development on an international scale. In 1994 in Padua ( Italy ), the "International Conference on Perspectives for the Interacting Boson Model on the Occasion of its 20th Anniversary" took place.

In recent years outstanding survey reports on the IBM have been published by Iachello and Arima ( 1987 ), Eisenberg and Greiner ( 1987 ), Talmi ( 1993 ), Frank and Van Isacker ( 1994 ) , and others. Unfortunately, there are few introductory books on the IBM available for the interested reader. The present publication might reduce this deficiency. It is directed towards physics students and experimental physicists interested in the main properties of the IBM. Knowledge of the elements of quantum mechanics, nuclear physics, and electrodynamics is a prerequisite.

The experienced reader may feel that some transformations and proofs have been rendered in too great detail. For the beginner, however, this is indispensable and may serve him as an exercise.

# 1 Introduction

A model of the atomic nucleus has to be able to describe nuclear properties such as spins and energies of the lowest levels, decay probabilities for the emission of gamma quantas, probabilities ( spectroscopic factors ) of transfer reactions, multipole moments and so forth. In this chapter those models are outlined from which the IBM comes.

The IBM is mainly rooted in the shell model, which has proved to be an excellent instrument for light nuclei ( up to 50 nucleons ). The larger the number of nucleons becomes the more shells have to be taken into account and the number of nuclear states soon becomes so colossal that the shell model will be intractable. For example the $2^+$ state ( spin 2 and positive parity ) of $^{154}$Sm shows $3 \cdot 10^{14}$ different configurations ( Casten, 1990, p. 198 ). The interacting boson model ( sometimes named interacting boson approximation IBA ) reduces the number of states heavily. It constitutes only 26 configurations for the $2^+$ state mentioned above.

The shell model reveals that the low-lying states of the even-even nuclei are made up predominantly by nucleon pairs with total spin 0 or 2. Higher spins of such pairs are rare for energy reasons ( Hess, 1983, p. 55 ). Particularly the spins of pairs of identical nucleons are even numbers because they constitute an antisymmetric state ( appendix A2 ). Furthermore, in the case of two identical nucleon pairs the total spin is strictly even, which follows from the fact that the pairs behave like bosons ( see appendix A2 ). This theoretical result is not far from the real situation of even-even nuclei, from which it is known that their total spin predominantly is even.

These and other arguments led to the basic assumption of the IBM which postulates that the nucleon pairs are represented by bosons with angular momenta $l$ = 0 or 2. The multitude of shells which appears in the shell model is reduced to the simple *s*-shell ( $l$ = 0 ) and the *d*-shell ( $l$ = 2 ) which is composed vectorially by *d*-bosons analogously to the shell model technique. The IBM builds on a closed shell i.e. the number of bosons depends on the number of active nucleon ( or hole ) pairs outside a closed shell. Each type of bosons, the *s*- and the *d*-boson, has its own binding energy with regard to the closed shell. Analogously to the standard shell model, the interacting potential of the bosons acts only in pairs.

As a peculiarity of the IBM there exist special cases in which certain linear combinations of matrix elements of this interaction potential vanish ( chapters 10 and 14 ). In these cases the energies of the nuclear states and the configurations can be expressed in a closed algebraic form. These special cases are named "dynamic symmetries". They correspond to the well-known "limits" allocated to the vibration, the rotation et cetera of the whole nucleus. However most nuclei have to be calculated by diagonalising the Hamilton matrix as is usual in quantum mechanics ( chapter 12 ).

The IBM is not only in connection with the shell model but also with the collective model of the atomic nucleus of Bohr and Mottelson ( 1953, 1975 ). In this model the deformation of the nuclear surface is represented by five parameters from which a Hamiltonian of a five dimensional oscillator results. It contains fivefold generating and annihilating operators for oscillator quanta. The operators of these bosons correspond to the operators of the *d*-shell in the IBM.

However, the handling of the collective model is laborious ( Jolos, 1985, p. 121 ). Moreover, the number of bosons is unlimited and is not a good quantum number in contrast to the situation in the IBM. The special cases mentioned above are reproduced by some versions of geometric models but they are not joined together continuously ( Barrett, 1981, p. 534 ). In the IBM these relations exist.

An additional relationship between both models consists in the fact that the form of the Hamilton operator ( after suitable transformations ) is similar to the one of the IBM ( Jolos, 1985, p. 124 ).

# 2 Characteristics of the IBM

The simplest versions of the IBM describe the even-even nucleus as an inert core combined with bosons which represent pairs of identical nucleons. Bosons behave symmetrically in the following way: supposing that each boson has a wave function, that can be attributed, the wave function of the total configuration does not alter if two bosons ( i.e. their variables) are interchanged. The analogy between nucleon pairs and bosons does not go so far that in the IBM the wave functions of the corresponding nucleons would appear. However, in the interacting boson-fermion model ( chapter 16 ) which deals with odd numbers of identical nucleons, bosons are coupled to nucleons. Bosons are taken as states without detailed structure and their symmetry properties result in commutation relations for the corresponding creation- and annihilation operators ( chapter 5 ).

The total spin of a boson is identical with its angular momentum i.e. one does not attribute an intrinsic spin to the bosons. Since the angular momenta of the bosons are even ( $l$ = 0, 2 ) their parity is positive. Although plausible arguments exist for these angular momenta mentioned in the foregoing chapter, this choice is arbitrary and constitutes a typical characteristic of the theory ( however, exotic variants have been developed with $l$ = 4 or odd values ). Only the success achieved by describing real nuclei justifies the assumption for the angular momenta.

The models IBM1 and IBM2 are restricted to nuclei with even numbers of protons and neutrons. In order to fix the number of bosons one takes into account that both types of nucleons constitute closed shells with particle numbers ..28, 50, 82 and 126 ( magic numbers ). Provided that the protons fill less than half of the furthest shell the number of the corresponding active protons has to be divided by two in order to obtain the boson number $N_p$ attributed to protons. If more than half of the shell is occupied the boson number reads $N_p$ = ( number of holes for protons )/2. By treating the neutrons in an analogous way, one obtains their number of bosons $N_n$ . In the IBM1 the boson number $N$ is calculated by adding the partial numbers i.e. $N = N_p + N_n$ . For example the nucleus $^{118}_{54}Xe_{64}$ shows the numbers $N_p$ = (54 - 50)/2 = 2, $N_n$ = (64 -50)/2 = 7 and for $^{128}_{54}Xe_{74}$ the values $N_p$ = (54 - 50)/2 = 2, $N_n$ = (82 - 74)/2 = 4 hold. Electromagnetic transitions don't alter the boson number but transfers of two identical nucleons lift or lower it by one.

Naturally the IBM has to take into account the fact that every nuclear state has a definite total nuclear angular momentum $J$ or rather that the eigenvalue of the angular momentum operator $\boldsymbol{J}^2$ is $J(J + 1)\cdot\mathbf{h}$. $J$ is an integer.

A boson interacts with the inert core of the nucleus ( having closed shells ) from which results its single boson energy $e$. Three-boson interactions are excluded in analogy with the assumptions of the standard shell model. In contrast to the collective model, in the IBM one does not obtain a semiclassical, vivid picture of the nucleus but one describes the algebraic structure of the Hamiltonian operator and of the states, for which reason it is named an algebraic model.

# 3 Many-body configurations

At the beginning of this chapter the representation of boson configurations will be outlined and in the second section completely symmetric states of a few *d*-bosons will be formulated explicitly. In the end the rules are put together which hold for the collective states in the seniority scheme. They are compared with the results of section 3.2. In this chapter vector coupling technique is being applied, which is reviewed in the appendices A1 up to A3.

## 3.1 Many-boson states

Here we introduce a formulation of completely symmetric states of $N$ bosons of which $n_d$ have a *d*-state and $N$ - $n_d$ bosons are in the *s*-state. Besides the total angular momentum $J$ and its projection $M$, for the most part additional ordering numbers are required in order to describe the collective state. One of these numbers is the seniority $t$, after which the most usual representation scheme is named.

For the moment we are leaving out the additional ordering number and write the completely symmetric configuration symbolically as

$$| \, (s^{N\text{-}n_d})^{(0)}{}_0 \, (d^{\,n_d})^{(J)}{}_M, \, J\,M \, \rangle. \tag{3.1}$$

The *s*-bosons are coupled to a $J$ = 0-state. In detail, the *d*-boson part is composed of single *d*-boson states having the angular momentum components 2, 1, 0, -1 and -2. These five single boson states $d_m$ appear in linear combinations as will be shown in the next section. The expression (3.1) is normalised to one.

## 3.2 Symmetric states of two and three *d*-bosons

In this section the *s*-bosons are left out of consideration and we will deal with the symmetrisation of configurations with a small number $n_d$ of *d*-bosons.

First we take $n_d$ = 2. According to the relation (A2.7) the configuration $|\, d^{\,2}, J\,M \,\rangle$ is symmetrical by itself if $J$ is an even number. It has the form (A1.1)

$$|\, d^{\,2}, J\,M \,\rangle = |\, \textstyle\sum_{m1\;m2} (2 \; m_1 \; 2 \; m_2 \,|\, J\,M\,) \; d_{m1} \, d_{m2} \,\rangle \;^{o}\; |\, [\, d \times d \,]^{(J)}{}_M \,\rangle, \quad J = 0, 2, 4. \tag{3.2}$$

In order to obtain a three-boson state we couple one *d*-boson to a boson pair which has an even angular momentum $J_0$ i.e. we form

$$|\, d^{\,3}, J_0 \, J\,M \,\rangle \;^{o}\; |\, [[\, d \times d \,]^{(J_0)} \times d \,]^{(J)}{}_M \,\rangle. \tag{3.3}$$

This expression is considered as a fully symmetrical three-*d*-boson state, which is obtained by carrying out a transposition procedure. In order to formulate this method, temporarily we are regarding bosons as distinguishable and we attribute an individual number to each single boson state. Supposing that such a state is described by a wave function, we have to label every variable with this boson number. We make use of the relation $[\, d(1) \times d(2)]^{(J_0)} = [\, d(2) \times d(1)\, ]^{(J_0)}$, which holds for an even $J_0$ according to ( A2.5 ) and ( A2.7 ). Starting from the partially symmetric ( p. s. ) form

$$|\, [\, d(1) \times d(2)\, ]^{(J_0)} \times d(3)\, ]^{(J)}{}_M \rangle_{\text{p.s.}},$$

we obtain a symmetric three-boson state by adding two analogous forms in which the last $d$-boson is substituted as follows

$$A^{-1}\, |\, [[\, d \times d\, ]^{(J_0)} \times d\, ]^{(J)}{}_M \rangle = |\, [[\, d(1) \times d(2)\, ]^{(J_0)} \times d(3)\, ]^{(J)}{}_M \rangle_{\text{p.s.}} +$$

$$|\, [[\, d(1) \times d(3)\, ]^{(J_0)} \times d(2)\, ]^{(J)}{}_M \rangle_{\text{p.s.}} + |\, [[\, d(3) \times d(2)\, ]^{(J_0)} \times d(1)\, ]^{(J)}{}_M \rangle_{\text{p.s.}}\,. \qquad (3.4)$$

$A$ is the normalisation factor of the right hand side of (3.4). This expression is symmetric because one reproduces it by interchanging two boson numbers ( for example 2 and 3 ). In the last but one term, we can interchange $d(1)$ and $d(3)$ because it is partially symmetric. We employ the recoupling procedure (A3.3) and (A3.6) to the last two terms in (3.4) and obtain

$$A^{-1}\, |\, [[\, d \times d\, ]^{(J_0)} \times d\, ]^{(J)}{}_M \rangle = |\, [[\, d(1) \times d(2)\, ]^{(J_0)} \times d(3)\, ]^{(J)}{}_M \rangle_{\text{p.s.}} + \qquad (3.5)$$

$$\Sigma_{J'}\, (-1)^J \cdot \surd(2J_0 + 1) \cdot \surd(2J' + 1) \cdot \{^2{}_2\ {}^2{}_J\ {}^{J_0}{}_{J'}\} \cdot |\, [\, d(3) \times [\, d(1) \times d(2)\, ]^{(J')}]^{(J)}{}_M \rangle_{\text{p.s.}} +$$

$$\Sigma_{J'}\, (-1)^J \cdot \surd(2J_0 + 1) \cdot \surd(2J' + 1) \cdot \{^2{}_2\ {}^2{}_J\ {}^{J_0}{}_{J'}\} \cdot |\, [\, d(3) \times [\, d(2) \times d(1)\, ]^{(J')}]^{(J)}{}_M \rangle_{\text{p.s.}}.$$

We now interchange $d(2)$ and $d(1)$ in the last term of (3.5), which yields the factor $(-1)^{J'}$ ((A1.4)). Both sums are added then, through which all terms with odd values $J'$ disappear. In the resulting sum we interchange $d(3)$ and $[\, d(2) \times d(1)\, ]^{(J')}$, which annihilates the factor $(-1)^J$ according to (A1.4), because $J'$ is even. For formal reasons the first term on the right hand side of (3.5) is replaced by $\Sigma_{J'\text{even}}\, d_{J'J_0}\, |\, [[\, d(1) \times d(2)\, ]^{(J')} \times d(3)\, ]^{(J)}{}_M \rangle_{\text{p.s.}}$. One obtains

$$A^{-1}\, |\, [[\, d \times d\, ]^{(J_0)} \times d\, ]^{(J)}{}_M \rangle = \qquad (3.6)$$

$$\Sigma_{J'\text{even}}\, (d_{J'J_0} + 2 \cdot \surd(2J_0 + 1) \cdot \surd(2J' + 1) \cdot \{^2{}_2\ {}^2{}_J\ {}^{J_0}{}_{J'}\})\, |\, [[\, d(1) \times d(2)\, ]^{(J')} \times d(3)\, ]^{(J)}{}_M \rangle.$$

The normalisation factor $A$ reads

$$A = (3 + 6(2J_0 + 1)\, \{^2{}_2\ {}^2{}_J\ {}^{J_0}{}_{J_0}\})^{-1/2}. \qquad (3.7)$$

It's a good exercise to derive this expression explicitly. The state $|\, [[\, d \times d\, ]^{(J_0)} \times d\, ]^{(J)}{}_M \rangle$ is regarded as normalised to one. Analogously to the two-boson states (A1.9) here the partially symmetric states $|[[d(1) \times d(2)]^{(J')} \times d(3)]^{(J)}{}_M\rangle$ with different $J'$ are orthogonal to each other. We employ a slightly modified form of (3.5)

$$A^{-1}\, |\, [[\, d \times d\, ]^{(J_0)} \times d\, ]^{(J)}{}_M \rangle = |\, [[\, d(1) \times d(2)\, ]^{(J_0)} \times d(3)\, ]^{(J)}{}_M \rangle_{\text{p.s.}} +$$

$$\surd(2J_0 + 1) \cdot \Sigma_{J'}\, (1 + (-1)^{J'}) \surd(2J' + 1) \cdot \{^2{}_2\ {}^2{}_J\ {}^{J_0}{}_{J'}\} \cdot |\, [[\, d(1) \times d(2)\, ]^{(J')} \times d(3)\, ]^{(J)}{}_M \rangle_{\text{p.s.}}\,.$$

and make up the following equation

$$A^{-2}\langle[[d\times d]^{(J_0)}\times \mathrm{d}]^{(J)}{}_{\mathrm{M}} \,|\, [[\,d\times d\,]^{(J_0)}\times d\,]^{(J)}{}_M\rangle = A^{-2} =$$

$$1 + 2\cdot 2\cdot(2J_0 + 1)\begin{Bmatrix}2 & 2 & J_0\\ 2 & J & J_0\end{Bmatrix} + (2J_0 + 1)\,\sum_{J'}(2 + 2\cdot(-1)^{J'})(2J' + 1)\begin{Bmatrix}2 & 2 & J_0\\ 2 & J & J'\end{Bmatrix}^2.$$

Due to (A3.8) and (A3.9) the relations

$$\sum_{J'}(2J' + 1)\begin{Bmatrix}2 & 2 & J_0\\ 2 & J & J'\end{Bmatrix}^2 = (2J_0 + 1)^{-1} \text{ and}$$

$$\sum_{J'}(-1)^{J'}(2J' + 1)\begin{Bmatrix}2 & 2 & J_0\\ 2 & J & J'\end{Bmatrix}^2 = \begin{Bmatrix}2 & 2 & J_0\\ 2 & J & J_0\end{Bmatrix} \text{ hold, from which we derive}$$

$$A^{-2} = 1 + 2 + (2 + 4)(2J_0 + 1)\begin{Bmatrix}2 & 2 & J_0\\ 2 & J & J_0\end{Bmatrix}, \text{ which is in agreement with (3.7).}$$

We now look into the $J$-values of symmetric three $d$-boson states represented in (3.6).

The case $J = 0$ is of some importance in the seniority scheme. The number of triplets with $J = 0$ is named $n_D$ i.e. in this case we have $n_D = 1$.

For $J \neq 0$ we insert the numerical values of the 6-$j$ symbols (A3.12 - 14) in the equation (3.6). For $J = 1$ the partial vectors can only show $J_0 = J' = 2$ and the expression (3.6) vanishes. For $J = 2$ the values $J_0 = 0, 2, 4$ have to be considered and the calculation yields

$$|\,[[\,d\times d\,]^{(0)}\times d\,]^{(2)}{}_M\rangle = |\,[[\,d\times d\,]^{(2)}\times d\,]^{(2)}{}_M\rangle = |\,[[\,d\times d\,]^{(4)}\times d\,]^{(2)}{}_M\rangle\,.$$

We take a special interest in states with $J_0 = 0$, that is why we treat the state ($J_0 = 2$, $J = 2$) mentioned above as equivalent to ($J_0 = 4$, $J = 2$) and to ($J_0 = 0$, $J = 2$) . Therefore we say, the configuration ($J_0 \neq 0$, $J = 2$) does not exist. In a similar way we see that the $J = 3$-states ($J_0 = 2, 4$) differ only in their signs. Both $J = 4$-states ($J_0 = 2, 4$) are identical. For $J = 5$ ($J_0 = 4$) the expression (3.6) vanishes. $J = 6$ characterises the so-called "stretched" state.

### 3.3 The seniority scheme, rules defining *J*

General symmetric states of $d$-bosons are constructed by vector coupling and complete symmetrisation using group theory ( Hamermesh, 1962 ), ( Bayman and Landé, 1966 ). Here we have a look at the seniority scheme, which is the most common version of this representation. The configuration of $n_d$ $d$-bosons is written as follows

$$|\,n_d\,,\,([d\times d]^{(0)}{}_0)^{n_p}\cdot([[d\times d]^{(2)}\times d]^{(0)}{}_0)^{n_D}\cdot(d^{l})^{(J)}{}_M\rangle\,. \qquad (3.8)$$

In the expression (3.8) the doublet $[d\times d]^{(0)}{}_0$ with angular momentum 0 appears $n_p$ times and the triplet $[[d\times d]^{(2)}\times d]^{(0)}{}_0$ exists $n_D$ times. The $l$ remaining $d$-bosons constitute a configuration with the total angular momentum $J$ ($M$) which contains neither a doublet nor a triplet with $J = 0$. Therefore the number of $d$-bosons reads $n_d = 2n_p + 3n_D + l$. The number $t = n_d - 2n_p = 3n_D + l$, which is left over after subtracting the doublets, is named seniority analogously to the description in the shell model. We name the configuration $(d^{\,l})^{(J)}{}_M$ the "reduced" state of the $l$ bosons. It is defined unambiguously by $l$, $J$ and $M$ (Talmi, 1993,

S. 763). Its total angular momentum $J$ is identical with the one of the whole configuration (3.8).

**In the seniority scheme the *d*-boson configurations are defined by the numbers** $n_d$, $n_p$, $n_D$, $J$, ($M$).

There exist restrictions for the $J$-values. It can be shown that in a "reduced" state of $l$ $d$-bosons the following values are permitted

$$J = l,\ l + 1,\ \dots,\ 2l - 3,\ 2l - 2,\ 2l, \tag{3.9}$$

i. e. $J < l$ and $J = 2l - 1$ are inadmissible. $J = 2l$ represents the "stretched" state.

The exclusion of $J = 2l - 1$ in (3.9) can be explained in the following way. We know that for the "stretched", symmetric and to $z$ orientated state of $l$ $d$-bosons the relation $J = M = 2l$ holds. We now construct the symmetric state with $M = 2l - 1$ and represent it using numbered bosons whose projections of the angular momentum is $m$ :

$$\begin{aligned} |(d^{\,l})_{M = 2l - 1}\rangle = A\ |(&d(1)_{m=1}\cdot d(2)_{m=2}\cdot d(3)_{m=2}\cdot\ \dots\ \cdot d(l)_{m=2} \\ &+\ d(1)_{m=2}\cdot d(2)_{m=1}\cdot d(3)_{m=2}\cdot\ \dots\ \cdot d(l)_{m=2} \\ &+\ \dots\dots\dots\dots \\ &+\ d(1)_{m=2}\cdot d(2)_{m=2}\cdot d(3)_{m=2}\cdot\ \dots\ \cdot d(l)_{m=1})\rangle . \end{aligned} \tag{3.10}$$

$A$ is the normalisation constant. The expression (3.10) reveals that there exists only one state with $M = 2l - 1$. On the other hand, if one turns the "stretched" state ( with $J = 2l$ ) relative to the z-axis in order to obtain the projection $M = 2l - 1$, the resulting state is still symmetric and must agree with the one of (3.10) because this is unique. For the same reason a state with $J = 2l - 1$ is not allowed because its maximal projection would be $M = 2l - 1$ which must not occur twice.

We now verify the rule (3.9) inspecting the boson states (3.2) and (3.6). For $l = 2$ the "reduced" state reads $|[d \times d]^{(J)}_M\rangle$ with $J \neq 0$. Owing to (3.9) only the values $J = 2, 4$ have to be considered which is in agreement with (3.2). For the "reduced" state with $l = 3$ according to (3.9) the values $J = 0, 1, 2$ are ruled out. In fact the discussion of equation (3.6) showed that $J = 1$ does not appear and that both other cases are equivalent to $J_0 = 0$ which is inconsistent with the term "reduced" state. The rule (3.9) excludes $J = 5$ which has been found to be true for $l = 3$. Thus, for $l = 2$ and 3 the selection rule (3.9) is confirmed.

In table 3.1 for several boson numbers $n_d$ the allowed values $n_p$ and $n_D$ are given. Accompanying values for $t$, $l$ and $J$ are in the columns 3, 5 and 6.

Table 3.1. Classification of the *d*-boson configurations in the seniority scheme.
$n_d$ : number of *d*-bosons,
$n_p$ : number of boson pairs with total angular momentum 0,
$n_D$ : number of boson triplets with total angular momentum 0,
$t$ : seniority, $l$ : number of bosons in the "reduced" state,
$J$ : total angular momentum

| $n_d$ | $n_p$ | $t = n_d - 2n_p$ | $n_D$ | $l = t - 3n_D$ | $J$ |
|---|---|---|---|---|---|
| 2 | 0 | 2 | 0 | 2 | 2, 4 |
| 2 | 1 | 0 | 0 | 0 | 0 |
| 3 | 0 | 3 | 0 | 3 | 3, 4, 6 |
| 3 | 0 | 3 | 1 | 0 | 0 |
| 3 | 1 | 1 | 0 | 1 | 2 |
| 4 | 0 | 4 | 0 | 4 | 4, 5, 6, 8 |
| 4 | 0 | 4 | 1 | 1 | 2 |
| 4 | 1 | 2 | 0 | 2 | 2, 4 |
| 4 | 2 | 0 | 0 | 0 | 0 |
| . | . | . | . | . | . |
| . | . | . | . | . | . |
| 7 | 0 | 7 | 0 | 7 | 7,8,9,10,11,12,14 |
| 7 | 0 | 7 | 1 | 4 | 4, 5, 6, 8 |
| 7 | 0 | 7 | 2 | 1 | 2 |
| 7 | 1 | 5 | 0 | 5 | 5,6,7,8,10 |
| 7 | 1 | 5 | 1 | 2 | 2, 4 |
| 7 | 2 | 3 | 0 | 3 | 3, 4, 6 |
| 7 | 2 | 3 | 1 | 0 | 0 |
| 7 | 3 | 1 | 0 | 1 | 2 |
| . | . | . | . | . | . |

Table 3.1 shows that for given $n_d > 3$ some angular momenta $J$ appear in more than one configuration. The value $J = 1$ is absent in the whole spectrum. Clearly it is missing also for $l = 1$ because this simplest "reduced" state consists of a single *d*-boson.

Among states with several *d*-bosons it happens that configurations with equal ( $n_d$, $t$, $J$ )-values differ in the quantity $n_D$ and are not orthogonal to one another. They have to be orthogonalised with the help of the well-known Schmidt procedure. By doing it, the number $n_D$ looses its character of an ordering number and it has to be replaced by an arbitrarily defined index.

Many-boson configurations in the seniority scheme stand out because they are eigenfunctions of the vibrational limit of the Hamilton operator ( chapter 10 and section 14.4 ). Since this special case correlates with the Lie algebra *u*(5) the states of the seniority scheme in addition are named *u*(5)-basis. "Spherical basis " is a further customary name. Besides this scheme there exist two less often used representations which are eigenfunctions of other limits of the Hamiltonian ( chapter 14 ).

# 4 Many-boson states with undefined angular momentum

In this chapter we will deal with the simplest representation of many-boson states. It is formed as a product of single-boson state functions. Symmetry is achieved by permuting the boson indices and adding up the resulting expressions. In contrast to (3.8), the single angular momenta are not coupled to a definit total quantity. We nam this representation "primitive" basis.

In the first section we will look into such representations of a few bosons and relate them to the formulations with defined angular momentum. The symmetric primitive representation of a multitude of different single-boson states is treated in the second section.

### 4.1 Two- and three-*d*-boson states

By way of introduction, we treat the "primitive" states of a few *d*-bosons. The single *d*-boson state is characterized by $d_m(n)$. The quantity *m* is the projection of the angular momentum *l* = 2 and *n* is the number of the bosons. Provisionally we are considering the bosons as distinguishable.

The symmetric and normalised two-*d*-boson state of this kind is represented by $| d\, m\, d\, m' \rangle$, in which single boson states such as $d_m(i)$ and $d_{m'}(k)$ are involved. It has the form

$$| d\, m\, d\, m' \rangle = (2(1 + \delta_{mm'}))^{-1/2}\, | d_m(1)\, d_{m'}(2) + d_m(2)\, d_{m'}(1) \rangle. \qquad (4.1)$$

Interchanging 1 and 2 has no influence and $\langle d\, m\, d\, m' | d\, m\, d\, m' \rangle = 1$ holds i.e. the expression (4.1) is symmetric and normalised.

Correspondingly, "primitive" three-*d*-boson states are constructed by summing up $| d_m(1)\, d_{m'}(2)\, d_{m''}(3) \rangle$, $| d_m(1)\, d_{m'}(3)\, d_{m''}(2) \rangle$ and $| d_m(2)\, d_{m'}(1)\, d_{m''}(3) \rangle$ etc and by normalising. One writes

$$| \mathrm{d}\, m\, d\, m'\, d\, m'' \rangle = A_p(m, m', m'')\cdot\Sigma_{\text{all permutations}}| d_m(1)\, d_{m'}(2)\, d_{m''}(3) \rangle. \qquad (4.2)$$

The permutations concern the numbers 1, 2 and 3. The normalisation factor $A_p(m, m', m'')$ is $1/\sqrt{6}$ if all quantities *m* are different. Supposing that only two of them agree, $A_p = 1/(2\sqrt{3})$ holds and for three identical single boson states we have $A_p = 1/6$.

Before we turn to the general representation of the "primitive" basis we show the connection between the states (4.1-2) and those of the seniority scheme (3.2), (3.3) and (3.8) whose angular momenta are good quantum numbers.

According to equation (3.2) the symmetric two-boson state of this kind reads

$$| d^2\, J\, M \rangle \equiv |[d \times d]^{(J)}{}_M \rangle = \Sigma_{m\, m'} (2\, m\, 2\, m' | J\, M)\, | d_m(1)\, d_{m'}(2) \rangle. \qquad (4.3)$$

The angular momentum $J$ is even. We introduce an operator $\boldsymbol{S}$ which interchanges in equation (4.3) the indices 1 and 2 and adds the new expression to the original one. Due to the symmetry in (4.3) the following relations hold

$$| d^2 J M \rangle = ½\cdot \boldsymbol{S} |d^2 J M \rangle =$$

$$½\cdot \sum_{m \neq m'} (2\ m\ 2\ m' |J\ M)\ (\sqrt{2}/\sqrt{2})\cdot|d_m(1)\ d_{m'}(2) + d_m(2)\ d_{m'}(1)\rangle +$$

$$½\cdot(2\ M/2\ 2\ M/2\ |J\ M)\cdot 2\cdot|d_{M/2}(1)\ d_{M/2}\ (2)\rangle\cdot\delta_{M\ \mathrm{even}} =$$

$$\sum_{m \neq m'} (2\ m\ 2\ m' |J\ M)(1/\sqrt{2})\cdot|d\ m\ d\ m'\rangle +$$

$$(2\ M/2\ 2\ M/2\ |J\ M)\cdot|d\ M/2\ d\ M/2\rangle\cdot\delta_{M\ \mathrm{even}} =$$

$$\sum_{mm'} \sqrt{((1 + \delta_{mm'})/2)}\ (2\ m\ 2\ m'\ |\ J\ M)\cdot|d\ m\ d\ m'\rangle \qquad (4.4)$$

The last expression in (4.4) is built with "primitive" states introduced in (4.1).

The symmetric states of three-$d$-bosons with defined angular momentum $J$ can be treated in a similar way. According to (3.4) we have

$$A^{-1} |\ d^3\ J_0\ J\ M\rangle = |\ [[d(1) \times d(2)]^{(J_0)} \times d(3)]^{(J)}_M\ \rangle +$$

$$|\ [[d(2) \times d(3)]^{(J_0)} \times d(1)]^{(J)}_M\ \rangle +$$

$$|\ [[d(3) \times d(1)]^{(J_0)} \times d(2)]^{(J)}_M\ \rangle. \qquad (4.5)$$

The quantity $J_0$ is even and the normalisation factor $A$ is given in (3.7). Using the Clebsch-Gordan coefficients defined in (A1.1) we rewrite (4.5) and in the following step we insert the "primitive" representation of (4.2) . We obtain

$$A^{-1}\ |\ d^3\ J_0\ J\ M\rangle = \sum_{M_0 m''} (J_0\ M_0\ 2\ m''\ |\ J\ M)\cdot \qquad (4.6)$$

$$\sum_{mm'}\ (2\ m\ 2\ m'\ |\ J_0\ M_0)\cdot ½\ \sum_{\text{all permutations}} |\ d_m(1)\ d_{m'}(2)\ d_{m''}(3)\ \rangle =$$

$$\sum_{M_0 m''} (J_0\ M_0\ 2\ m''\ |J\ M)\cdot\sum_{mm'}\ (2\ m\ 2\ m'\ |\ J_0\ M_0)\cdot½\cdot A_p(m,m',m'')^{-1}\cdot|\ d\ m\ d\ m'\ d\ m''\ \rangle.$$

**In (4.4) and (4.6) the states of the seniority scheme are written as a linear combination of "primitive" states**. It's obvious that this is true for all many-boson states. This stimulates further research on the "primitive" basis. In doing so, we will encounter the well-known creation- and annihilation operators for bosons ( chapter 5 ), which facilitate an elegant representation of the boson interactions.

### 4.2 General "primitive" many-boson states

In the interacting boson model the bosons occupy the following six single states $s$, $d_2$, $d_1$, $d_0$, $d_{-1}$ and $d_{-2}$ . For the sake of simplicity, we have a look at an arbitrary number of single boson states, which we represent by $\psi_a$, $\psi_b$, .. , $\psi_f$, .. . In essence, we follow the report of Landau (86).

A single boson state can be occupied by several bosons. The number of bosons which are in the single state $y_f$ is named $N_f$ . The total number of bosons $N$ amounts to the sum of these partial numbers

$$N = N_a + N_b + ... + N_f + .. \quad . \tag{4.7}$$

The ansatz for the state $Y$ of $N$ bosons is written as a product of $N$ single boson states which contains $N_a$ times the factor $y_a$, $N_b$ times the factor $y_b$, etc. Each single state is designed by a different boson number as follows

$$Y(N_a, N_b, .. ,N_f, .) = P_{i=1}^{N_a} y_a(i) \cdot P_{i=Na+1}^{N_a+N_b} y_b(i) \cdot ... \cdot P_{i=Ne+1}^{N_a+..+N_f} y_f(i) \cdot ... \ . \tag{4.8}$$

The index $i$ in the partial products marks the bosons occupying the affiliated single state. The function $Y$ does not alter if the labels of bosons are interchanged which belong to the same single state. In order to obtain complete symmetry we interchange bosons in all possible ways, which belong to different single states. All these configurations are generated by arranging the $N$ numbered bosons in the single states $a, b, ..,f,..$ with the partial numbers $N_a$, $N_b$, .. ,$N_f$ ..in all different ways. In other words, we have here the combinatorial task to calculate how often $N$ different balls can be put in vessels dissimilarly in such a way that the first one contains $N_a$ balls, the second one $N_b$ balls, etc. From the combinatorial analysis we take that the number of possibilities or permutations reads

$$N!/(N_a!\cdot N_b!\cdot .. N_f!\cdot ..). \tag{4.9}$$

We label these permutations by the index $r$ and introduce the operator $\mathbf{P'}_r$ which performs the $r$th permutation when acting on the state $Y(N_a, N_b,.., N_f, .)$. We will show that the state

$$A^{-1} \mid N_a, N_b, ... ,N_f, .. \rangle = \sum_r \mathbf{P'}_r Y(N_a, N_b,.., N_f,..) \tag{4.10}$$

is completely symmetric with regard to interchanging numbered bosons. We can see this by interchanging two bosons which belong to different single states. Since every summand in (4.10) has a counterpart which contains the mentioned bosons in the interchanged positions, the transposition has no effect on the whole sum. This is true for all kinds of exchange.

We now turn to the normalisation constant $A$ in (4.10). We assume that the single states $y_f(i)$ depend on variables which we represent by the symbol $x_i$ . We regard these states as orthonormalised i. e.

$$\int y_g{}^*(x_i) y_f(x_i)\, dx_i = d_{gf} \ . \tag{4.11}$$

We maintain that all functions $\mathbf{P'}_r Y(N_a, N_b,.., N_f, .)$ are orthonormal with respect to each other as follows

$$\int [\, \mathbf{P'}_r Y(N_a, N_b,.., N_g,..)]^* \, \mathbf{P'}_{r'} \, Y(N_a, N_b,.., N_f, .) dx_1 .. dx_N = d_{r r'}. \tag{4.12}$$

At least one boson ( say the $i$th ) can be found, namely, which is assigned to different single states ( for instance the states $f$ and $g$ ) in the functions $\boldsymbol{P'_r} Y(N_a, N_b,.., N_g,..)$ and $\boldsymbol{P'_{r'}}\ Y(N_a, N_b,.., N_f, .)$. As a result, among other things the partial integral $\int y_g{*}(x_i)\,y_f(x_i)dx_i$ $(g\ ^{1}\ f)$ vanishes. With that, the orthogonality of the functions $\boldsymbol{P'_r} Y(N_a, N_b,.., N_f,..)$ is shown. Taking into account the number of special permutations $\boldsymbol{P'_{r'}}$ (4.9) and using (4.12) we obtain

$$\langle\ N_a, N_b,.., N_f, .\ |\ N_a, N_b,.., N_f,..\ \rangle = A^2 \cdot N!/(N_a! \cdot N_b! \cdot\ ..\ N_f! \cdot\ ..). \qquad (4.13)$$

The left-hand side of (4.13) has to be 1 i. e.

$$A = (N_a! \cdot N_b! \cdot\ ..\ N_f! \cdot\ ../N!)^{½} \qquad (4.14)$$

and $$|\ N_a, N_b,.., N_f,..\ \rangle = (N_a! \cdot N_b! \cdot\ ..\ N_f! \cdot\ ../N!)^{½} \cdot \sum_r \boldsymbol{P'_r} Y(N_a, N_b,.., N_f,..). \qquad (4.15)$$

This function describes the " primitive" state of $N$ bosons with single states *a*, *b*, .. , *f*, .. containing the boson numbers $N_a, N_b,.., N_f$ , .. . For $N$ = 2 and 3 one can show directly the agreement between (4.15) and (4.1) or (4.2). Doing so, one has to take into account that in (4.2) **all** permutations appear.

# 5 Operators and matrix elements

The interacting boson model and other many-particle models deal with operators which act on single boson states or on pair states. Because in our model the representation of the collective state is symmetric for all $N$ bosons the mentioned operators cannot be directed towards individual, labelled bosons. However, they act on the collective of bosons and correspondingly they have a symmetric form. The single-boson operator $\boldsymbol{F}^{(1)}$ reads

$$\boldsymbol{F}^{(1)} = \sum_{i=1}^{N} \boldsymbol{f}^{(1)}(\underline{x}_i). \tag{5.1}$$

The operator $\boldsymbol{f}^{(1)}$ acts on a single boson which we consider to be distinguishable provisionally and to which we assign the index $i$. The argument $\underline{x}_i$ describes vectorial, local and other variables of the $i$th boson on which $\boldsymbol{f}^{(1)}(\underline{x}_i)$ acts. For example, the operator of the kinetic energy or the operator of a fixed potential has the form of $\boldsymbol{F}^{(1)}$.

The two-boson operator reads

$$\boldsymbol{F}^{(2)} = \sum_{i>j=1}^{N} \boldsymbol{f}^{(2)}(\underline{x}_i,\underline{x}_j). \tag{5.2}$$

The partial operator $\boldsymbol{f}^{(2)}$ acts on the variables of a boson pair. For example, the potential operating in twos between bosons has the form (5.2). In simple models $\boldsymbol{f}^{(2)}$ is only a function of the difference $|\underline{x}_i - \underline{x}_j|$. The Hamilton operator, which is decisive for quantum mechanical problems, contains the operator types (5.1) and (5.2) in a linear combination.

In the following section we will deal with matrix elements of single boson operators using the representation of the “primitive“ basis. Almost unavoidably, one reaches the definition of creation and annihilation operators ( section 5.2 ). In the third section we will show how these operators contribute to the form of the matrix elements.

## 5.1 Matrix elements of the single-boson operator

In order to calculate eigenfunctions and eigenvalues of the Hamilton operator, matrix elements of the operators (5.1) and (5.2) have to be made up. Represented in terms of the “primitive“ basis these matrix elements have the following form

$$\langle N_a, N_b, \ldots ,N_f, .. \,|\, \boldsymbol{F} \,|\, N_a', N_b', \ldots ,N_f', .. \rangle\ . \tag{5.3}$$

First, we deal with the single-boson operator $\boldsymbol{F}^{(1)}$. For the sake of simplicity we admit only two single states $a$ and $b$. In the matrix element

$$\langle N_a, N_b \,|\, \boldsymbol{F}^{(1)} \,|\, N_a', N_b' \rangle \tag{5.4}$$

the condition $N_a + N_b = N_a' + N_b'$ holds because both total states belong to the same basis only if the total number of bosons is equal. We begin with the **diagonal** matrix element, of which both functions are identical as follows

$$\sum_{i=1}^{N} \langle N_a, N_b \,|\, \boldsymbol{f}^{(1)}(\underline{x}_i) \,|\, N_a, N_b \rangle. \qquad (5.5)$$

First we take $i = 1$ and ask how many terms exist in the state $|\, N_a, N_b \rangle$ according to (4.15) and (4.9) in which the first boson is in the single state $a$. It is the number of permutations $\boldsymbol{P}_r'$ ( see section 4.2 ) of the residual $N – 1$ bosons. Therefore there are $(N - 1)!/((N_a - 1)!N_b!)$ terms in the state function $|\, N_a, N_b \rangle$ with the first boson in the state $a$. In this case in (5.5) the states $y_a(\underline{x}_1)$ and $y_a^*(\underline{x}_1)$ constitute a separate integral together with $\boldsymbol{f}^{(1)}$ . The integration over the residual factors containing the variables $\underline{x}_2, \underline{x}_3, .. \underline{x}_N$ yields the value 1 ( apart from the normalisation constant ) if both factors agree in the allocation of bosons, otherwise it vanishes. Therefore the part of the diagonal matrix element in which the first boson is in the state $a$ reads

$$\sqrt{(N_a!N_b!/N!)}\cdot\sqrt{(N_a!N_b!/N!)}\cdot\int y_a^*(\underline{x}_1)\boldsymbol{f}^{(1)}(\underline{x}_1)y_a(\underline{x}_1)\, d\underline{x}_1 \,(N - 1)!/((N_a - 1)!N_b!) =$$

$$(N_a/N) \int y_a^*(\underline{x}_1)\boldsymbol{f}^{(1)}(\underline{x}_1)y_a(\underline{x}_1)\, d\underline{x}_1 \,. \qquad (5.6)$$

The first factors in (5.6) are normalisation factors according to (4.14). Furthermore, in diagonal matrix elements the first boson must not appear in different states ( $a$ and $b$ ) together with $\boldsymbol{f}^{(1)}(\underline{x}_1)$ in the partial integral, because then the residual states are orthogonal. On the other hand, one obtains an expression analogous to (5.6) if the first boson is twice in the state $b$ . Consequently for $i = 1$ the diagonal matrix element reads

$$N^{-1}\cdot(N_a \cdot\int y_a^*(\underline{x}_1)\boldsymbol{f}^{(1)}(\underline{x}_1)y_a(\underline{x}_1)\, d\underline{x}_1 + N_b \cdot\int y_b^*(\underline{x}_1)\boldsymbol{f}^{(1)}(\underline{x}_1)y_b(\underline{x}_1)\, d\underline{x}_1). \qquad (5.7)$$

Since according to (5.1) the operator $\boldsymbol{F}^{(1)}$ contains a sum including all $i$ $(1 £ i £ N)$ and because the expression

$$\int y_f^*(\underline{x}_i)\boldsymbol{f}^{(1)}(\underline{x}_i)y_f(\underline{x}_i)\, d\underline{x}_i \;o\; f_{ff}^{(1)} \qquad (5.8)$$

does not depend on the boson number $i$, the expression (5.7) appears $N$ times and the diagonal element reads

$$\langle N_a, N_b \,|\, \boldsymbol{F}^{(1)} \,|\, N_a, N_b \rangle = N_a f_{aa}^{(1)} + N_b f_{bb}^{(1)}. \qquad (5.9)$$

If several single states $a$, $b$, $c$, .. , $f$, .. are involved, correspondingly we have

$$\langle N_a, N_b, .., N_f, ..|\, \boldsymbol{F}^{(1)} \,|\, N_a, N_b, .., N_f, ..\rangle = N_a f_{aa}^{(1)} + N_b f_{bb}^{(1)} + .. + N_f f_{ff}^{(1)}.. \qquad (5.10)$$

We now deal with **non-diagonal** matrix elements of the single-boson operator $\boldsymbol{F}^{(1)}$. Again we restrict the single states to $a$ and $b$. We bring the total states in the following sequence in which the occupation numbers $N_a$ and $N_b$ are stressed

| occupation numbers | | |
|---|---|---|
| state *a* | state *b* | |
| 0 | $N$ | |
| 1 | $N$ - 1 | |
| . | . | |
| $N_a$ - 1 | $N_b$ + 1 | (5.11) |
| $N_a$ | $N_b$ | |
| $N_a$ + 1 | $N_b$ - 1 | |
| . | . | |
| $N$ | 0 | |

with $N = N_a + N_b$. It turns out that non-diagonal ( belonging to two different lines in (5.11)) matrix elements are different from zero only if the occupation numbers of one single state on both sides of (5.4) differ in 1. In order to show this we treat

$$\sum_{i=1}^{N} \langle N_a, N_b \,|\, \boldsymbol{f}^{(1)} (\underline{x_i}) \,|\, N_a + 1, N_b - 1 \rangle. \qquad (5.12)$$

We begin also with $i = 1$ and determine the number of summands on the right hand side of (5.12) in which the first boson occupies the **state *a***. Again it results from the number of permutations $\boldsymbol{P_r'}$ of the residual $N$ - 1 bosons i. e. there are $(N - 1)!/(N_a!(N_b - 1)!)$ terms of this kind. On the left hand side of (5.12) the number of terms showing the first boson in the **single state *b*** is $(N - 1)!/(N_a!(N_b - 1)!)$ as well. These are the only terms which we have to take into account if we form the integral in (5.12) in demanding $i = 1$. We separate the factors $y_b^*(\underline{x_1})$, $y_a(\underline{x_1})$ and $\boldsymbol{f}^{(1)}$ from the rest and carry out the integral over $\underline{x_1}$. Analogously to (5.6) the integration of the residual function depending on $\underline{x_2}$, $\underline{x_3}$.. $\underline{x_N}$ yields the value 1 ( apart from the normalisation constant ) if both factors agree in the allocation of bosons which happens $(N - 1)!/(N_a!(N_b - 1)!)$ times. The integral vanishes if the residues disagree. Therefore the part of the matrix element in which the first boson is on the left hand side in the state *b* and on the right hand side in *a* reads as follows

$$\langle N_a, N_b \,|\, \boldsymbol{f}^{(1)} (\underline{x_1}) \,|\, N_a +1, N_b -1 \rangle =$$

$$\sqrt{(N_a!N_b!/N!)}\cdot\sqrt{((N_a +1)!(N_b -1)!/N!)} \int y_b^*(\underline{x_1})\boldsymbol{f}^{(1)}(\underline{x_1})y_a(\underline{x_1})\, d\underline{x_1}\cdot(N -1)!/((N_a!(N_b -1)!) =$$

$$\sqrt{(N_a +1)}\sqrt{N_b}/N\cdot\int y_b^*(\underline{x_1})\boldsymbol{f}^{(1)}(\underline{x_1})y_a(\underline{x_1})\, d\underline{x_1}. \qquad (5.13)$$

We take into account that $\boldsymbol{F}^{(1)}$ contains a sum over all $N$ bosons and that the integral

$$\int y_b^*(\underline{x_1})\boldsymbol{f}^{(1)}(\underline{x_1})y_a(\underline{x_1})\, d\underline{x_1} \;\mathrm{o}\; f_{ba}^{(1)} \qquad (5.14)$$

does not depend on the boson number *i*. Thus we obtain

$$\langle N_a, N_b \,|\, \boldsymbol{F}^{(1)} \,|\, N_a +1, N_b -1 \rangle = \sqrt{(N_a +1)}\sqrt{N_b}\cdot f_{ba}^{(1)} \,. \qquad (5.15)$$

The matrix elements of a basis are arranged usually in a quadratic matrix in which the *x*-direction indicates the increasing order of the right hand side part of (5.4) and the falling *y*-direction is associated with the left hand side state. In thisrepresentation the elements given by (5.15) lie directly beside the diagonal which is why we name them off diagonal. The matrix elements lying directly on the other side of the diagonal are formulated analogously to the equation (5.15).

Matrix elements of the type $\langle N_a, N_b \mid \boldsymbol{F}^{(1)} \mid N_a + 2, N_b - 2 \rangle$ and those which are more distant from the diagonal all vanish because the residual functions mentioned above never agree i. e. they constitute an orthogonal pair. The result of (5.15) remains the same if there are further single states *c*, *d, .. , f,* .. on both sides of the matrix element with corresponding occupation numbers as follows

$$\langle N_a, N_b, N_c, .., N_f, .. \mid \boldsymbol{F}^{(1)} \mid N_a +1, N_b -1, N_c, .., N_f, .. \rangle = \sqrt{(N_a +1)}\sqrt{N_b} \cdot f_{ba}^{(1)}$$

or generally (5.16)

$$\langle N_a, .. , N_d, .., N_g, .. \mid \boldsymbol{F}^{(1)} \mid N_a, .., N_d +1, .., N_g - 1, .. \rangle = \sqrt{(N_d +1)}\sqrt{N_g} \cdot f_{gd}^{(1)}.$$

Thus, in the non-vanishing matrix elements of $\boldsymbol{F}^{(1)}$ at most one occupation number on each side may exceed the corresponding one at most by 1.

## 5.2 Creation and annihilation operators

We employ a trick which permits to formulate the matrix elements of $\boldsymbol{F}^{(1)}$ in a simple way, and to this end, we introduce the annihilation operator $\boldsymbol{b_f}$ which reduces the boson number. Its definition reads: If the annihilation operator $\boldsymbol{b_f}$ acts on the boson state $\mid N_a, N_b, .., N_f, .. \rangle$ this one is replaced by the normalised and completely symmetric state $\mid N_a, N_b, .., N_f - 1, .. \rangle$ in which the occupation number $N_f$ is reduced to $N_f - 1$. Furthermore the factor $\sqrt{N_f}$ has to be attached i. e.

$$\boldsymbol{b_f} \mid N_a, N_b, .., N_f, .. \rangle = \sqrt{N_f} \mid N_a, N_b, .., N_f - 1, .. \rangle. \quad (5.17)$$

We form the following matrix element

$$\langle N_a, N_b, .., N_f - 1, .. \mid \boldsymbol{b_f} \mid N_a, N_b, .., N_f, .. \rangle =$$

$$\langle N_a, N_b, .., N_f - 1, .. \mid N_a, N_b, .., N_f - 1, .. \rangle \cdot \sqrt{N_f} = \sqrt{N_f}. \quad (5.18)$$

It is known that the left hand side of the first expression in (5.18) represents the conjugate complex form of the given state function. Now we go to the conjugate complex of the matrix element (5.18). We postulate that there exists an operator, say $\boldsymbol{b_f}^+$ which generates just this conjugate matrix element if it is placed between the states $\langle N_a, N_b, .., N_f, .. \mid$ and $\mid N_a, N_b, .., N_f - 1, .. \rangle$ i. e.

$$\sqrt{N_f} = \langle N_a, N_b, .., N_f - 1, .. \mid \boldsymbol{b_f} \mid N_a, N_b, .., N_f, .. \rangle^* =$$

$$\langle N_a, N_b, .., N_f, .. \mid \boldsymbol{b_f}^+ \mid N_a, N_b, .., N_f - 1, .. \rangle. \quad (5.19)$$

We learn from (5.19) that the so-called adjoint operator $\boldsymbol{b_f}^+$ has the following effect:

$$\boldsymbol{b_f}^{+} \mid N_a, N_b, .., N_f - 1, ..\rangle = \sqrt{N_f} \mid N_a, N_b, .., N_f, ..\rangle. \qquad (5.20)$$

The operator $\boldsymbol{b_f}^{+}$ raises the occupation number by 1 and attaches the root of the new occupation number as a factor. Consequently, $\boldsymbol{b_f}^{+}$ is named creation operator.

Let's act the operators $\boldsymbol{b_f}^{+}$ and $\boldsymbol{b_p}$ ( $f$ ¹ $p$ ) one after the other:

$$\boldsymbol{b_f}^{+} \boldsymbol{b_p} \mid N_{a\,..}, N_f, .., N_p, ..\rangle =$$

$$\boldsymbol{b_f}^{+} \sqrt{N_p} \mid N_{a\,..}, N_f, .., N_p - 1, ..\rangle = \sqrt{(N_f + 1)}\sqrt{N_p} \mid N_{a\,..}, N_f + 1, .., N_p - 1, ..\rangle, \qquad (5.21)$$

or $$\boldsymbol{b_p}\, \boldsymbol{b_f}^{+} \mid N_{a\,..}, N_f, .., N_p, ..\rangle = \sqrt{(N_f + 1)}\sqrt{N_p} \mid N_{a\,..}, N_f + 1, .., N_p - 1, ..\rangle. \qquad (5.22)$$

If both operators act on the same single boson state ( $f = p$ ) we have

$$\boldsymbol{b_f}^{+}\, \boldsymbol{b_f} \mid N_{a\,..}, N_f, ..\rangle = \boldsymbol{b_f}^{+} \sqrt{N_f} \mid N_{a,\,..}, N_f - 1, ..\rangle = N_f \mid N_{a,\,..}, N_f, ..\rangle. \qquad (5.23)$$

In (5.23) the total state remains unchanged but $\boldsymbol{b_f}^{+} \boldsymbol{b_f}$ attaches the factor $N_f$ i. e. the effect of the operator

$$\boldsymbol{N_f} \;{}^{o}\; \boldsymbol{b_f}^{+}\, \boldsymbol{b_f} \qquad (5.24)$$

is simply multiplicative.

## 5.3 Single- and two-boson operators represented by creation and annihilation operators

First, we will show that calculating a matrix element we can replace the single-boson operator $\boldsymbol{F}^{(1)}$ by a combination of creation and annihilation operators in the following way

$$\boldsymbol{F}^{(1)} = \Sigma_{p,q}\, f_{pq}^{(1)} \cdot \boldsymbol{b_p}^{+}\, \boldsymbol{b_q}. \qquad (5.25)$$

The integrals $f_{pq}^{(1)}$ are defined in (5.8) and (5.14). First, we check the relation (5.25) for a diagonal matrix element. For $p = q = d$ we obtain according to (5.24)

$$\langle\, N_a, ..\, N_d, .. \mid f_{dd}^{(1)} \cdot \boldsymbol{b_d}^{+}\, \boldsymbol{b_d} \mid N_a, ..\, N_d, ..\, \rangle = f_{dd}^{(1)} \cdot N_d. \qquad (5.26)$$

Terms with $p$ ¹ $q$ vanish because in this case the resulting state function $|N_a,..\,, N_p + 1, ..\,, N_q - 1, ..\,\rangle$ is orthogonal to $\langle N_a, ..\,, N_p, ..\,, N_q, ..\,|$. Consequently, in order to form the diagonal matrix element of $\boldsymbol{F}^{(1)}$ we have to sum up the terms of the type (5.26). The result is in agreement with the directly calculated expression (5.10).

We now turn to the off diagonal matrix element. It reads

$$\Sigma_{pq}\, \langle N_a, ..\,, N_d, ..\,, N_g, .. \mid f_{pq}^{(1)} \cdot \boldsymbol{b_p}^{+}\, \boldsymbol{b_q} \mid N_a, ..\,, N_d + 1, ..\,, N_g - 1, ..\,\rangle \qquad (5.27)$$

Only the term with $p = g$ and $q = d$ differs from zero and yields the expression $f_{dg}^{(1)}\sqrt{(N_d + 1)}\sqrt{N_g}$ in agreement with (5.16). Namely, if $\boldsymbol{b}$-operators with other $p$ and $q$ values act on the right hand side of (5.27) the resulting state is orthogonal to the one on the left-hand side.

In matrix elements lying farther from the diagonal, the operator $\boldsymbol{b_p^+} \boldsymbol{b_q}$ is not able to bring both sides in agreement. Therefore, such elements vanish as we have realised directly in section 5.1. Thus, the assertion (5.25) is proved.

Creation and annihilation operators are also employed in order to represent matrix elements of two-boson operators. The original form of these operators is given by the equation (5.2)

$$\boldsymbol{F}^{(2)} = \sum_{i>j=1}^{N} \boldsymbol{f}^{(2)}(\underline{x}_i,\underline{x}_j).$$

Similar consideration as those for $\boldsymbol{F}^{(1)}$ lead to the following expression for the operator $\boldsymbol{F}^{(2)}$

$$\boldsymbol{F}^{(2)} = ½\cdot\sum_{dgpq} f^{(2)}_{dgpq}\, \boldsymbol{b_d^+}\, \boldsymbol{b_g^+}\, \boldsymbol{b_p}\, \boldsymbol{b_q} \qquad (5.28)$$

with $f^{(2)}_{dgpq} = \int y^*_d(\underline{x})\cdot y^*_g(\underline{z})\cdot f^{(2)}(\underline{x},\underline{z})\; y_p(\underline{x})\cdot y_q(\underline{z})\; d\underline{x}\; d\underline{z}$.

Correspondingly to (5.25), in (5.28) the sum extends over the single states. Diagonal matrix elements read now

$$\langle N_a, .. ,N_p, .. ,N_q, ..| \boldsymbol{F}^{(2)} | N_a, .. ,N_p, .. ,N_q, .. \rangle = ½ \sum_{pq} f^{(2)}_{pqpq}\cdot N_p\cdot N_q =$$

$$½ \sum_r f^{(2)}_{rrrr} N_r^2 + \sum_{p<q} f^{(2)}_{pqpq} N_p N_q. \qquad (5.29)$$

The remaining not vanishing matrix elements of the two-boson operator have the following form

$$\langle N_a, .., N_d, ..,N_g, ..,N_p, ..,N_q, ..| \boldsymbol{F}^{(2)} | N_a, .., N_d -1, ..,N_g-1, ..,N_p+1, ..,N_q+1, ..\rangle =$$

$$½ f^{(2)}_{dgpq} \sqrt{N_d} \sqrt{N_g} \sqrt{(N_p + 1)} \sqrt{(N_q + 1)}. \qquad (5.30)$$

In the broader sense they can be named off diagonal.

We have found matrix elements of $\boldsymbol{F}^{(1)}$ and $\boldsymbol{F}^{(2)}$ on the “primitive“ basis of total states (4.15). **Because every state of the seniority scheme (3.8) can be built with “primitive“ states ( section 4.1 ) the operator representations (5.25) and (5.28) hold also in the seniority scheme**. Although the annihilation and creation operators were introduced as a formal aid, they have a central importance in the interacting boson model.

We put together further properties of the operators $\boldsymbol{b_p^+}$ and $\boldsymbol{b_q}$. According to (5.24) the equation

$$\boldsymbol{b_p^+}\, \boldsymbol{b_p} = \boldsymbol{N_p} \text{ holds.}$$

Analogously one finds $\boldsymbol{b_p}\, \boldsymbol{b_p^+} = \boldsymbol{N_p} + 1.$ (5.31)

Subtracting the equation (5.24) from (5.31) one obtains the commutation rules for the operators $\boldsymbol{b_p}$ and $\boldsymbol{b_p^+}$ :

$$\boldsymbol{b}_p \boldsymbol{b}_p^+ - \boldsymbol{b}_p^+ \boldsymbol{b}_p = 1. \qquad (5.32)$$

In pairs of operators with different indices *p* and *q* the operators can be interchanged according to (5.21) and (5.22) which yields the following relation

$$\boldsymbol{b}_p \boldsymbol{b}_q^+ - \boldsymbol{b}_q^+ \boldsymbol{b}_p = 0. \qquad (p \neq q) \qquad (5.33)$$

We sum up

$$\boldsymbol{b}_p \boldsymbol{b}_q^+ - \boldsymbol{b}_q^+ \boldsymbol{b}_p = \delta_{pq}. \qquad (5.34)$$

If there are only annihilation or creation operators the following relations hold

$$\boldsymbol{b}_p \boldsymbol{b}_q - \boldsymbol{b}_q \boldsymbol{b}_p = 0,$$

$$\boldsymbol{b}_p^+ \boldsymbol{b}_q^+ - \boldsymbol{b}_q^+ \boldsymbol{b}_p^+ = 0. \qquad (5.35)$$

We write the commutation rules for boson operators especially for *s*- and *d*-operators :

$$[\boldsymbol{d}_m , \boldsymbol{d}^+_n] \equiv \boldsymbol{d}_m \boldsymbol{d}^+_n - \boldsymbol{d}^+_n \boldsymbol{d}_m = \delta_{m n}$$

$$[\boldsymbol{s} , \boldsymbol{s}^+] \equiv \boldsymbol{s}\boldsymbol{s}^+ - \boldsymbol{s}^+ \boldsymbol{s} = 1 \qquad (5.36)$$

All other commutators of these operators vanish.

Following the corresponding methods of the quantum electrodynamics, the use of the creation and annihilation operators occasionally is named second quantisation. It is also used for the description of the harmonic oscillator or in the shell model of the atomic nucleus.

# 6 Applications of the creation and annihilation operators

In the first section the formation of “primitive“ many-boson states and states with defined angular momentum will be described in terms of creation operators. In the following section we will show how the number of bosons can be increased with the aid of operators leaving the seniority unchanged. In section 6.3 the annihilation operator is modified in order to fulfil the rules of angular momentum coupling. The inverse of the pair-generating operator is treated in section 6.4. In the end the boson counting operators are put together.

### 6.1 Many-boson configurations represented by operators

The creation operators can act not only on fully symmetric many-boson states but also on the so-called vacuum state |⟩ which is realised in the case the protons and neutrons fill closed shells and there are no active bosons. Starting from the vacuum state we now construct “primitive“ states i. e. **states without defined angular momentum**.

We employ a creation operator $\boldsymbol{b_a}^+$ whose index *a* represents the angular momentum of the single state and its projection. It raises the occupation number $N_a$ of this state according to (5.20) as follows

$$\boldsymbol{b_a}^+ | N_a - 1 \rangle = \sqrt{N_a} \, | N_a \rangle. \qquad (6.1)$$

The effect of $\boldsymbol{b_a}^+$ on the vacuum state |⟩ is defined like this

$$\boldsymbol{b_a}^+ | \rangle = | N_a = 1 \rangle. \qquad (6.2)$$

If a number $N_a$ of operators $\boldsymbol{b_a}^+$ acts together on |⟩ one obtains the following symmetric many-boson state

$$(\boldsymbol{b_a}^+)^{N_a} | \rangle = \sqrt{(N_a!)} \, | N_a \rangle. \qquad (6.3)$$

A normalised, symmetric state with occupation numbers $N_a$, $N_b$, .. ,$N_f$ , .. reads therefore

$$(N_a! N_b! \,..\, N_f!)^{-½} \, (\boldsymbol{b_a}^+)^{N_a} (\boldsymbol{b_b}^+)^{N_b} \,..\, (\boldsymbol{b_f}^+)^{N_f} | \rangle = | N_a \, N_b \,..\, N_f \rangle. \qquad (6.4)$$

The symmetry of the state (6.4) comes from the definition of the creation operator ( section 5.2 ). Any sequence of the operators on the left-hand side of (6.4) can be chosen because they commute according to (5.35).

We now turn to **many-boson states with defined angular momentum**. From states with a few bosons, we learn how the creation operators have to be arranged.

First, we will look at two *d*-bosons which are coupled to *J*. According to (4.4) we have

$$| d^2 J M \rangle = (1/\sqrt{2})\cdot\sum_{m \neq m'} (2\ m\ 2\ m' \,|J\ M)\cdot| d\ m\ d\ m' \rangle +$$

$$(2\ m\ 2\ m \,|\, J\ M)\cdot| d\ m\ d\ m \rangle\cdot\ \delta_{M\ \text{even}}.$$

With $| d\ m\ d\ m' \rangle = \boldsymbol{d_m^+}\ \boldsymbol{d_{m'}^+} \,|\, \rangle$ ( for $m \neq m'$ ) and $| d\ m\ d\ m \rangle = (1/\sqrt{2})\ (\boldsymbol{d_m^+})^2\ |\rangle$ we obtain

$$| [d \times d]^{(J)}_M \rangle \equiv | d^2 J M \rangle =$$

$$(1/\sqrt{2})\cdot\sum_{mm'} (2\ m\ 2\ m' \,|J\ M)\cdot\ \boldsymbol{d_m^+}\ \boldsymbol{d_{m'}^+} \,|\, \rangle \equiv (1/\sqrt{2})\cdot[\boldsymbol{d^+} \times \boldsymbol{d^+}]^{(J)}_M\ |\rangle \qquad (6.5)$$

Thus, in order to construct the symmetric state with defined angular momentum the creation operators have to be coupled in the same way as the single states. A normalisation factor arises ( here it is $1/\sqrt{2}$ ). Analogously to (A2.7) the operator relation

$$[\boldsymbol{d}^+ \times \boldsymbol{d}^+]^{(J)}_M = 0 \quad \text{for odd } J \qquad (6.6)$$

holds because the operators $\boldsymbol{d}^+_{\boldsymbol{m}}$ and $\boldsymbol{d}^+_{\boldsymbol{m'}}$ can be interchanged in the same way as the state functions in (A2.2).

We now look into the combination of three *d*-bosons. According to (4.6) we have

$$| d^3 J_0\ J\ M \rangle = A\cdot \sum_{M_0 m''} (J_0\ M_0\ 2\ m'' \,|\, J\ M)\cdot \qquad (6.7)$$

$$\sum_{mm'} (2m\ 2\ m' \,|\, J_0\ M_0)\cdot(2A_p(m,m',m''))^{-1}\ | d\ m\ d\ m'\ d\ m'' \rangle.$$

The expressions for $A_p$ are given in section 4.1. The "primitive" states can be written as follows

$$\boldsymbol{d_m^+}\ \boldsymbol{d_{m'}^+}\ \boldsymbol{d_{m''}^+}\ |\rangle = | d\ m\ d\ m'\ d\ m'' \rangle \qquad \text{with } m \neq m' \neq m'' \neq m,$$

$$(1/\sqrt{2})\ (\boldsymbol{d_m^+})^2\ \boldsymbol{d_{m''}^+}\ |\rangle = | d\ m\ d\ m\ d\ m'' \rangle \qquad \text{with } m \neq m''$$

and $\qquad (1/\sqrt{(3!)})\ (\boldsymbol{d_m^+})^3 = | d\ m\ d\ m\ d\ m\rangle.$

Using the function $A_p(m,m',m'')$ we summarise

$$\sqrt{6}\cdot A_p(m,m',m'')\ \boldsymbol{d_m^+}\ \boldsymbol{d_{m'}^+}\ \boldsymbol{d_{m''}^+}\ |\rangle = | d\ m\ d\ m'\ d\ m'' \rangle \qquad (6.8)$$

and insert the result in (6.7).

$$| [[d \times d]^{(J_0)} \times d]^{(J)}_M \rangle \equiv | d^3 J_0\ J\ M \rangle =$$

$$A\cdot\sqrt{(3/2)}\ \sum_{M_0 m''} (J_0\ M_0\ 2\ m'' \,|\, J\ M)\cdot \sum_{mm'} (2\ m\ 2\ m' \,|\, J_0\ M_0)\cdot\ \boldsymbol{d_m^+}\ \boldsymbol{d_{m'}^+}\ \boldsymbol{d_{m''}^+}\ |\rangle =$$

$$A\cdot\sqrt{(3/2)}\cdot\ [[\boldsymbol{d^+} \times \boldsymbol{d^+}]^{(J_0)} \times \boldsymbol{d^+}]^{(J)}_M\ |\rangle. \qquad (6.9)$$

Again, the creation operators have to be coupled in the same way as the single boson states. Here the normalisation factor reads $A\sqrt{(3/2)}$. This coupling rule holds generally for symmetric states with defined angular momentum. Operators like $\boldsymbol{d}^+$ which obey the angular coupling rules are named tensor operators.

### 6.2 Generating boson pairs with total angular momentum zero

Here we deal again with many *d*-boson states in the seniority scheme introduced in (3.8). These states are built with boson pairs having $J = 0$ which are coupled to a state with $t$ *d*-bosons which is characterised by $n_D$, $J$ and $M$ ( section 3.3 ). We investigate how this coupling procedure forming symmetric states can be represented by operators. We symbolise the operator which generates the normalised state of $t$ boson by $[(\boldsymbol{d}^+)^t{}_{nD}]^{(J)}{}_M$ i. e.

$$| n_d = t,\ t\ \ n_D\ J\ M \rangle = [(\boldsymbol{d}^+)^t{}_{nD}]^{(J)}{}_M\ |\rangle. \tag{6.10}$$

According to the statements of section 6.1 a $J$=0-boson pair coupled up to the state (6.10) is described in the following way

$$\begin{aligned} | n_d = t + 2,\ t\ \ n_D\ J\ M \rangle &= A'\,[[\boldsymbol{d}^+ \times \boldsymbol{d}^+]^{(0)} \times [(\boldsymbol{d}^+)^t{}_{nD}]^{(J)}\ ]^{(J)}{}_M\ |\rangle = \\ & A'\,[\boldsymbol{d}^+ \times \boldsymbol{d}^+]^{(0)} \cdot [(\boldsymbol{d}^+)^t{}_{nD}]^{(J)}{}_M\ |\rangle = \\ & A'\,(1/\sqrt{5})\ \sqrt{5}[\boldsymbol{d}^+ \times \boldsymbol{d}^+]^{(0)} \cdot\ | n_d = t,\ t\ \ n_D\ J\ M \rangle = \\ & A'\,(1/\sqrt{5})\ \boldsymbol{P}^+\ | n_d = t,\ t\ \ n_D\ J\ M \rangle. \end{aligned} \tag{6.11}$$

In the last line of (6.11) the so-called pair creation operator $\boldsymbol{P}^+$ has been introduced which is defined as follows

$$\boldsymbol{P}^+ = \sum_m (-1)^m \boldsymbol{d}^+{}_m\, \boldsymbol{d}^+{}_{-m} = \sqrt{5}\ \sum_m (1/\sqrt{5})\ (-1)^m \boldsymbol{d}^+{}_m\, \boldsymbol{d}^+{}_{-m} = \sqrt{5}[\boldsymbol{d}^+ \times \boldsymbol{d}^+]^{(0)}. \tag{6.12}$$

Several authors attach the factor ½ to the right hand side of (6.12). Sometimes the symbol $\boldsymbol{P}^+ = \boldsymbol{d}^+ \bullet \boldsymbol{d}^+$ is used.

We now generalise the relation (6.11) starting from the state $|n_d - 2,\ t\ \ n_D\ J\ M \rangle$ in which the seniority $t$ may fall below its maximal value $n_d – 2$ i. e. $n_d - 2\ ^3\ t$, on condition that $n_d - 2 - t$ is even. In chapter 10 the following relation will be shown ( see 10.15 )

$$\begin{aligned} | n_d\ t\ \ n_D\ J\ M \rangle &= (n_d(n_d + 3) - t(t + 3))^{-½} \cdot \boldsymbol{P}^+\ |n_d - 2,\ t\ \ n_D\ J\ M \rangle = \\ & ((n_d + t + 3)(\ n_d - t))^{-½} \cdot \boldsymbol{P}^+\ |n_d - 2,\ t\ \ n_D\ J\ M \rangle. \end{aligned} \tag{6.13}$$

Replacing $n_d$ by the expression $t + 2$ in (6.13) and comparing with (6.11) we obtain $A' = 1/\sqrt{2}$. Analogously to (6.13) the relation

$$\begin{aligned} &|n_d - 2,\ t\ \ n_D\ J\ M \rangle = \\ &((n_d + t\ + 3 - 2)(\ n_d - t - 2))^{-½} \cdot \boldsymbol{P}^+\ |n_d - 4,\ t\ \ n_D\ J\ M \rangle \end{aligned} \tag{6.14}$$

holds which we insert in (6.13). We continue this procedure until the boson number in the state on the right hand side is $t$ and this state has the form (6.10). Finally, one obtains

$$| n_d\ t\ \ n_D\ J\ M \rangle = A_{n_d t}\ (\boldsymbol{P}^+)^{(n_d\ -\ t)/2}\ |\ t\ t\ \ n_D\ J\ M \rangle \tag{6.15}$$

with $$A_{n_d t} = ((n_d + t + 3)(n_d + t + 3 - 2) \,..\, (2\ t + 5)\cdot(\ n_d - t)(\ n_d - t - 2)\cdot..2)^{-½}$$

$$= ((2\ t + 3)!!/((n_d + t + 3)!!\ (n_d - t)!!))^{½}. \tag{6.16}$$

The symbol $n$!! stands for $n(n - 2)(n - 4)\cdot..\cdot 2$ or 1.

To complete the picture we write the states of the seniority scheme including the $s$-bosons whose number is $n_s$. According to (6.3) we have

$$(\boldsymbol{s}^+)^{n_s}\, |\rangle = \sqrt{(n_s!)}\ |\ n_s \rangle. \tag{6.17}$$

We let the operator $(\boldsymbol{s}^+)^{n_s}\ /\sqrt{(n_s!)}$ act on (6.15) creating a normalised and completely symmetric state as follows

$$| n_s\ n_d\ t\ n_D\ J\ M \rangle =$$

$$((2t + 3)!!/(n_s!(n_d + t + 3)!!(n_d - t)!!))^{½}\, (\boldsymbol{s}^+)^{n_s}\, (\boldsymbol{P}^+)^{(n_d\ -\ t)/2}\ |\ t\ t\ n_D\ J\ M \rangle. \tag{6.18}$$

The structure of the $d$-boson states $|\ t\ t\ n_D\ J\ M \rangle$ with maximal seniority may be complex but seldom they have to be formulated explicitly. For the most part it is sufficient to give the so-called "coefficients of fractional parentage" which permit a representation by states with $\tau$ - 1 $d$-bosons. In this report we don't deal with this method.

## 6.3 Tensor operators annihilating bosons

In section 6.1 we have shown that the creation operators $\boldsymbol{d}^+_m$ are components of the tensor operator $\boldsymbol{d}^+$ i. e. they obey the rules of angular momentum coupling using Clebsch-Gordan coefficients. This property, which correlates with a special behaviour on the occasion of rotations of the co-ordinate system is very important for treating matrix elements. Owing to the Wigner-Eckart theorem ( appendix A5 ) the matrix elements of tensor operators can be simplified heavily by writing them as a product of a reduced matrix element and a Clebsch-Gordan coefficient.

It can be shown that the annihilation operators $\boldsymbol{d}_m$ cannot be coupled vectorially but operators of the form

$$\boldsymbol{d}^{\sim}_m = (-1)^{m}\,\boldsymbol{d}_{-m} \tag{6.19}$$

have this property. They are named modified hermitian adjoint operators. Thus, we maintain that the operator $\boldsymbol{d}^{\sim}$ whose components are defined in (6.19) is a tensor operator. We verify this statement on two simple examples.

If the claim holds, an operator expression such as $[\boldsymbol{d}^{\sim} \times \boldsymbol{d}^{+}]^{(0)}$ can be built which we let act on the vacuum state $|\rangle$. In analogy with the operator $\boldsymbol{P}^{+}$ ((6.12)) a state with angular momentum 0 must result which contains no bosons because the creation and the annihilation operator cancel out in every term. So we expect

$$[\boldsymbol{d}^{\sim} \times \boldsymbol{d}^{+}]^{(0)}\,|\rangle = B\cdot|\rangle \tag{6.20}$$

with a constant $B$. Employing (A1.1) and (A1.16) we remodel the left hand side of (6.20)

$$[\boldsymbol{d}^{\sim} \times \boldsymbol{d}^{+}]^{(0)}\,|\rangle = \Sigma_m (2\ m\ 2, -m\ |\ 0\ 0)\ \boldsymbol{d}^{\sim}_m\, \boldsymbol{d}^{+}_{-m}\,|\rangle =$$

$$\Sigma_m (-1)^{(2\,-\,m)}\,(5)^{-½}(-1)^m\,\boldsymbol{d}_{-m}\,\boldsymbol{d}^{+}_{-m}\,|\rangle = \Sigma_m\,(5)^{-½}\,\boldsymbol{d}_{-m}|\,d_{-m}\rangle = (5)^{-½}\,\Sigma_m\,|\rangle = 5(5)^{-½}|\rangle. \tag{6.21}$$

Therefore the relation (6.20) is satisfied with $B = \sqrt{5}$.

In the second example the operator $\boldsymbol{d}^{\sim}$ is coupled up to the state $[\boldsymbol{d}^{+} \times \boldsymbol{d}^{+}]^{(J_0)}$ in such a way that the angular momentum 2 with the component $n$ results. Because in every term one annihilating and two creating operators are opposite we expect that the resulting state consists in one boson in the state $|\ d_n\ \rangle$ i. e.

$$[\boldsymbol{d}^{\sim} \times [\boldsymbol{d}^{+} \times \boldsymbol{d}^{+}]^{(J_0)}]^{(J\,=\,2)}{}_n\,|\rangle = B'\times\boldsymbol{d}^{+}{}_n|\rangle. \tag{6.22}$$

$J_0$ is even because the original two-boson state is symmetric. We write the left-hand side of (6.22) as

$$[\boldsymbol{d}^{\sim} \times [\boldsymbol{d}^{+} \times \boldsymbol{d}^{+}]^{(J_0)}]^{(J\,=\,2)}{}_n\,|\rangle =$$

$$\Sigma_{M_0\ m}\,(2\ m\ J_0\,M_0\,|\ 2\ n)\ \Sigma_{m'\,m''}\,(2\ m'\ 2\ m''\ |\ J_0\,M_0\,)\ \boldsymbol{d}^{\sim}_m\boldsymbol{d}^{+}{}_{m'}\,\boldsymbol{d}^{+}{}_{m''}|\rangle. \tag{6.23}$$

With (6.19) and the commutation rules (5.36) we obtain

$$(-1)^{-m}\,\boldsymbol{d}^{\sim}_m\,\boldsymbol{d}^{+}{}_{m'}\,\boldsymbol{d}^{+}{}_{m''}|\rangle = \boldsymbol{d}_{-m}\,\boldsymbol{d}^{+}{}_{m'}\,\boldsymbol{d}^{+}{}_{m''}|\rangle = (d_{m'\text{-}m}\times\boldsymbol{d}^{+}{}_{m''} + \boldsymbol{d}^{+}{}_{m'}\,\boldsymbol{d}_{-m}\,\boldsymbol{d}^{+}{}_{m''})|\rangle =$$

$$(d_{m',-m}\times\boldsymbol{d}^{+}{}_{m''} + d_{m'',-m}\times\boldsymbol{d}^{+}{}_{m'} + \boldsymbol{d}^{+}{}_{m'}\,\boldsymbol{d}^{+}{}_{m''}\ \boldsymbol{d}_{-m})|\rangle = (d_{m',-m}\times\boldsymbol{d}^{+}{}_{m''} + d_{m'',-m}\times\boldsymbol{d}^{+}{}_{m'})|\rangle \tag{6.24}$$

which we insert in (6.23) as follows

$$[\boldsymbol{d}^{\sim} \times [\boldsymbol{d}^{+} \times \boldsymbol{d}^{+}]^{(J_0)}]^{(J\,=\,2)}{}_n\,|\rangle =$$

$$\Sigma_{\,m\,M_0\ m''}(2\ m\ J_0\,M_0\,|\ 2\ n)(2, -m\ 2\ m''\ |\ J_0\,M_0)(-1)^{\,m}\,\boldsymbol{d}^{+}{}_{m''}|\rangle\ +$$

$$\Sigma_{\,m\,M_0\ \ m'}(2\ m\ J_0\,M_0\,|\ 2\ n)(2\ m'\ 2, -m\ |\ J_0\,M_0)(-1)^{\,m}\,\boldsymbol{d}^{+}{}_{m'}\,|\rangle = \tag{6.25}$$

$$2\cdot\Sigma_{\,m}\,(\Sigma_{M_0\ \ m'}\,(2\ m\ J_0\,M_0\,|\ 2\ n)(2, -m\ 2\ m'\ |\ J_0\,M_0))(-1)^m\,\boldsymbol{d}^{+}{}_{m'}\,|\rangle.$$

We made use of the symmetry of the last Clebsch-Gordan coefficient. This is rewritten using (A1.6) like this

$$(2,-m\ 2\ m' \mid J_0\ M_0) = (2,\ -m\ J_0\ M_0 \mid 2,\ -m')((2J_0 + 1)/5)^{½}\cdot(-1)^m =$$

$$(2\ m\ J_0\ M_0 \mid 2\ m')((2J_0 + 1)/5)^{½}\cdot(-1)^m\ . \qquad (6.26)$$

Owing to (A1.10) we obtain

$$[\boldsymbol{d}^{\sim} \times [\boldsymbol{d}^{+} \times \boldsymbol{d}^{+}]^{(J_0)}]^{(J = 2)}{}_n \,|\rangle = 2\cdot((2J_0 + 1)/5)^{½}\cdot\Sigma_{m'}\ d_{m'n}\ \boldsymbol{d}^{+}{}_{m'}|\rangle =$$

$$2\cdot((2J_0 + 1)/5)^{½}\cdot\boldsymbol{d}^{+}{}_n|\rangle \qquad (6.27)$$

in agreement with our forecast (6.22). Both examples illustrate that $\boldsymbol{d}^{\sim}$ is a tensor operator.

## 6.4 Annihilating boson pairs with total angular momentum zero

Analogously to (6.12) we define the pair annihilation operator $\boldsymbol{P}^{-}$ as follows

$$\boldsymbol{P}^{-} = \Sigma_m(-1)^{m}\chi\ \boldsymbol{d}_m\,\boldsymbol{d}_{-m} = \surd 5\cdot\Sigma_m(-1)^m\ (5)^{-½}\ \boldsymbol{d}^{\sim}{}_m\,\boldsymbol{d}^{\sim}{}_{-m} = \surd 5\cdot\Sigma_m(2\ m\ 2,\ -m \mid 0\ 0)\boldsymbol{d}^{\sim}{}_m\,\boldsymbol{d}^{\sim}{}_{-m} =$$

$$\surd 5\cdot[\boldsymbol{d}^{\sim} \times \boldsymbol{d}^{\sim}]^{(0)}. \qquad (6.28)$$

Several authors attach the factor ½ to the right hand side of (6.28). Sometimes the symbol $\boldsymbol{P}^{-} = \boldsymbol{d}^{\sim}\bullet\boldsymbol{d}^{\sim}$ is used. From (10.16) we have

$$\boldsymbol{P}^{-}|n_d\ t\ n_D\ J\ M\rangle = (n_d(n_d + 3) - t(t + 3))^{½}\ |n_d\ \text{-}2,\ t\ n_D\ J\ M\rangle. \qquad (6.29)$$

The operator

$$(n_d(n_d + 3) - t(t + 3))^{-½}\cdot\ \boldsymbol{P}^{-} \qquad (6.30)$$

is normalised. If $\boldsymbol{P}^{-}$ acts on the state $|\ t\ t\ n_D\ J\ M\rangle$ of the seniority scheme ( with $n_d = t$ ), from (6.29) follows

$$\boldsymbol{P}^{-}|n_d = t,\ t\ n_D\ J\ M\rangle = 0. \qquad (6.31)$$

## 6.5 Number operators for bosons

We let the operator $\boldsymbol{d}^{+}{}_m\ \boldsymbol{d}_m$ ( with 2 ³ $m$ ³ -2 ) act on the "primitive" state $|\ (d_2)^{n_2}\ (d_1)^{n_1}\ (d_0)^{n_0}\ (d_{-1})^{n_{-1}}\ (d_{-2})^{n_{-2}}\rangle$ º $|\ n_2\ n_1\ n_0\ n_{-1}\ n_{-2}\rangle$. Owing to (5.23) we obtain

$$\boldsymbol{d}^{+}{}_m\,\boldsymbol{d}_m\,|\ n_2\ n_1\ n_0\ n_{-1}\ n_{-2}\rangle = n_m\,|\ n_2\ n_1\ n_0\ n_{-1}\ n_{-2}\rangle. \qquad (6.32)$$

Therefore the eigenvalue of $\boldsymbol{d}^{+}{}_m\,\boldsymbol{d}_m$ is the number $n_m$ of $d$-bosons in the state $m$. Consequently the operator $\Sigma_m\ \boldsymbol{d}^{+}{}_m\,\boldsymbol{d}_m$ yields

$$\Sigma_m\ \boldsymbol{d}^{+}{}_m\,\boldsymbol{d}_m\,|\ n_2\ n_1\ n_0\ n_{-1}\ n_{-2}\rangle = \Sigma_m n_m\,|\ n_2\ n_1\ n_0\ n_{-1}\ n_{-2}\rangle = n_d\,|\ n_2\ n_1\ n_0\ n_{-1}\ n_{-2}\rangle. \ (6.33)$$

Thus its eigenvalue is the total number $n_d$ of $d$-bosons. The operator $\Sigma_m\ \boldsymbol{d}^{+}{}_m\,\boldsymbol{d}_m$ can be applied also to states with defined total angular momentum because they can be written as a linear combinations of "primitive" states with the same number of *d*-bosons ( section 4.1 ). This number operator is characterized by $\boldsymbol{n_d}$,

$$\boldsymbol{n_d} = \sum_m \boldsymbol{d}^+_m \boldsymbol{d}_m \ . \qquad (6.34)$$

With the help of $\boldsymbol{d}^{\sim}_{-m}\,(-1)^m = \boldsymbol{d}_m$ (6.19) and of (A1.16) we write also

$$\boldsymbol{n_d} = \sqrt{5}\ \sum_m (-1)^m (5)^{-½} \boldsymbol{d}^+_m \boldsymbol{d}^{\sim}_{-m} = \sqrt{5}\ \sum_m (2\ m\ 2, -\mu \mid 0\ 0) \boldsymbol{d}^+_m \boldsymbol{d}^{\sim}_{-m} =$$

$$\sqrt{5}\cdot[\boldsymbol{d}^+ \times \boldsymbol{d}^{\sim}]^{(0)}. \qquad (6.35)$$

The number operator for *s*-bosons is built correspondingly of creation and annihilation operators for *s*-bosons, $\boldsymbol{s}^+$ and $\boldsymbol{s}$, as follows

$$\boldsymbol{n_s} = \boldsymbol{s}^+ \boldsymbol{s}. \qquad (6.36)$$

Its eigenvalue is the number of *s*-bosons, $n_s$. The operator

$$\boldsymbol{N} = \boldsymbol{n_d} + \boldsymbol{n_s} \qquad (6.37)$$

has the eigenvalue $n_s + n_d$ .i. e. the total number of bosons of the collective state.

As mentioned above, the use of creation and annihilation operators for bosons frequently is named second quantisation. An advantage of this method consists in the fact that the matrix elements of the Hamilton operator and of the operators describing the electromagnetic transition can be written completely with the boson operators $\boldsymbol{d}^+_m$, $\boldsymbol{d}^{\sim}_n$, $\boldsymbol{s}^+$ and $\boldsymbol{s}$ . In a word, not only the operators of the interactions (5.25) and (5.28) but also the boson states are represented by the single boson operators. For the algebraic treatment of these elements the commutation rules (5.36) are applied.

# 7 The Hamilton operator of the IBM1

In the first section of this chapter, the components of the Hamilton operator will be sketched in. The part with boson-boson interactions is transformed in the second section by coupling the tensor operators. In section 7.3 the Hamiltonian is written explicitly with *d*- and *s*-operators. It must conserve the number of bosons i. e. it must commute with the number operator, which is verified in the fourth section. In the last one the Hamiltonian will be brought in a more compact form.

## 7.1 The components of the Hamiltonian

In chapter 2 the number $N$ of active bosons in the interacting boson model is determined. Formulating the energy of the system, we don't go into the structure and eigenenergy of the inner part of the nucleus i. e. of the closed shell ( core ). The kinetic energy $T(i)$ and the potential energy $U(i)$ of an active boson have the character of operators and constitute together the Hamilton operator $\boldsymbol{H}^{(1)}$ of a single state $| b_{lm} \rangle$ . It generates the eigenenergy $e_{lm}$ owing to the equation

$$(\boldsymbol{T}^{(1)} + \boldsymbol{U}^{(1)})| b_{lm} \rangle = \boldsymbol{H}^{(1)} | b_{lm} \rangle = e_{lm} \; | b_{lm} \rangle \tag{7.1}$$

The singe-boson states $| b_{lm} \rangle$ have a defined angular momentum $l$ with the projection $m$ i. e. we suppose that the angular momentum operator commutes with the Hamiltonian $\boldsymbol{H}^{(1)}$. We mentioned that there are only six *lm*-states in the IBM1.

Because no spatial axis prevails, the single-boson energy $e_{lm}$ does not depend on *m.* Therefore there are only two energies of this kind, namely $e_s$ and $e_d$ . If the bosons were independent of one another, a system of $n_s$ s-bosons and $n_d$ *d*-boson would have the energy $n_s e_s + n_d e_d$ . Consequently the Hamilton operator of the whole system would contain counting operators ( section 6.5 ) in the following way

$$e_s \boldsymbol{n_s} + e_d \boldsymbol{n_d} \,. \tag{7.2}$$

From (6.34) and (6.36) we see that this operator has the structure of a single-boson operator (5.25) with $f^{(1)}_{p{}^1q} = 0$ ( see section 5.3). Owing to (6.37) the form

$$e_s \boldsymbol{N} + (e_d - e_s)\boldsymbol{n_d} \,. \tag{7.3}$$

can be used also. We now introduce the interaction between the active bosons from which we assume that it takes place in twos. According to (5.2) we have to add a two-boson operator such as

$$\tfrac{1}{2}\textstyle\sum_{1=i<j}^{N} \boldsymbol{W}(ij) \;\text{o}\; \boldsymbol{W} \tag{7.4}$$

to the Hamiltonian. From (5.28), we take the corresponding form

$$\tfrac{1}{2}\sum_{i,j=1}^{N} \boldsymbol{W}(ij) = \tfrac{1}{2}\sum_{f\,g\,p,q=1}^{6} \langle\, f\, g \,|\, \boldsymbol{w} \,|\, p\, q \,\rangle\; \boldsymbol{b}^{+}_{f}\boldsymbol{b}^{+}_{g}\boldsymbol{b}_{p}\boldsymbol{b}_{q}\ . \qquad (7.5)$$

The indices on the right hand side of (7.5) denote the *s*- and five *d*-states.

## 7.2 The operator of the boson-boson interaction formulated with defined angular momentum

Before we write the expression (7.5) down for the IBM, we convert it into a form containing operator configurations and matrix elements with defined angular momenta. In (7.5) we replace the indices *f*, *g*, .. by pairs of quantum numbers $l_f\, m_f$, $l_g\, m_g$, .. and make the following claim

$$\boldsymbol{W}\ ^{\mathrm{o}}\ \tfrac{1}{2}\sum_{l_f m_f l_g m_g l_p m_p l_q m_q} \langle\, l_f\, m_f\, l_g\, m_g \,|\, \boldsymbol{w} \,|\, l_p\, m_p\, l_q\, m_q \,\rangle\; \boldsymbol{b}^{+}_{l_f m_f}\boldsymbol{b}^{+}_{l_g m_g}\boldsymbol{b}_{l_p m_p}\boldsymbol{b}_{l_q m_q} = \qquad (7.6)$$

$$\tfrac{1}{2}\sum_{l_f l_g l_p l_q, J=\mathrm{even}} \langle\, l_f\, l_g\, J \,|\, \boldsymbol{w} \,|\, l_p\, l_q\, J \,\rangle\, \sqrt{(2J+1)}\cdot[[\boldsymbol{b}^{+}_{f}\times\boldsymbol{b}^{+}_{g}]^{(J)}\times[\boldsymbol{b}^{\sim}_{p}\times\boldsymbol{b}^{\sim}_{q}]^{(J)}]^{(0)}.$$

We start the proof by adding the factor $d_{m_f m_f'}\, d_{m_g m_g'}\, d_{m_p m_p'}\, d_{m_q m_q'}$ on the left hand side of (7.6) and by summing over $m_f'$, $m_g'$, $m_p'$ and $m_q'$ which yields the original expression. It reads

$$\boldsymbol{W} = \tfrac{1}{2}\sum_{l_f l_g l_p l_q}\ \sum_{m_f m_g m_p m_q m_f' m_g' m_p' m_q'} \langle\, l_f\, m_f\, l_g\, m_g \,|\, \boldsymbol{w} \,|\, l_p\, m_p\, l_q\, m_q \,\rangle\ \cdot$$

$$d_{m_f m_f'}\, d_{m_g m_g'}\, d_{m_p m_p'}\, d_{m_q m_q'}\ \boldsymbol{b}^{+}_{l_f m_f'}\, \boldsymbol{b}^{+}_{l_g m_g'}\, \boldsymbol{b}_{l_p m_p'}\, \boldsymbol{b}_{l_q m_q'}\ , \qquad (7.7)$$

in which the quantum numbers *m* of the ***b***-operators have been supplied with primes. With the help of (A1.15), i. e. of

$$\sum_{JM} (l_f\, m_f\, l_g\, m_g \,|\, J\, M)(l_f\, m_f'\, l_g\, m_g' \,|\, J\, M) = d_{m_f m_f'}\, d_{m_g m_g'} \qquad (7.8)$$

and of an analogous expression for *p* and *q* we obtain

$$\boldsymbol{W} = \tfrac{1}{2}\sum_{l_f l_g l_p l_q}\ \sum_{J\, M\, J''\, M''}\ \sum_{m_f m_g m_p m_q m_f' m_g' m_p' m_q'} \langle\, l_f\, m_f\, l_g\, m_g \,|\, \boldsymbol{w} \,|\, l_p\, m_p\, l_q\, m_q \,\rangle\ \cdot$$

$$(l_f\, m_f\, l_g\, m_g \,|\, J\, M)\ (l_p\, m_p\, l_q\, m_q \,|\, J''\, M'')\ (l_f\, m_f'\, l_g\, m_g' \,|\, J\, M)\ (l_p\, m_p'\, l_q\, m_q' \,|\, J''\, M'')$$

$$\boldsymbol{b}^{+}_{l_f m_f'}\, \boldsymbol{b}^{+}_{l_g m_g'}\, \boldsymbol{b}_{l_p m_p'}\, \boldsymbol{b}_{l_q m_q'}\ . \qquad (7.9)$$

We make use of (A1.1) for the following equations

$$\sum_{m_f' m_g'} (l_f\, m_f'\, l_g\, m_g' \,|J\, M)\ \boldsymbol{b}^{+}_{l_f m_f'}\boldsymbol{b}^{+}_{l_g m_g'} = [\boldsymbol{b}^{+}_{f}\times\boldsymbol{b}^{+}_{g}]^{(J)}_{\mathrm{M}} \qquad (7.10)$$

and $\sum_{m_p' m_q'} (l_p\, m_p'\, l_q\, m_q' \,|\, J''\, M'')\ \boldsymbol{b}_{l_p m_p'}\, \boldsymbol{b}_{l_q m_q'} =$

$$\sum_{m_p' m_q'} (l_p\, m_p'\, l_q\, m_q' \,|\, J''\, M'')\ (-1)^{-l_p-l_q+m_p'+m_q'}\boldsymbol{b}^{\sim}_{l_p\,-m_p'}\, \boldsymbol{b}^{\sim}_{l_q\,-m_q'} =$$

$$\sum_{m_p' m_q'} (l_p,-m_p',\, l_q,-m_q' \,|\, J'',-M'')\ (-1)^{M''-J''}\ \boldsymbol{b}^{\sim}_{l_p\,-m_p'}\, \boldsymbol{b}^{\sim}_{l_q\,-m_q'} =$$

$$(-1)^{M''-J''}\ [\boldsymbol{b}^{\sim}_{p}\times\boldsymbol{b}^{\sim}_{q}]^{(J'')}_{-M''}\ , \qquad (7.11)$$

in which (A1.5) is used and (6.19) is put in the original form $\boldsymbol{b}_{lm} = (-1)^{-l+m}\boldsymbol{b}^{\sim}_{l,\,-m}$. We obtain

$$\boldsymbol{W} = ½\sum_{l_f\, l_g\, l_p\, l_q} \sum_{J\, M\, J“\, M“} \sum_{m_f\, m_g\, m_p\, m_q} \langle\, l_f\, m_f\, l_g\, m_g \mid \boldsymbol{w} \mid l_p\, m_p\, l_q\, m_q \,\rangle \cdot \qquad (7.12)$$

$$(l_f\, m_f\, l_g\, m_g \mid J\, M)\, (l_p\, m_p\, l_q\, m_q \mid J“\, M“)\, (-1)^{M“-J“}\, [\boldsymbol{b}^{+}{}_{f} \times \boldsymbol{b}^{+}{}_{g}]^{(J)}{}_{M} [\boldsymbol{b}^{\sim}{}_{p} \times \boldsymbol{b}^{\sim}{}_{q}]^{(J“)}{}_{-M“}\,.$$

The sum over $m_f$, $m_g$, $m_p$ and $m_q$ yields

$$\boldsymbol{W} = ½\sum_{l_f\, l_g\, l_p\, l_q} \sum_{J\, M\, J“\, M“} \langle\, l_f\;\; l_g\, J\, M \mid \boldsymbol{w} \mid l_p\, l_q\, J“\, M“ \,\rangle \cdot$$

$$(-1)^{M“-J“} [\boldsymbol{b}^{+}{}_{f} \times \boldsymbol{b}^{+}{}_{g}]^{(J)}{}_{M}\, [\boldsymbol{b}^{\sim}{}_{p} \times \boldsymbol{b}^{\sim}{}_{q}]^{(J“)}{}_{-M“}\,. \qquad (7.13)$$

The matrix element in (7.13) is constructed in terms of two-boson states with defined total angular momenta $J$ and $J“$ which, however, must agree. Namely, if we assume that $\boldsymbol{w}$ is a tensor operator with rank 0 ( see appendix A5 ) the Clebsch-Gordan coefficient in (A5.8) vanishes for $(J,M)$ ¹ $(J“,M“)$. Moreover, owing to (A5.8), (A1.6) and (A1.16) the matrix element in (7.13) is independent of $M$. Using (A1.16) we obtain

$$\boldsymbol{W} = ½\sum_{l_f\, l_g\, l_p\, l_q\, ,J=\text{even}} \langle\, l_f\;\; l_g\, J \mid \boldsymbol{w} \mid l_p\, l_q\, J \,\rangle \cdot \sqrt{(2J + 1)}$$

$$[[\boldsymbol{b}^{+}{}_{f} \times \boldsymbol{b}^{+}{}_{g}]^{(J)} \times [\boldsymbol{b}^{\sim}{}_{p} \times \boldsymbol{b}^{\sim}{}_{q}]^{(J)}]^{(0)}\,. \qquad (7.14)$$

No odd values occur for $J$ if the $l$ 's are 0 or 2, under which condition the equation (7.6) is proved.

### 7.3 The basic form of the Hamilton operator

Now for the operators of the $f$-, $g$-, $p$- and $q$-states in (7.6) specially we put in the $s$- and $d$-operators respectively and make use of the symmetry properties of the matrix elements. The boson-boson interaction operator obtains the following form

$$\boldsymbol{W} = ½\sum_{J=0,2,4} \langle\, d\, d\, J \mid \boldsymbol{w} \mid d\, d\, J \,\rangle\, \sqrt{(2J + 1)}\, [[\boldsymbol{d}^{+} \times \boldsymbol{d}^{+}]^{(J)} \times [\boldsymbol{d}^{\sim} \times \boldsymbol{d}^{\sim}]^{(J)}]^{(0)} +$$

$$½ \langle\, d\, d, J = 2 \mid \boldsymbol{w} \mid d\, s, J = 2 \,\rangle \sqrt{5} \cdot 2([[\boldsymbol{d}^{+} \times \boldsymbol{d}^{+}]^{(2)} \times \boldsymbol{d}^{\sim}\boldsymbol{s}]^{(0)} + [\boldsymbol{s}^{+}\boldsymbol{d}^{+} \times [\boldsymbol{d}^{\sim} \times \boldsymbol{d}^{\sim}]^{(2)}]^{(0)}) +$$

$$½ \langle\, d\, d, J = 0 \mid \boldsymbol{w} \mid s\, s, J = 0 \,\rangle ([[\boldsymbol{d}^{+} \times \boldsymbol{d}^{+}]^{(0)} \times \boldsymbol{ss}]^{(0)} + [\boldsymbol{s}^{+}\boldsymbol{s}^{+} \times [\boldsymbol{d}^{\sim} \times \boldsymbol{d}^{\sim}]^{(0)}]^{(0)}) +$$

$$½ \langle\, d\, s, J = 2 \mid \boldsymbol{w} \mid d\, s, J = 2 \,\rangle\, \sqrt{5}\; 4[\boldsymbol{d}^{+}\boldsymbol{s}^{+} \times \boldsymbol{d}^{\sim}\boldsymbol{s}]^{(0)} + \qquad (7.15)$$

$$½ \langle\, s\, s, J = 0 \mid \boldsymbol{w} \mid s\, s, J = 0 \,\rangle\, [\boldsymbol{s}^{+}\boldsymbol{s}^{+} \times \boldsymbol{ss}]^{(0)}\,.$$

We introduce usual symbols for the matrix elements as follows

$$\langle\, d\, d\, J \mid \boldsymbol{w} \mid d\, d\, J \,\rangle = c_J$$

$$\langle\, d\, d, J = 2 \mid \boldsymbol{w} \mid d\, s, J = 2 \,\rangle \sqrt{(2 \cdot 5)} = v_2$$

$$\langle\, d\, d, J = 0 \mid \boldsymbol{w} \mid s\, s, J = 0 \,\rangle = v_0 \qquad (7.16)$$

$$\langle\, d\, s, J = 2 \mid \boldsymbol{w} \mid d\, s, J = 2 \,\rangle 2\sqrt{5} = u_2$$

$$\langle\, s\, s, J = 0 \mid \boldsymbol{w} \mid s\, s, J = 0 \,\rangle = u_0\,.$$

In the following the quantities $c_J$ , $v_i$ and $u_i$ will play the role of parameters. Finally, we put the Hamilton operator together from (7.3), (7.15) and (7.16)

$$\begin{aligned} \boldsymbol{H} = e_s \boldsymbol{N} + ( e_d - e_s ) \boldsymbol{n_d} + ½ \textstyle\sum_{J=0,2,4} c_J \sqrt{(2J + 1)}\, [[\boldsymbol{d}^+ \times \boldsymbol{d}^+]^{(J)} \times [\boldsymbol{d}^\sim \times \boldsymbol{d}^\sim]^{(J)}]^{(0)} + \\ \sqrt{(½)}\, v_2 ([[\boldsymbol{d}^+ \times \boldsymbol{d}^+]^{(2)} \times \boldsymbol{d}^\sim \boldsymbol{s}]^{(0)} + [\boldsymbol{s}^+\boldsymbol{d}^+ \times [\boldsymbol{d}^\sim \times \boldsymbol{d}^\sim]^{(2)}]^{(0)}) + \\ ½\, v_0 ([[\boldsymbol{d}^+ \times \boldsymbol{d}^+]^{(0)} \times \boldsymbol{ss}]^{(0)} + [\boldsymbol{s}^+\boldsymbol{s}^+ \times [\boldsymbol{d}^\sim \times \boldsymbol{d}^\sim]^{(0)}]^{(0)}) + \\ u_2 [\boldsymbol{d}^+\boldsymbol{s}^+ \times \boldsymbol{d}^\sim \boldsymbol{s}]^{(0)} + \qquad (7.17) \\ ½\, u_0 [\boldsymbol{s}^+\boldsymbol{s}^+ \times \boldsymbol{ss}]^{(0)} . \end{aligned}$$

We name the representation (7.17) the basic form of the Hamilton operator in the IBM1. In the next section but one we will see that the 9 parameters can be brought to 6 by reducing the field of applicability.

### 7.4 The conservation of boson number

We expect that collective states with definite energy have also a definite number $N$ of bosons. In quantum mechanical terms this means that the Hamilton operator $\boldsymbol{H}$ and the boson number operator, $\boldsymbol{N} = \boldsymbol{s}^+\boldsymbol{s} + \sum_m \boldsymbol{d}^+_m \boldsymbol{d}_m$ (6.37), commute

$$[ \boldsymbol{H}, \boldsymbol{N} ] = 0. \qquad (7.18)$$

In order to prove (7.18) we fall back upon the representation (7.5) and write the Hamilton operator this way

$$\boldsymbol{H} = e_s \boldsymbol{N} + (e_d - e_s)\, \boldsymbol{n_d} + ½\textstyle\sum_{f\,g\,p\,q=1}^{6} \langle\, f\, g \mid \boldsymbol{w} \mid p\, q \,\rangle\, \boldsymbol{b}^+_f \boldsymbol{b}^+_g \boldsymbol{b_p} \boldsymbol{b_q}. \qquad (7.19)$$

From (6.34) up to (6.37) we take that the number operator reads

$$\boldsymbol{N} = \textstyle\sum_{n=1}^{6} \boldsymbol{b}^+_n \boldsymbol{b_n} . \qquad (7.20)$$

In order to see if $\boldsymbol{N}$ commutes with the group of the four $b$-operators in (7.19) we push $\boldsymbol{N}$ step by step from the left to the right hand side of this group as follows

$$\begin{aligned} \boldsymbol{N}\, \boldsymbol{b}^+_f \boldsymbol{b}^+_g \boldsymbol{b_p} \boldsymbol{b_q} = \textstyle\sum_n \boldsymbol{b}^+_n \boldsymbol{b_n}\, \boldsymbol{b}^+_f \boldsymbol{b}^+_g \boldsymbol{b_p} \boldsymbol{b_q} = \sum_n \boldsymbol{b}^+_n ( d_{n,f} + \boldsymbol{b}^+_f \boldsymbol{b_n})\boldsymbol{b}^+_g \boldsymbol{b_p} \boldsymbol{b_q} = \\ \boldsymbol{b}^+_f \boldsymbol{b}^+_g \boldsymbol{b_p} \boldsymbol{b_q} + \textstyle\sum_n \boldsymbol{b}^+_f \boldsymbol{b}^+_n \boldsymbol{b_n}\boldsymbol{b}^+_g \boldsymbol{b_p} \boldsymbol{b_q} = 2\, \boldsymbol{b}^+_f \boldsymbol{b}^+_g \boldsymbol{b_p} \boldsymbol{b_q} + \sum_n \boldsymbol{b}^+_f \boldsymbol{b}^+_g \boldsymbol{b}^+_n \boldsymbol{b_n} \boldsymbol{b_p} \boldsymbol{b_q} = \\ 2\, \boldsymbol{b}^+_f \boldsymbol{b}^+_g \boldsymbol{b_p} \boldsymbol{b_q} + \textstyle\sum_n \boldsymbol{b}^+_f \boldsymbol{b}^+_g (-d_{n,p} + \boldsymbol{b_p} \boldsymbol{b}^+_n )\, \boldsymbol{b_n} \boldsymbol{b_q} = \qquad (7.21) \\ \boldsymbol{b}^+_f \boldsymbol{b}^+_g \boldsymbol{b_p} \boldsymbol{b_q} + \textstyle\sum_n \boldsymbol{b}^+_f \boldsymbol{b}^+_g \boldsymbol{b_p} \boldsymbol{b}^+_n \boldsymbol{b_q} \boldsymbol{b_n} = \sum_n \boldsymbol{b}^+_f \boldsymbol{b}^+_g \boldsymbol{b_p} \boldsymbol{b_q} \boldsymbol{b}^+_n \boldsymbol{b_n} = \boldsymbol{b}^+_f \boldsymbol{b}^+_g \boldsymbol{b_p} \boldsymbol{b_q} \boldsymbol{N}. \end{aligned}$$

It is not difficult to extend the proof of (7.18) on the other terms in (7.19). Thus, $\boldsymbol{N}$ commutes with $\boldsymbol{H}$, i. e. $\boldsymbol{N}$ is a so-called constant of motion. Provided that the Hamiltonian does not contain $s$-operators ( see chapter 10 ), implying the condition $v_2 = v_0 = u_2 = u_0 = 0$, the $d$-number operator $\boldsymbol{n_d} = \sum_m \boldsymbol{d}^+_m \boldsymbol{d}_m$ commutes with $\boldsymbol{H}$ . Then the number of $d$-bosons is definite.

## 7.5 A simplified form of the Hamilton operator

Since the boson number $N$ is a good quantum number, it's obvious to replace the operators $\boldsymbol{s}$ and $\boldsymbol{s}^+$ as far as possible by $\boldsymbol{N}$. Making use of (5.34) we write

$$[\boldsymbol{s}^+\boldsymbol{s}^+ \times \boldsymbol{ss}]^{(0)} = \boldsymbol{s}^+\boldsymbol{s}^+\boldsymbol{ss} = \boldsymbol{s}^+\boldsymbol{s}\,\boldsymbol{s}^+\boldsymbol{s} - \boldsymbol{s}^+\boldsymbol{s} = \boldsymbol{n_s}(\boldsymbol{n_s} - 1) = (\boldsymbol{N} - \boldsymbol{n_d})(\boldsymbol{N} - \boldsymbol{n_d} - 1). \quad (7.22)$$

In addition, owing to (6.35) we obtain

$$[\boldsymbol{d}^+\boldsymbol{s}^+ \times \boldsymbol{d}^{\sim}\boldsymbol{s}]^{(0)} = [\boldsymbol{d}^+ \times \boldsymbol{d}^{\sim}]^{(0)}\, \boldsymbol{s}^+\boldsymbol{s} = \sqrt{(1/5)}\, \boldsymbol{n_d}\, \boldsymbol{n_s} = \sqrt{(1/5)}\, \boldsymbol{n_d}\, (\boldsymbol{N} - \boldsymbol{n_d}). \quad (7.23)$$

In (7.22) and (7.23) the operator $\boldsymbol{n_d}^2$ appears, for which we give the expression

$$\boldsymbol{n_d}^2 = \boldsymbol{n_d} + \Sigma_{J=0,2,4} \sqrt{(2J + 1)}[[\boldsymbol{d}^+ \times \boldsymbol{d}^+]^{(J)} \times [\boldsymbol{d}^{\sim} \times \boldsymbol{d}^{\sim}]^{(J)}]^{(0)}. \quad (7.24)$$

Namely, according to (6.35) we write

$$\boldsymbol{n_d}^2 = 5[\boldsymbol{d}^+ \times \boldsymbol{d}^{\sim}]^{(0)} \cdot [\boldsymbol{d}^+ \times \boldsymbol{d}^{\sim}]^{(0)} = 5[[\boldsymbol{d}^+ \times \boldsymbol{d}^{\sim}]^{(0)} \times [\boldsymbol{d}^+ \times \boldsymbol{d}^{\sim}]^{(0)}]^{(0)}. \quad (7.25)$$

From (A4.3) we learn that (7.25) can be rebuilt as follows

$$\boldsymbol{n_d}^2 = 5\Sigma_J (2J + 1) \{^2{}_{2J}\, ^2{}_{2J}\, ^0{}_{00}\} \cdot \Sigma_M (J\, M\, J, -M \mid 0\, 0) \cdot \quad (7.26)$$

$$\Sigma_m (2\, m\, 2, M - m \mid J\, M) \cdot \Sigma_{m'} (2\, m'\, 2, -M - m' \mid J, -M) \cdot \boldsymbol{d}^+_m \boldsymbol{d}^{\sim}_{m'} \boldsymbol{d}^+_{M-m} \boldsymbol{d}^{\sim}_{-M-m'}.$$

We employ (6.19) and the commutation rules (5.36) and obtain

$$\boldsymbol{d}^+_m \boldsymbol{d}^{\sim}_{m'} \boldsymbol{d}^+_{M-m} \boldsymbol{d}^{\sim}_{-M-m'} = (-1)^{m'} \delta_{-m',M-m} \boldsymbol{d}^+_m \boldsymbol{d}^{\sim}_{-M-m'} + \boldsymbol{d}^+_m \boldsymbol{d}^+_{M-m} \boldsymbol{d}^{\sim}_{m'} \boldsymbol{d}^{\sim}_{-M-m'}. \quad (7.27)$$

This leads the expression (7.26) to fall into two parts. Because $J$ is an integer, from (A4.2), (A3.12) and (A1.16) we obtain

$$\{^2{}_{2J}\, ^2{}_{2J}\, ^0{}_{00}\} = (1/5)/\sqrt{(2J + 1)} \text{ and } (J\, M\, J, -M \mid 0\, 0) = (-1)^{J-M}/\sqrt{(2J + 1)}\,. \quad (7.28)$$

The first expression of (7.26) reads now

$$\Sigma_{J=0}^4 (-1)^J \Sigma_m \boldsymbol{d}^+_m \boldsymbol{d}_m \Sigma_M (2\, m\, 2, M - m \mid J\, M)^2 = \quad (7.29)$$

$$\Sigma_{J=0}^4 (-1)^J \Sigma_m \boldsymbol{d}^+_m \boldsymbol{d}_m (2J + 1)/5 = \Sigma_m \boldsymbol{d}^+_m \boldsymbol{d}_m (1 - 3 + 5 - 7 + 9)/5 = \boldsymbol{n_d}$$

where (A1.11) has been applied. With the aid of (A4.1) and (A1.3), we bring the second expression of (7.26) in the following form

$$\Sigma_{J=0,2,4} \sqrt{(2J + 1)}[[\boldsymbol{d}^+ \times \boldsymbol{d}^+]^{(J)} \times [\boldsymbol{d}^{\sim} \times \boldsymbol{d}^{\sim}]^{(J)}]^{(0)}. \quad (7.30)$$

Owing to (6.6), the quantity $J$ is even. With that, the equation (7.24) is proven. We now insert (7.22) up to (7.24) in (7.17) and obtain

$$\boldsymbol{H} = e_s N + \tfrac{1}{2}u_0 N(N - 1) + e\, \boldsymbol{n_d} +$$

$$\tfrac{1}{2} \sum_{J=0,2,4} c_J' \sqrt{(2J + 1)}\, [[\boldsymbol{d}^+ \times \boldsymbol{d}^+]^{(J)} \times [\boldsymbol{d}^{\sim} \times \boldsymbol{d}^{\sim}]^{(J)}]^{(0)} +$$

$$\sqrt{(\tfrac{1}{2})}\, v_2 ([[\boldsymbol{d}^+ \times \boldsymbol{d}^+]^{(2)} \times \boldsymbol{d}^{\sim}\boldsymbol{s}]^{(0)} + [\boldsymbol{s}^+\boldsymbol{d}^+ \times [\boldsymbol{d}^{\sim} \times \boldsymbol{d}^{\sim}]^{(2)}]^{(0)}) + \qquad (7.31)$$

$$\tfrac{1}{2} v_0 ([[\boldsymbol{d}^+ \times \boldsymbol{d}^+]^{(0)} \times \boldsymbol{ss}]^{(0)} + [\boldsymbol{s}^+\boldsymbol{s}^+ \times [\boldsymbol{d}^{\sim} \times \boldsymbol{d}^{\sim}]^{(0)}]^{(0)})$$

with $e = e_d - e_s + (u_2/\sqrt{5} - u_0)(N - 1)$,

$$c_J' = c_J + u_0 - 2u_2/\sqrt{5}. \qquad (7.32)$$

Since the operators $\boldsymbol{H}$ and $\boldsymbol{N}$ commute (7.21), the operator $\boldsymbol{N}$ has been replaced by the number $N$. The simplified form (7.31) of the Hamilton operator contains one free parameter less than the basic form (7.17). However, if only the spectrum of a single nucleus has to be investigated neglecting the binding energy relative to other nuclei one can disregard the first two terms in $\boldsymbol{H}$ (7.31). They contribute the same amount to every energy level. The following 6 parameters are left over

$$e,\ c_0',\ c_2',\ c_4',\ v_0 \text{ and } v_2. \qquad (7.33)$$

From (7.32) we see that these parameters don't depend directly from one another because there are enough parameters in (7.17).

All *d*- and *s*-operators in (7.17) and (7.31) couple to the angular momentum 0. Therefore they don't prefer a spatial axis and are insensitive to rotations of the co-ordinate system. Consequently, we expect that $\boldsymbol{H}$ commutes with the components of the angular momentum operator $\underline{\boldsymbol{J}}$ which will be treated in the next chapter.

# 8 The angular momentum operator of the IBM1

In the first section the commutation properties of the quantum mechanical angular momentum operators and their eigenvalues will be put together. In the second section the angular momentum operator of the IBM1 is introduced and its properties are checked. Finally, the commutator of the Hamiltonian and the angular momentum operator is investigated.

### 8.1 The angular momentum operator in quantum mechanics

Introducing the spin of particles quantum mechanics postulates spin operators $\boldsymbol{J_x}$, $\boldsymbol{J_y}$ and $\boldsymbol{J_z}$ which satisfy the same commutation relations as the angular momentum operators $\boldsymbol{L_x} = \boldsymbol{y\, p_z - z\, p_y}$, $\boldsymbol{L_y}$ and $\boldsymbol{L_z}$. Their commutation rules read

$$[\, \boldsymbol{J_x} , \boldsymbol{J_y} \,] = i\hbar \boldsymbol{J_z} \quad \text{with cyclic permutations.} \tag{8.1}$$

From these one derives the existence of quantum states $|\, j\ m \,\rangle$ where $j$ is an integer or half integer and $m$ is its $z$-component. The following eigenvalue equations hold

$$\boldsymbol{J_z}\, |\, j\ m \,\rangle = \hbar m\, |\, j\ m \,\rangle,$$

$$\boldsymbol{J}^{\mathbf{2}}\, |\, j\ m \,\rangle = (\boldsymbol{J_x}^2 + \boldsymbol{J_y}^2 + \boldsymbol{J_z}^2)\, |\, j\ m \,\rangle = \hbar^2\, j(j+1)\, |\, j\ m \,\rangle\ . \tag{8.2}$$

We define new spin- or angular momentum operators which differ in the factor $1/\hbar$ as follows

$$\boldsymbol{J_x}(\text{new}) = \boldsymbol{J_x}(\text{old})/\hbar \quad \text{etc.} \tag{8.3}$$

We rewrite equations (8.1) and (8.2) in terms of the new operators like this

$$[\, \boldsymbol{J_x} , \boldsymbol{J_y} \,] = i\boldsymbol{J_z} \quad \text{with cyclic permutations,} \tag{8.4}$$

$$\boldsymbol{J_z}\, |\, j\ m \,\rangle = m\, |\, j\ m \,\rangle, \tag{8.5}$$

$$\boldsymbol{J}^{\mathbf{2}}\, |\, j\ m \,\rangle = (\boldsymbol{J_x}^2 + \boldsymbol{J_y}^2 + \boldsymbol{J_z}^2)\, |\, j\ m \,\rangle = j\,(j+1)\, |\, j\ m \,\rangle.$$

From now on, we keep the new $J$-operators. The frequently used “ladder“ operators read

$$\boldsymbol{J_1} = -(\boldsymbol{J_x} + i\boldsymbol{J_y})/\sqrt{2},$$

$$\boldsymbol{J_{-1}} = (\boldsymbol{J_x} - i\boldsymbol{J_y})/\sqrt{2}, \tag{8.6}$$

$$\boldsymbol{J_0} = \boldsymbol{J_z}\ .$$

They satisfy the following commutation relations

$[\,\boldsymbol{J_{-1}}, \boldsymbol{J_0}\,] = \boldsymbol{J_{-1}}$,

$[\,\boldsymbol{J_{-1}}, \boldsymbol{J_1}\,] = \boldsymbol{J_0}$, (8.7)

$[\,\boldsymbol{J_0}, \boldsymbol{J_1}\,] = \boldsymbol{J_1}$ .

They are proved by inserting (8.6) and comparing with (8.4). Making use of (8.6) and (8.7) the operator $\boldsymbol{J}^2$ can be written this way

$$\boldsymbol{J}^2 = \boldsymbol{J_x}^2 + \boldsymbol{J_y}^2 + \boldsymbol{J_z}^2 = -\boldsymbol{J_1}\,\boldsymbol{J_{-1}} - \boldsymbol{J_{-1}}\,\boldsymbol{J_1} + \boldsymbol{J_0}^2 \text{ o } \sum_m (-1)^m \boldsymbol{J}_m \boldsymbol{J}_{-m} = \boldsymbol{J_0}^2 - \boldsymbol{J_0} - 2\boldsymbol{J_1}\boldsymbol{J_{-1}}. \quad (8.8)$$

## 8.2 The angular momentum operator expressed in terms of *d*-boson operators

We will construct operators in terms of creation and annihilation operators of *d*-bosons which satisfy the conditions (8.7) and which we name $\boldsymbol{J_1}$, $\boldsymbol{J_{-1}}$ and $\boldsymbol{J_0}$ . At the same time, making use of (8.6) one obtains operators $\boldsymbol{J_x}$, $\boldsymbol{J_y}$ and $\boldsymbol{J_z}$, which fulfil the rules (8.4). The operator $\boldsymbol{J}^2$ is constructed, which satisfies an eigenvalue equation simultaneously with $\boldsymbol{J_z}$ (8.5).

In the IBM1 the following expressions for the “ladder“ operators are used

$$\boldsymbol{J}_m = \sqrt{10}\,[\boldsymbol{d}^+ \times \boldsymbol{d}^{\sim}]^{(1)}_m\,,\ \ m = 1, 0, -1. \quad (8.9)$$

With the help of (A8.4), one recognizes that these operators satisfy the commutation relations (8.7). In order to show that, we put *J'* = *J* “ = 1 in the equation (A8.4) which yields *k* = 1. We obtain

$$10[[\boldsymbol{d}^+ \times \boldsymbol{d}^{\sim}]^{(1)}_{M'}\,, [\boldsymbol{d}^+ \times \boldsymbol{d}^{\sim}]^{(1)}_{M''}] =$$

$$10[\boldsymbol{d}^+ \times \boldsymbol{d}^{\sim}]^{(1)}_{M'+M''}\, 3(1\, M'\, 1\, M'' \mid 1, M' + M'')\{^1{}_2\, ^1{}_2\, ^1{}_2\}(-2).$$

Because of $\{^1{}_2\, ^1{}_2\, ^1{}_2\} = 1/(6\sqrt{5})$ (see A3.13), (8.9) and

$(1, -1\ 1\ 0 \mid 1, -1) = (1, -1\ 1\ 1 \mid 1\ 0) = (1\ 0\ 1\ 1 \mid 1\ 1) = -1/\sqrt{2}$ we have

$$10[[\boldsymbol{d}^+ \times \boldsymbol{d}^{\sim}]^{(1)}_{M'}\,, [\boldsymbol{d}^+ \times \boldsymbol{d}^{\sim}]^{(1)}_{M''}] = \sqrt{10}[\boldsymbol{d}^+ \times \boldsymbol{d}^{\sim}]^{(1)}_{M'+M''} \text{ with } M' < M'' \quad (8.10)$$

in agreement with (8.7).

As an example we check the effect of the operators $\boldsymbol{J_0}$ and $\boldsymbol{J}^2$ on the single-boson state $\boldsymbol{d}^+_n\ |\rangle$.

$$\boldsymbol{J_0}\,\boldsymbol{d}^+_n\ |\rangle = \sqrt{10}\,[\boldsymbol{d}^+ \times \boldsymbol{d}^{\sim}]^{(1)}_0\,\boldsymbol{d}^+_n\ |\rangle =$$

$$\sqrt{10}\,\sum_m (2\ m\ 2, -m \mid 1\ 0)\boldsymbol{d}^+_m (-1)^{-m}\boldsymbol{d}_m \boldsymbol{d}^+_n\,|\rangle. \quad (8.11)$$

Because of $\boldsymbol{d}_m\boldsymbol{d}^+_n = d_{mn} + \boldsymbol{d}^+_n\,\boldsymbol{d}_m$ and $(2\ n\ 2, -n \mid 1\ 0) = (-1)^{-n} n/\sqrt{10}$ (A1.16) the relation

$$\boldsymbol{J_0}\,\boldsymbol{d}^+_n\ |\rangle = \sqrt{10}\ (2\ n\ 2, -n \mid 1\ 0)(-1)^{-n}\boldsymbol{d}^+_n\,|\rangle = n\,\boldsymbol{d}^+_n\,|\rangle \quad (8.12)$$

follows in agreement with (8.5). Furthermore, with the aid of (8.8) we calculate

$$\boldsymbol{J}^2\boldsymbol{d}^+_n\,|\rangle = (\boldsymbol{J_0}^2 - \boldsymbol{J_0} - 2\boldsymbol{J_1}\boldsymbol{J_{-1}})\,\boldsymbol{d}^+_n\,|\rangle =$$

$$(n^2 - n - 2\cdot 10\,[\boldsymbol{d}^+\times\boldsymbol{d}^{\sim}]^{(1)}{}_1[\boldsymbol{d}^+\times\boldsymbol{d}^{\sim}]^{(1)}{}_{-1})\,\boldsymbol{d}^+_n\,|\rangle.$$

Due to $(2,\, n - 1,\, 2,\, -n\;|\;1,\,-1)^2 = (3 - n)(2 + n)/20$ (A1.16) we obtain

$2\cdot 10\,[\boldsymbol{d}^+\times\boldsymbol{d}^{\sim}]^{(1)}{}_1[\boldsymbol{d}^+\times\boldsymbol{d}^{\sim}]^{(1)}{}_{-1}\,\boldsymbol{d}^+_n\,|\rangle =$

$2\cdot 10\,\Sigma_{l_1\,l_2}\Sigma_{m_1\,m_2}(2\;l_1\;2\;l_2\,|\,1\;1)(2\;m_1\;2\;m_2\,|\,1,\,-1)\,\cdot$

$\boldsymbol{d}^+_{l_1}(-1)^{\,l_2}\boldsymbol{d}_{-l_2}\boldsymbol{d}^+_{m_1}(-1)^{m_2}\boldsymbol{d}_{-m_2}\boldsymbol{d}^+_n\,|\rangle =$

$2\cdot 10\,\Sigma_{l_1\,l_2}(2\;l_1\;2\;l_2\,|\,1\;1)(2,\,n-1,\,2,\,-n\,|\,1,\,-1)\boldsymbol{d}^+_{l_1}(-1)^{\,l_2}\boldsymbol{d}_{-l_2}\boldsymbol{d}^+_{n-1}(-1)^{-n}\,|\rangle =$

$-2\cdot 10(2,\,n-1,\,2,\,-n\;|\;1,\,-1)^2\,\boldsymbol{d}^+_n\,|\rangle = -(3 - n)(2 + n)\boldsymbol{d}^+_n\,|\rangle$ .

That is $\boldsymbol{J}^2\boldsymbol{d}^+_n\,|\rangle = (n^2 - n + 6 + n - n^2)\,{}^2\boldsymbol{d}^+_n\,|\rangle = 2(2 + 1)\boldsymbol{d}^+_n\,|\rangle$ . (8.13)

The result (8.13) agrees with (8.5). According to (A1.22) and (A1.26) $\boldsymbol{J_0}$ and $\boldsymbol{J}^2$ act analogously on every coupled state like $[\boldsymbol{d}^+\times\boldsymbol{d}^+]^{(J)}{}_M\,|\rangle$ which can be generalised to all states of the spherical basis (3.8, 6.18).

With the help of (8.8) and (8.9) we write the operators $\boldsymbol{J_z}$ and $\boldsymbol{J}^2$ explicitly making use of (6.19), (8.11) and (A1.16) as follows

$\boldsymbol{J_z} = \boldsymbol{J_0} = \sqrt{10}\,[\boldsymbol{d}^+\times\boldsymbol{d}^{\sim}]^{(1)}{}_0 = \sqrt{10}\,\Sigma_m(2\;m\;2,\,-m\,|\,1\;0)\boldsymbol{d}^+_m\boldsymbol{d}^{\sim}_{-m} = \Sigma_m m\,\boldsymbol{d}^+_m\boldsymbol{d}_m$ ,

$\boldsymbol{J}^2 = \Sigma_m(-1)^m\,\boldsymbol{J}_m\boldsymbol{J}_{-m} = -\sqrt{3}\Sigma_m(-1)^{1-m}/\sqrt{3}\;\boldsymbol{J}_m\boldsymbol{J}_{-m} =$ (8.14)

$-\sqrt{3}\Sigma_m(1\;m\;1,\,-m\,|\,0\;0)\,\boldsymbol{J}_m\boldsymbol{J}_{-m} = -\sqrt{3}[\boldsymbol{J}\times\boldsymbol{J}]^{(0)} = -10\sqrt{3}\cdot[[\boldsymbol{d}^+\times\boldsymbol{d}^{\sim}]^{(1)}\times[\boldsymbol{d}^+\times\boldsymbol{d}^{\sim}]^{(1)}]^{(0)}$ .

## 8.3 The conservation of angular momentum

At the end of the last chapter, we made a supposition which can be written as

$[\,\boldsymbol{J_z}\,,\,\boldsymbol{H}\,] = 0$ and $[\,\boldsymbol{J}^2\,,\,\boldsymbol{H}\,] = 0.$ (8.15)

We check the first commutator in (8.15) and pick just the most complex part of $\boldsymbol{H}$ (7.31) out, namely the terms with the coefficients $c_J$. They contain linear combinations of expressions like

$$\boldsymbol{d}^+_m\boldsymbol{d}^+_{m'}\boldsymbol{d}_n\,\boldsymbol{d}_{n'} \qquad (8.16)$$

with the condition $m + m' - n - n' = 0$ (8.17)

which results from (6.19) and from the coupling yielding the angular momentum zero. We let $\boldsymbol{J_z}$ act on a term like (8.16) and push it step by step to the right hand side as follows

$$\boldsymbol{J_z}\, \boldsymbol{d}^+_m \boldsymbol{d}^+_{m'} \boldsymbol{d}_n\, \boldsymbol{d}_{n'} = \sum_{mo} m_O\, \boldsymbol{d}^+_{mo} \boldsymbol{d}_{mo} \boldsymbol{d}^+_m \boldsymbol{d}^+_{m'} \boldsymbol{d}_n\, \boldsymbol{d}_{n'} =$$

$$\sum_{mo} m_O\, (\boldsymbol{d}^+_{mo} \boldsymbol{d}^+_{m'} \boldsymbol{d}_n\, \boldsymbol{d}_{n'}\, d_{mo\, m} + \boldsymbol{d}^+_m \boldsymbol{d}^+_{mo} \boldsymbol{d}_{mo} \boldsymbol{d}^+_{m'} \boldsymbol{d}_n\, \boldsymbol{d}_{n'}) , \qquad (8.18)$$

where the commutation relations

$$\boldsymbol{d}_{mo}\, \boldsymbol{d}^+_m = d_{mo\, m} + \boldsymbol{d}^+_m \boldsymbol{d}_{mo} \qquad (8.19)$$

have been applied. The first sum on the right hand side of (8.18) yields one single term. An analogous step results in

$$\boldsymbol{J_z}\, \boldsymbol{d}^+_m \boldsymbol{d}^+_{m'} \boldsymbol{d}_n\, \boldsymbol{d}_{n'} = \qquad (8.20)$$

$$m\, \boldsymbol{d}^+_m \boldsymbol{d}^+_{m'} \boldsymbol{d}_n\, \boldsymbol{d}_{n'} + \sum_{mo} m_O\, (\boldsymbol{d}^+_m \boldsymbol{d}^+_{mo}\, \boldsymbol{d}_n\, \boldsymbol{d}_{n'}\, d_{mo\, m'} + \boldsymbol{d}^+_m \boldsymbol{d}^+_{m'} \boldsymbol{d}^+_{mo}\, \boldsymbol{d}_n\, \boldsymbol{d}_{n'}\, \boldsymbol{d}_{mo} ).$$

The following step produces

$$\boldsymbol{J_z}\, \boldsymbol{d}^+_m \boldsymbol{d}^+_{m'} \boldsymbol{d}_n\, \boldsymbol{d}_{n'} = (m + m' - n)\ \boldsymbol{d}^+_m \boldsymbol{d}^+_{m'} \boldsymbol{d}_n\, \boldsymbol{d}_{n'} + \sum_{mo} m_O\, \boldsymbol{d}^+_m \boldsymbol{d}^+_{m'} \boldsymbol{d}_n\, \boldsymbol{d}^+_{mo}\, \boldsymbol{d}_{n'}\, \boldsymbol{d}_{mo} =$$

$$(m + m' - n\ - n')\ \boldsymbol{d}^+_m \boldsymbol{d}^+_{m'} \boldsymbol{d}_n\, \boldsymbol{d}_{n'} + \boldsymbol{d}^+_m \boldsymbol{d}^+_{m'} \boldsymbol{d}_n\, \boldsymbol{d}_{n'}\, \boldsymbol{J_z}. \qquad (8.21)$$

The last term but one in (8.21) vanishes because of (8.17). Therefore $\boldsymbol{J_z}$ commutes with $\boldsymbol{d}^+_m \boldsymbol{d}^+_{m'} \boldsymbol{d}_n\, \boldsymbol{d}_{n'}$ and consequently with the first sum in (7.31). In the same way one shows that $\boldsymbol{J_z}$ commutes with expressions such as

$$\boldsymbol{d}^+_m \boldsymbol{d}_m\, ,\ \boldsymbol{d}^+_m \boldsymbol{d}^+_{-m}\, ,\ \boldsymbol{d}_m\, \boldsymbol{d}_{-m}\, ,\ \boldsymbol{d}^+_m \boldsymbol{d}^+_{m'} \boldsymbol{d}_n\ \ (m + m' - n = 0) \text{ etc.}$$

Therefore [ $\boldsymbol{J_z}$ , $\boldsymbol{H}$ ] = 0 (8.15) is satisfied.

Since in $\boldsymbol{H}$ no spatial axis prevails the analogous relations [ $\boldsymbol{J_x}$ , $\boldsymbol{H}$ ] = [ $\boldsymbol{J_y}$ , $\boldsymbol{H}$ ] = 0 and consequently [ $\boldsymbol{J}^2$ , $\boldsymbol{H}$ ] = 0 ( see 8.15) are true. From this, the well-known fact follows that the eigenstates of $\boldsymbol{H}$ have a defined angular momentum *J.*

# 9 The Hamiltonian expressed in terms of Casimir operators

In this chapter the Hamilton operator (7.17) will be transformed to a linear combination of Casimir operators. These are characteristic of Lie algebras ( chapter 13 ) and permit to express the eigenenergies of the IBM in a closed form in some special cases. Essentially, we follow the reports of Castaños et al. (1979) and Eisenberg and Greiner (1987).

In chapter 14 we will show that the number operator $\boldsymbol{n_d}$ of $d$-bosons (6.35), the number operator $\boldsymbol{N}$ of all bosons (6.37) and the angular momentum operator $\boldsymbol{J}^2$ belong to the type of Casimir operators. We give them again

$$\boldsymbol{n_d} = \sqrt{5}\,[\boldsymbol{d}^+ \times \boldsymbol{d}^\sim]^{(0)} = \sum_m \boldsymbol{d}^+_m \boldsymbol{d}_m,$$

$$\boldsymbol{N} = \boldsymbol{s}^+\boldsymbol{s} + \boldsymbol{n_d}, \qquad (9.1)$$

$$\boldsymbol{J^2} = -\sqrt{3\cdot 10}\,[[\boldsymbol{d}^+ \times \boldsymbol{d}^\sim]^{(1)} \times [\boldsymbol{d}^+ \times \boldsymbol{d}^\sim]^{(1)}\,]^{(0)}\,.$$

The operator

$$\boldsymbol{T}^2 = \boldsymbol{n_d}\,(\boldsymbol{n_d} + 3) - 5\cdot[\boldsymbol{d}^+ \times \boldsymbol{d}^+]^{(0)}\,[\boldsymbol{d}^\sim \times \boldsymbol{d}^\sim]^{(0)} \qquad (9.2)$$

is also of this kind and we name it seniority operator. It shows a certain relationship to the following operator

$$\boldsymbol{R}^2 = \boldsymbol{N}(\boldsymbol{N} + 4) - (\sqrt{5}\cdot[\boldsymbol{d}^+ \times \boldsymbol{d}^+]^{(0)} - \boldsymbol{s}^+\boldsymbol{s}^+)(\sqrt{5}[\boldsymbol{d}^\sim \times \boldsymbol{d}^\sim]^{(0)} - \boldsymbol{ss}). \qquad (9.3)$$

$\boldsymbol{R}^2$ is identical with $\boldsymbol{C_{o(6)}}$ (14.53) but it differs from the corresponding expression employed by Eisenberg and Greiner (1987, p.398) in the signs of $\boldsymbol{s}^+\boldsymbol{s}^+$ and $\boldsymbol{ss}$. The operator

$$\boldsymbol{Q}^2 = \sqrt{5}\cdot[\boldsymbol{Q} \times \boldsymbol{Q}]^{(0)} = \sum_m (-1)^m\, \boldsymbol{Q}_m \boldsymbol{Q}_{-m}$$

with $\quad \boldsymbol{Q}_m = \boldsymbol{d}^+_m \boldsymbol{s} + \boldsymbol{s}^+\boldsymbol{d}^\sim_m - (\sqrt{7}/2)\,[\boldsymbol{d}^+ \times \boldsymbol{d}^\sim]^{(2)}_m \qquad (9.4)$

is related to the quadruple moment of the nucleus. Several authors place a positive sign in front of the last term in (9.4).

Now we replace all components of the Hamilton operator (7.17) step by step by expressions constituted by the operators (9.1) up to (9.4). First, we deal with terms containing exclusively $d$-operators. Besides $\boldsymbol{n_d}$ in (7.17) we have

$$\tfrac{1}{2}\, c_J \sqrt{(2J + 1)}\,[[\boldsymbol{d}^+ \times \boldsymbol{d}^+]^{(J)} \times [\boldsymbol{d}^\sim \times \boldsymbol{d}^\sim]^{(J)}]^{(0)}, \qquad J = 0, 2, 4. \qquad (9.5)$$

It is easy to treat the term with $J = 0$ in (9.5). The relation (9.2) reveals

$$[[\boldsymbol{d}^+ \times \boldsymbol{d}^+]^{(0)} \times [\boldsymbol{d}^\sim \times \boldsymbol{d}^\sim]^{(0)}]^{(0)} = (1/5)(\boldsymbol{n_d}(\boldsymbol{n_d} + 3) - \boldsymbol{T}^2). \qquad (9.6)$$

Since the operators $\boldsymbol{n_d}$, $\boldsymbol{J}^2$ and $\boldsymbol{Q}_m$ are built with terms of the type $[\boldsymbol{d}^+ \times \boldsymbol{d}^{\sim}]^{(J)}_M$ , representing the operators $[[\boldsymbol{d}^+ \times \boldsymbol{d}^+]^{(J)} \times [\boldsymbol{d}^{\sim} \times \boldsymbol{d}^{\sim}]^{(J)}]^{(0)}$ by the same terms is advisable. According to (A4.2) and (A4.3) and analogously to (7.26) the following relation holds

$$[[\boldsymbol{d}^+ \times \boldsymbol{d}^+]^{(J)} \times [\boldsymbol{d}^{\sim} \times \boldsymbol{d}^{\sim}]^{(J)}]^{(0)} = \sum_{J'} (-1)^{J'} \sqrt{(2J+1)}\sqrt{(2J'+1)}\{^{2}_{2}\,^{2}_{2}\,^{J'}_{J}\} \cdot$$

$$\sum_{Mm\,m} (J'\,M\,J', -M \mid 0\,0)(2\,m\,2, M - m \mid J'\,M)(2\,m\,2, -M - m \mid J', -M) \cdot$$

$$\boldsymbol{d}^+_m \boldsymbol{d}^+_m \boldsymbol{d}^{\sim}_{M-m} \boldsymbol{d}^{\sim}_{-M-m} \,. \qquad (9.7)$$

Almost with the same method which yielded the equations (7.29) and (7.30), starting from (9.7) one obtains

$$[[\boldsymbol{d}^+ \times \boldsymbol{d}^+]^{(J)} \times [\boldsymbol{d}^{\sim} \times \boldsymbol{d}^{\sim}]^{(J)}]^{(0)} =$$

$$\sum_{J'} (-1)^{J'} \sqrt{(2J+1)}\sqrt{(2J'+1)}\{^{2}_{2}\,^{2}_{2}\,^{J'}_{J}\} \cdot [[\boldsymbol{d}^+ \times \boldsymbol{d}^{\sim}]^{(J')} \times [\boldsymbol{d}^+ \times \boldsymbol{d}^{\sim}]^{(J')}]^{(0)} -$$

$$(1/5)\sqrt{(2J+1)}\boldsymbol{n}_d \,. \qquad (9.8)$$

Consequently, we now treat the terms

$$\boldsymbol{W}_{J'} \text{ o } [[\boldsymbol{d}^+ \times \boldsymbol{d}^{\sim}]^{(J')} \times [\boldsymbol{d}^+ \times \boldsymbol{d}^{\sim}]^{(J')}]^{(0)} \qquad (9.9)$$

in order to represent the expressions (9.5) showing $J$ = 2 and 4 by means of Casimir operators. First, we take from (9.1) the following relations

$$\boldsymbol{W_0} \text{ o } [[\boldsymbol{d}^+ \times \boldsymbol{d}^{\sim}]^{(0)} \times [\boldsymbol{d}^+ \times \boldsymbol{d}^{\sim}]^{(0)}]^{(0)} = (1/5)\boldsymbol{n_d}^2 \qquad (9.10)$$

and $$\boldsymbol{W_1} \text{ o } [[\boldsymbol{d}^+ \times \boldsymbol{d}^{\sim}]^{(1)} \times [\boldsymbol{d}^+ \times \boldsymbol{d}^{\sim}]^{(1)}]^{(0)} = -\,(10\sqrt{3})^{-1}\boldsymbol{J}^2. \qquad (9.11)$$

For the operators $\boldsymbol{W_2}$, $\boldsymbol{W_3}$ and $\boldsymbol{W_4}$ we draw up a system of linear equations. For that purpose we represent the left-hand side of (9.6) with (9.8) too and get by means of (A3.12)

$$\boldsymbol{n_d}(\boldsymbol{n_d} + 3) - \boldsymbol{T}^2 + \boldsymbol{n_d} = \sum_{J'} \sqrt{(2J'+1)} \cdot [[\boldsymbol{d}^+ \times \boldsymbol{d}^{\sim}]^{(J')} \times [\boldsymbol{d}^+ \times \boldsymbol{d}^{\sim}]^{(J')}]^{(0)} \,. \qquad (9.12)$$

We obtain further equations from the fact that the terms $\boldsymbol{W}_J$ are linearly dependent on one another owing to their symmetry. In order to make this visible, first we interchange the $d$-operators in $[\boldsymbol{d}^+ \times \boldsymbol{d}^{\sim}]^{(J')}_M$ :

$$[\boldsymbol{d}^+ \times \boldsymbol{d}^{\sim}]^{(J)}_M = \sum_m (2\,m\,2, M - m \mid J\,M)(-1)^{M-m} \boldsymbol{d}^+_m \boldsymbol{d}_{m-M} = \qquad (9.13)$$

$$-\sum_m (2\,m\,2, M - m \mid J\,M)(-1)^{M-m}\, d_{m,\,m-M} + (-1)^J [\boldsymbol{d}^{\sim} \times \boldsymbol{d}^+]^{(J)}_M \,.$$

In particular the expression

$$\sum_m (2\,m\,2, M - m \mid J\,M)(-1)^{M-m}\, d_{m,\,m-M} =$$

$$\sum_m (2\,m\,2, -m \mid J\,0)(-1)^{-m} = \sqrt{5}\sum_m (2\,m\,2, -m \mid J\,0)(-1)^{-m}/\sqrt{5} = \qquad (9.14)$$

$$\sqrt{5}\sum_m (2\,m\,2, -m \mid J\,0)(2\,m\,2, -m \mid 0\,0) = \sqrt{5}\cdot d_{J0}$$

can be inserted in (9.13) yielding the following equation

$$[\boldsymbol{d}^{+} \times \boldsymbol{d}^{\sim}]^{(J)}{}_{M} = -\sqrt{5}\cdot d_{J0} + (-1)^{J}[\boldsymbol{d}^{\sim} \times \boldsymbol{d}^{+}]^{(J)}{}_{M}. \qquad (9.15)$$

The inverse relation reads $[\boldsymbol{d}^{\sim} \times \boldsymbol{d}^{+}]^{(J)}{}_{M} = \sqrt{5}\cdot d_{J0} + (-1)^{J}[\boldsymbol{d}^{+} \times \boldsymbol{d}^{\sim}]^{(J)}{}_{M}$ . (9.16)

Bringing (9.15) in the expression (9.9) we obtain

$$[[\boldsymbol{d}^{+} \times \boldsymbol{d}^{\sim}]^{(J')} \times [\boldsymbol{d}^{+} \times \boldsymbol{d}^{\sim}]^{(J')}]^{(0)} = (-1)^{J'} [[\boldsymbol{d}^{+} \times \boldsymbol{d}^{\sim}]^{(J')} \times [\boldsymbol{d}^{\sim} \times \boldsymbol{d}^{+}]^{(J')}]^{(0)} -$$

$$\sqrt{5}[\boldsymbol{d}^{+} \times \boldsymbol{d}^{\sim}]^{(0)}\, d_{J'0}\,. \qquad (9.17)$$

Now we interchange both facing operators $\boldsymbol{d}^{\sim}$ on the right hand side of (9.17) by means of (A4.1) and (A4.2) which results in

$$[[\boldsymbol{d}^{+} \times \boldsymbol{d}^{\sim}]^{(J')} \times [\boldsymbol{d}^{+} \times \boldsymbol{d}^{\sim}]^{(J')}]^{(0)} = -\sqrt{5}[\boldsymbol{d}^{+} \times \boldsymbol{d}^{\sim}]^{(0)}\, d_{J',0} + \qquad (9.18)$$

$$(-1)^{J'} \sum_{J''} \sqrt{(2J'+1)}\sqrt{(2J''+1)}\{^{2}{}_{2}\,{}^{2}{}_{2}\,{}^{J'}{}_{J''}\}(-1)^{J'+J''}\, [[\boldsymbol{d}^{+} \times \boldsymbol{d}^{\sim}]^{(J'')} \times [\boldsymbol{d}^{\sim} \times \boldsymbol{d}^{+}]^{(J'')}]^{(0)}.$$

In (9.18) we interchange the boson operators of the expression $[\boldsymbol{d}^{\sim} \times \boldsymbol{d}^{+}]^{(J'')}$ (9.16) and obtain

$$[[\boldsymbol{d}^{+} \times \boldsymbol{d}^{\sim}]^{(J')} \times [\boldsymbol{d}^{+} \times \boldsymbol{d}^{\sim}]^{(J')}]^{(0)} = -\sqrt{5}[\boldsymbol{d}^{+} \times \boldsymbol{d}^{\sim}]^{(0)}\, d_{J',0} +$$

$$\sqrt{(2J'+1)}\sum_{J''}\sqrt{(2J''+1)}\{^{2}{}_{2}\,{}^{2}{}_{2}\,{}^{J'}{}_{J''}\}[[\boldsymbol{d}^{+} \times \boldsymbol{d}^{\sim}]^{(J'')} \times [\boldsymbol{d}^{+} \times \boldsymbol{d}^{\sim}]^{(J'')}]^{(0)} +$$

$$\sqrt{(2J'+1)}\sum_{J''}\sqrt{(2J''+1)}\{^{2}{}_{2}\,{}^{2}{}_{2}\,{}^{J'}{}_{J''}\}\sqrt{5}[\boldsymbol{d}^{+} \times \boldsymbol{d}^{\sim}]^{(0)}\, d_{J'',0} =$$

$$-\sqrt{5}[\boldsymbol{d}^{+} \times \boldsymbol{d}^{\sim}]^{(0)}\, d_{J',0} +$$

$$\sqrt{(2J'+1)}\sum_{J''}\sqrt{(2J''+1)}\{^{2}{}_{2}\,{}^{2}{}_{2}\,{}^{J'}{}_{J''}\}[[\boldsymbol{d}^{+} \times \boldsymbol{d}^{\sim}]^{(J'')} \times [\boldsymbol{d}^{+} \times \boldsymbol{d}^{\sim}]^{(J'')}]^{(0)} +$$

$$\sqrt{(2J'+1)}(-1)^{J'}(1/\sqrt{5})[\boldsymbol{d}^{+} \times \boldsymbol{d}^{\sim}]^{(0)}. \qquad (9.19)$$

In the last term we made use of (A3.12). Employing (A3.12) we write (9.19) for $J' = 0$ and 1 in the following form

$$\boldsymbol{W_0} \circ [[\boldsymbol{d}^{+} \times \boldsymbol{d}^{\sim}]^{(0)} \times [\boldsymbol{d}^{+} \times \boldsymbol{d}^{\sim}]^{(0)}]^{(0)} =$$

$$(1/5)\sum_{J''}(-1)^{J''}\sqrt{(2J''+1)}[[\boldsymbol{d}^{+} \times \boldsymbol{d}^{\sim}]^{(J'')} \times [\boldsymbol{d}^{+} \times \boldsymbol{d}^{\sim}]^{(J'')}]^{(0)} +$$

$$(1/\sqrt{5})[\boldsymbol{d}^{+} \times \boldsymbol{d}^{\sim}]^{(0)} - \sqrt{5}[\boldsymbol{d}^{+} \times \boldsymbol{d}^{\sim}]^{(0)}\,, \qquad (9.20)$$

$$\boldsymbol{W_1} \circ [[\boldsymbol{d}^{+} \times \boldsymbol{d}^{\sim}]^{(1)} \times [\boldsymbol{d}^{+} \times \boldsymbol{d}^{\sim}]^{(1)}]^{(0)} =$$

$$\sqrt{3}\sum_{J''}\sqrt{(2J''+1)}\{^{2}{}_{2}\,{}^{2}{}_{2}\,{}^{1}{}_{J''}\}[[\boldsymbol{d}^{+} \times \boldsymbol{d}^{\sim}]^{(J'')} \times [\boldsymbol{d}^{+} \times \boldsymbol{d}^{\sim}]^{(J'')}]^{(0)} - \sqrt{(3/5)}[\boldsymbol{d}^{+} \times \boldsymbol{d}^{\sim}]^{(0)}. \qquad (9.21)$$

We calculate the 6-*j* symbols with the aid of (A3.13) and insert (9.9) up to (9.11) in (9.12), (9.20) and (9.21). Doing so and making use of (A3.13), we obtain an inhomogeneous and linear system of equations for $\boldsymbol{W_2}$, $\boldsymbol{W_3}$ and $\boldsymbol{W_4}$ (9.9). We write down the solutions of the system and add the expressions (9.10) and (9.11) as follows

$$W_0 = (1/5)n_d^2,$$

$$W_1 = -(10\sqrt{3})^{-1}J^2,$$

$$W_2 = 2(7\sqrt{5})^{-1}(n_d^2 + 5n_d - T^2 + (1/4)J^2), \qquad (9.22)$$

$$W_3 = -(2\sqrt{7})^{-1}(T^2 - (1/5)J^2),$$

$$W_4 = (1/7)((6/5)n_d^2 + 6n_d - (1/2)T^2 - (1/6)J^2).$$

Inserting these expressions in (9.8) yields for $J = 2$

$$[[d^+ \times d^+]^{(2)} \times [d^\sim \times d^\sim]^{(2)}]^{(0)} = (7\sqrt{5})^{-1}(2n_d^2 - 4n_d + 2T^2 - J^2). \qquad (9.23)$$

In order to calculate the corresponding expression for $J = 4$ we take the 6-$j$ symbols from (A3.14) and obtain

$$[[d^+ \times d^+]^{(4)} \times [d^\sim \times d^\sim]^{(4)}]^{(0)} =$$

$$(1/7)((6/5)n_d^2 - (12/5)n_d - (1/5)T^2 + (1/3)J^2). \qquad (9.24)$$

Thus we have treated all terms in the Hamiltonian (7.17) which are constituted exclusively by $d$-operators and we are turning to the other terms. Owing to (7.22) and (7.23) we have

$$[s^+s^+ \times ss]^{(0)} = (N - n_d)(N - n_d - 1), \qquad (9.25)$$

$$[d^+s^+ \times d^\sim s]^{(0)} = (1/\sqrt{5})(N - n_d)n_d\,. \qquad (9.26)$$

The preceding term in (7.17) reads

$$[[d^+ \times d^+]^{(0)} \times ss]^{(0)} + [s^+s^+ \times [d^\sim \times d^\sim]^{(0)}]^{(0)} =$$

$$[d^+ \times d^+]^{(0)}ss + s^+s^+[d^\sim \times d^\sim]^{(0)} =$$

$$-(1/\sqrt{5})(-R^2 + N(N + 4) + T^2 - n_d(n_d + 3) - s^+s^+ss) = \qquad (9.27)$$

$$-(1/\sqrt{5})(-R^2 + T^2 + N + (N - n_d)(2n_d + 4))$$

where (9.2), (9.3) and (9.25) have been employed. In order to work on the remaining term in $H$ (7.17) we write (9.4) this way

$$Q^2/\sqrt{5} = [(d^+s + s^+d^\sim - (\sqrt{7}/2)[d^+ \times d^\sim]^{(2)}) \times (d^+s + s^+d^\sim - (\sqrt{7}/2)[d^+ \times d^\sim]^{(2)})]^{(0)}$$

which we transform to

$$Q^2/\sqrt{5} = [d^+ \times d^+]^{(0)}ss + s^+s^+[d^\sim \times d^\sim]^{(0)} + (7/4)[[d^+ \times d^\sim]^{(2)} \times [d^+ \times d^\sim]^{(2)}]^{(0)} +$$

$$ss^+[d^+ \times d^\sim]^{(0)} + [d^\sim \times d^+]^{(0)}s^+s -$$

$$(\sqrt{7}/2)s([d^+ \times [d^+ \times d^\sim]^{(2)}]^{(0)} + [[d^+ \times d^\sim]^{(2)} \times d^+]^{(0)}) - \qquad (9.28)$$

$$(\sqrt{7}/2)s^+([d^\sim \times [d^+ \times d^\sim]^{(2)}]^{(0)} + [[d^+ \times d^\sim]^{(2)} \times d^\sim]^{(0)}).$$

The first line in (9.28) contains the relation (9.27). We write the following term by means of (9.22) as

$$(7/4)[[\boldsymbol{d}^+\times\boldsymbol{d}^\sim]^{(2)}\times[\boldsymbol{d}^+\times\boldsymbol{d}^\sim]^{(2)}]^{(0)} = (2\sqrt{5})^{-1}(\boldsymbol{n_d}^2 + 5\boldsymbol{n_d} - \boldsymbol{T}^2 + (1/4)\boldsymbol{J}^2). \qquad (9.29)$$

The second line in (9.28) is represented with the help of (9.16) and of the relation $\boldsymbol{ss}^+ = 1 + \boldsymbol{s}^+\boldsymbol{s}$ like this

$$\boldsymbol{ss}^+[\boldsymbol{d}^+\times\boldsymbol{d}^\sim]^{(0)} + [\boldsymbol{d}^\sim\times\boldsymbol{d}^+]^{(0)}\boldsymbol{s}^+\boldsymbol{s} =$$

$$[\boldsymbol{d}^+\times\boldsymbol{d}^\sim]^{(0)} + [\boldsymbol{d}^+\times\boldsymbol{d}^\sim]^{(0)}\,\boldsymbol{s}^+\boldsymbol{s} + \sqrt{5}\,\boldsymbol{s}^+\boldsymbol{s} + [\boldsymbol{d}^+\times\boldsymbol{d}^\sim]^{(0)}\,\boldsymbol{s}^+\boldsymbol{s} =$$

$$(1/\sqrt{5})\boldsymbol{n_d} + \sqrt{5}(\boldsymbol{N} - \boldsymbol{n_d}) + (2/\sqrt{5})\boldsymbol{n_d}(\boldsymbol{N} - \boldsymbol{n_d}) = \qquad (9.30)$$

$$(1/\sqrt{5})(\boldsymbol{n_d} + (5 + 2\boldsymbol{n_d})(\boldsymbol{N} - \boldsymbol{n}_\mathrm{d})).$$

The third line in (9.28) contains the expression

$$[[\boldsymbol{d}^+\times\boldsymbol{d}^\sim]^{(2)}\times\boldsymbol{d}^+]^{(0)} = \sum_{m_2\,m'\,m_0}(2\;m_2\;2\;m'\,|\,2\;m_0)(2\;m_0\;2,-m_0\,|\,0\;0)\;\boldsymbol{d}^+_{m_2}\boldsymbol{d}^\sim_{m'}\boldsymbol{d}^+_{-m_0}.$$

With the aid of $\boldsymbol{d}^+_{m_2}\,\boldsymbol{d}^\sim_{m'}\,\boldsymbol{d}^+_{-m_0} = (-1)^{m'}\,(\boldsymbol{d}^+_{m_2}\,d_{m'\,m_0} + \boldsymbol{d}^+_{-m_0}\,\boldsymbol{d}^+_{m_2}\,\boldsymbol{d}_{-m'})$, $(2\;m_0\;2,-m_0\,|\,0\;0) = (-1)^{-m_0}/\sqrt{5}$, $(2\;m_2\;2\;m_0\,|\,2\;m_0) = d_{m_2\,0}\,(m_0^2 - 2)/\sqrt{14}$ and of (A1.16) we obtain

$$[[\boldsymbol{d}^+\times\boldsymbol{d}^\sim]^{(2)}\times\boldsymbol{d}^+]^{(0)} =$$

$$\sum_{m_0=-2}^{2}\sum_{m_2}(2\;m_2\;2\;m_0\,|\,2\;m_0)(1/\sqrt{5})\boldsymbol{d}^+_{m_2} + [\boldsymbol{d}^+\times[\boldsymbol{d}^+\times\boldsymbol{d}^\sim]^{(2)}]^{(0)} =$$

$$0 \qquad + [\boldsymbol{d}^+\times[\boldsymbol{d}^+\times\boldsymbol{d}^\sim]^{(2)}]^{(0)} \qquad (9.31)$$

and recouple according to (A3.5) this way

$$[[\boldsymbol{d}^+\times\boldsymbol{d}^\sim]^{(2)}\times\boldsymbol{d}^+]^{(0)} = [\boldsymbol{d}^+\times[\boldsymbol{d}^+\times\boldsymbol{d}^\sim]^{(2)}]^{(0)} = \qquad (9.32)$$

$$\sum_{J_2}\sqrt{5}\,\sqrt{(2J_2+1)}\begin{Bmatrix}2 & 2 & 2\\ 2 & 0 & J_2\end{Bmatrix}[[\boldsymbol{d}^+\times\boldsymbol{d}^+]^{(J_2)}\boldsymbol{d}^\sim]^{(0)} = [[\boldsymbol{d}^+\times\boldsymbol{d}^+]^{(2)}\times\boldsymbol{d}^\sim]^{(0)}.$$

The value of the 6-*j* symbol for $J_2 = 2$ has been taken from (A3.12). Analogously the relation

$$[\boldsymbol{d}^\sim\times[\boldsymbol{d}^+\times\boldsymbol{d}^\sim]^{(2)}]^{(0)} = [[\boldsymbol{d}^+\times\boldsymbol{d}^\sim]^{(2)}\times\boldsymbol{d}^\sim]^{(0)} = [\boldsymbol{d}^+\times[\boldsymbol{d}^\sim\times\boldsymbol{d}^\sim]^{(2)}]^{(0)} \qquad (9.33)$$

holds so that the third and the fourth line in (9.28) can be written as follows

$$-(\sqrt{7}/2)(2\boldsymbol{s}[[\boldsymbol{d}^+\times\boldsymbol{d}^+]^{(2)}\times\boldsymbol{d}^\sim]^{(0)} + 2\boldsymbol{s}^+[\boldsymbol{d}^+\times[\boldsymbol{d}^\sim\times\boldsymbol{d}^\sim]^{(2)}]^{(0)}) =$$

$$-\sqrt{7}([[\boldsymbol{d}^+\times\boldsymbol{d}^+]^{(2)}\times\boldsymbol{d}^\sim\boldsymbol{s}]^{(0)} + [\boldsymbol{s}^+\boldsymbol{d}^+\times[\boldsymbol{d}^\sim\times\boldsymbol{d}^\sim]^{(2)}]^{(0)}). \qquad (9.34)$$

Apart from a factor, the expression (9.34) is just the remaining term in ***H*** (7.17). By means of (9.27) up to (9.30) and (9.34), we can write it this way

$$-\sqrt{(1/2)}\, v_2\, ([[\boldsymbol{d}^+ \times \boldsymbol{d}^+]^{(2)} \times \boldsymbol{d}^{\sim}\boldsymbol{s}]^{(0)} + [\boldsymbol{s}^+\boldsymbol{d}^+ \times [\boldsymbol{d}^{\sim} \times \boldsymbol{d}^{\sim}]^{(2)}]^{(0)}) =$$

$$(1/\sqrt{14})\, v_2\, ((1/\sqrt{5})\boldsymbol{Q}^2 + (1/\sqrt{5})(-\boldsymbol{R}^2 + \boldsymbol{T}^2 + \boldsymbol{N} + (\boldsymbol{N} - \boldsymbol{n_d})(2\boldsymbol{n_d} + 4)) -$$

$$(1/\sqrt{20})(\boldsymbol{n_d}^2 + 5\boldsymbol{n_d} - \boldsymbol{T}^2 + (1/4)\boldsymbol{J}^2) -$$

$$(1/\sqrt{5})(\boldsymbol{n_d} + (5 + 2\boldsymbol{n_d})(\boldsymbol{N} - \boldsymbol{n_d}))) = \qquad (9.35)$$

$$v_2\, (1/\sqrt{70})(\boldsymbol{Q}^2 + (3/2)\boldsymbol{T}^2 - \boldsymbol{R}^2 - (1/8)\boldsymbol{J}^2 - (1/2)\boldsymbol{n_d}^2 - (5/2)\, \boldsymbol{n_d}).$$

Now we are able to represent the Hamilton operator (7.17) by the operators cited in (9.1) up to (9.4) making use of (6.35), (6.37), (9.6), (9.23) up to (9.25), (9.27) and (9.35) like this

$$\boldsymbol{H} = e_s \boldsymbol{N} + (e_d - e_s)\boldsymbol{n_d} + c_0\, (1/10)(-\boldsymbol{T}^2 + \boldsymbol{n_d}\, (\boldsymbol{n_d} + 3)) +$$

$$c_2\, (1/14)(2\boldsymbol{n_d}^2 + 2\boldsymbol{T}^2 - 4\boldsymbol{n_d} - \boldsymbol{J}^2) +$$

$$c_4 (3/14)((6/5)\boldsymbol{n_d}^2 - (12/5)\boldsymbol{n_d} - (1/5)\boldsymbol{T}^2 + (1/3)\boldsymbol{J}^2) -$$

$$v_2\, (1/\sqrt{70})(\boldsymbol{Q}^2 + (3/2)\boldsymbol{T}^2 - \boldsymbol{R}^2 - (1/8)\boldsymbol{J}^2 - (1/2)\boldsymbol{n_d}^2 - (5/2)\boldsymbol{n_d}) -$$

$$v_o\, (1/\sqrt{20})(-\boldsymbol{R}^2 + \boldsymbol{T}^2 + 5\boldsymbol{N} + 2\boldsymbol{N}\boldsymbol{n_d} - 2\boldsymbol{n_d}^2 - 4\boldsymbol{n_d}) + \qquad (9.36)$$

$$u_2\, (1/\sqrt{5})(\boldsymbol{N}\boldsymbol{n}_{\mathrm{d}} - \boldsymbol{n_d}^2) + u_0 (1/2)(\boldsymbol{N}^2 - 2\boldsymbol{N}\boldsymbol{n_d} - \boldsymbol{N} + \boldsymbol{n_d}^2 + \boldsymbol{n_d}).$$

We summarise

$$\boldsymbol{H} = e_n \boldsymbol{N} + e_d{}' \boldsymbol{n}_d + v_n \boldsymbol{N}^2 + v_{n\,d} \boldsymbol{N}\boldsymbol{n_d} + v_d \boldsymbol{n_d}^2 + v_r \boldsymbol{R}^2 + v_t \boldsymbol{T}^2 + v_j \boldsymbol{J}^2 + v_q \boldsymbol{Q}^2 \qquad (9.37)$$

with

$$e_n = e_s - (\sqrt{5}/2)v_0 - (1/2)u_0\,,$$

$$e_d{}' = e_d - e_s + (3/10)c_0 - (2/7)c_2 - (18/35)c_4 + (5/\sqrt{280})v_2 + (2/\sqrt{5})v_0 + (1/2)u_0\,,$$

$$v_n = (1/2)u_0\,, \qquad (9.38)$$

$$v_{nd} = -\, (1/\sqrt{5})v_0 + (1/\sqrt{5})u_2 - u_0\,,$$

$$v_d = (1/10)c_0 + (1/7)c_2 + (9/35)c_4 + \sqrt{(1/\sqrt{280})}v_2 + (1/\sqrt{5})v_0 - (1/\sqrt{5})u_2 + (1/2)u_0\,,$$

$$v_r = (1/\sqrt{70})v_2 + (1/\sqrt{20})v_0\,,$$

$$v_t = -\, (1/10)c_0 + (1/7)c_2 - (3/70)c_4 - (3/\sqrt{280})v_2 - (1/\sqrt{20})v_0\,,$$

$$v_j = -\, (1/14)c_2 + (1/14)c_4 + (1/\sqrt{4480})v_2,$$

$$v_q = -\, (1/\sqrt{70})v_2\,.$$

All summands with a coefficient $v_2$ change their sign if $\boldsymbol{Q}_m$ is defined with a plus sign in front of the last term in (9.4).

In (9.37) the Hamilton operator of the IBM1 is represented by the Casimir operators (9.1) up to (9.4). This is the so-called multipole form. In the same way as in (7.21) we replace the operator ***N*** by its eigenvalue *N* and obtain

$$\boldsymbol{H} = e_n N + v_n N^2 + (e_d' + v_{nd} N)\boldsymbol{n_d} + v_d \boldsymbol{n_d}^2 + v_r \boldsymbol{R}^2 + v_t \boldsymbol{T}^2 + v_j \boldsymbol{J}^2 + v_q \boldsymbol{Q}^2. \quad (9.39)$$

If only a single nucleus is investigated the term $e_n N + v_n N^2$ is omitted. Then the eigenstates depend solely on the following six parameters

$$e_d'' = e_d' + v_{nd} N\ ,\ v_d\ ,\ v_r\ ,\ v_t\ ,\ v_j \text{ and } v_q \quad (9.40)$$

analogously to (7.33).

In order to solve eigenvalue problems usually the Hamilton matrix is diagonalised ( chapter 12 ). However, the interacting boson model has the peculiarity that for special cases (limits) the eigenenergies can be given analytically. These cases are found by allocating the value zero to several coefficients in (9.39). In chapters 10 and 14 we will deal with the physical significance of this measure and of the operators (9.1) up to (9.4) and look at the analytical solutions.

# 10 The $u(5)$- or vibrational limit

At the end of the last chapter the fact was mentioned briefly that the eigenvalues of the Hamilton operator $\boldsymbol{H}$ (9.39) can be written in a closed form if certain coefficients of this operator are zero. We deal here with a special case characterised by vanishing coefficients $v_r$ and $v_q$ , which is realised approximately by a group of real nuclei.

In the first section we will write down this special operator and remember the spherical basis of the boson model. The second section reveals that this basis contains the eigenfunctions of the seniority operator $\boldsymbol{T}^2$. In section 10.3 calculated nuclear spectra of the relevant special case will be compared with measured ones.

## 10.1 The Hamiltonian of the vibrational limit. The spherical basis

We will formulate now the first special case of the Hamiltonian by putting $v_r = v_q = 0$. This has the consequence that $v_0$ and $v_2$ vanish (see 9.38) and that in $\boldsymbol{H}$ (7.17) remain only the terms which conserve both the number of $d$-bosons and the one of the $s$-bosons ($\boldsymbol{d}^+$ is coupled to $\boldsymbol{d}^\sim$ and $\boldsymbol{s}^+$ is connected with $\boldsymbol{s}$). The Hamilton operator of this approximation reads according to (9.39)

$$\boldsymbol{H}^{(I)} = e_n N + v_n N^2 + (e_d' + v_{nd} N)\boldsymbol{n_d} + v_d \boldsymbol{n_d}^2 + v_t \boldsymbol{T}^2 + v_j \boldsymbol{J}^2 \tag{10.1}$$

with $e_n = e_s - (1/2)u_0$ ,

$$\begin{aligned} e_d' &= e_d - e_s + (3/10)c_0 - (2/7)c_2 - (18/35)c_4 + (1/2)u_0 , \\ v_n &= (1/2)u_0 , \\ v_{nd} &= (1/\sqrt{5})u_2 - u_0 , \\ v_d &= (1/10)c_0 + (1/7)c_2 + (9/35)c_4 - (1/\sqrt{5})u_2 + (1/2)u_0 , \\ v_t &= -(1/10)c_0 + (1/7)c_2 - (3/70)c_4 , \\ v_j &= -(1/14)c_2 + (1/14)c_4 . \end{aligned} \tag{10.2}$$

The coefficients in (10.2) result from the equations (9.38) taking into account the condition $v_0 = v_2 = 0$. Because the operators in (10.1) act only on $d$-boson states, the eigenfunctions of $\boldsymbol{H}^{(I)}$ are products of pure $s$-boson states and $d$-boson states undergoing the symmetrisation procedure. The states of the spherical basis (3.8) and (6.18) have this structure and we will show in this chapter that they are really eigenfunctions of $\boldsymbol{H}^{(I)}$. They are characterised by the integers $N$ ( number of bosons ), $n_d$ ( number of $d$-bosons ), $t$ ( seniority ), $n_D$ (number of $d$-boson triples resulting in the angular momentum zero ) and $J$ ( angular momentum ) and are written like this

$$|N\, n_d\; t\; n_D\, J\,\rangle. \qquad (10.3)$$

In this chapter we drop the projection $M$ of $J$. According to (6.33), (6.34) and to the section 8.2 the following eigenvalue equations hold

$$\boldsymbol{n_d}\,|N\, n_d\; t\; n_D\, J\,\rangle = n_d\,|N\, n_d\; t\; n_D\, J\,\rangle \qquad (10.4)$$

$$\boldsymbol{J}^2\,|N\, n_d\; t\; n_D\, J\,\rangle = J(J+1)\,|N\, n_d\; t\; n_D\, J\,\rangle.$$

Owing to (3.8) and (3.9) for $J$ only the following values are allowed

$$J = l,\; l+1,\; \ldots ,2l-3,\; 2l-2,\; 2l \;\text{ with }\; l = t - 3n_D \text{ and } t \leq n_d. \qquad (10.5)$$

## 10.2 Eigenvalues of the seniority operator

We now investigate the effect of the operator $\boldsymbol{T}^2$ (9.2) on a state of the spherical basis (10.3). By means of (9.2), (6.12) and (6.28) we write

$$\boldsymbol{T}^2 = \boldsymbol{n_d}(\boldsymbol{n_d}+3) - \boldsymbol{P}^+\boldsymbol{P}^- \text{ with the following details} \qquad (10.5a)$$

$$\boldsymbol{n_d} = \Sigma_m \boldsymbol{d}^+_m \boldsymbol{d}_m,\;\; \boldsymbol{P}^+ = \Sigma_m (-1)^m \boldsymbol{d}^+_m \boldsymbol{d}^+_{-m},\;\; \boldsymbol{P}^- = \Sigma_m (-1)^m \boldsymbol{d}_m \boldsymbol{d}_{-m}.$$

We look into several commutators of these operators. First we claim

$$[\boldsymbol{n_d}, \boldsymbol{P}^+] = 2\boldsymbol{P}^+, \qquad (10.6)$$

which we prove with the help of (5.36)

$$[\,\Sigma_m \boldsymbol{d}^+_m \boldsymbol{d}_m,\, 2\boldsymbol{d}^+_2\boldsymbol{d}^+_{-2} - 2\boldsymbol{d}^+_1\boldsymbol{d}^+_{-1} + \boldsymbol{d}^{+2}_0] = [\boldsymbol{d}^+_2\boldsymbol{d}_2 + \boldsymbol{d}^+_{-2}\boldsymbol{d}_{-2},\, 2\boldsymbol{d}^+_2\boldsymbol{d}^+_{-2}] -$$

$$[\boldsymbol{d}^+_1\boldsymbol{d}_1 + \boldsymbol{d}^+_{-1}\boldsymbol{d}_{-1},\, 2\boldsymbol{d}^+_1\boldsymbol{d}_{-1}] + [\boldsymbol{d}^+_0\boldsymbol{d}_0,\, \boldsymbol{d}^{+2}_0] =$$

$$2(2\boldsymbol{d}^+_2\boldsymbol{d}^+_{-2} - 2\boldsymbol{d}^+_1\boldsymbol{d}^+_{-1} + \boldsymbol{d}^{+2}_0) = 2\boldsymbol{P}^+.$$

The validity of the relation

$$[\boldsymbol{P}^-, \boldsymbol{P}^+] = 2(2\boldsymbol{n_d}+5) \qquad (10.7)$$

is shown this way

$$[\,\Sigma_m (-1)^m \boldsymbol{d}_m\boldsymbol{d}_{-m},\, \Sigma_m (-1)^m \boldsymbol{d}^+_m\boldsymbol{d}^+_{-m}\,] =$$

$$4[\boldsymbol{d}_2\boldsymbol{d}_{-2},\, \boldsymbol{d}^+_2\boldsymbol{d}^+_{-2}] + 4[\boldsymbol{d}_1\boldsymbol{d}_{-1},\, \boldsymbol{d}^+_1\boldsymbol{d}^+_{-1}] + [\boldsymbol{d}_0^2,\, \boldsymbol{d}^{+2}_0] =$$

$$4(1 + \boldsymbol{d}^+_2\boldsymbol{d}_2 + \boldsymbol{d}^+_{-2}\boldsymbol{d}_{-2}) + 4(1 + \boldsymbol{d}^+_1\boldsymbol{d}_1 + \boldsymbol{d}^+_{-1}\boldsymbol{d}_{-1}) + 2(1 + 2\boldsymbol{d}^+_0\boldsymbol{d}_0) =$$

$$10 + 4\boldsymbol{n_d} = 2(2\boldsymbol{n_d}+5).$$

The third commutator reads $[\boldsymbol{P}^-, (\boldsymbol{P}^+)^n]$. Owing to (10.7) the relation

$$\boldsymbol{P}^-(\boldsymbol{P}^+)^n = 2(2\boldsymbol{n_d}+5)(\boldsymbol{P}^+)^{n-1} + \boldsymbol{P}^+\boldsymbol{P}^-(\boldsymbol{P}^+)^{n-1} \qquad (10.8)$$

holds. We carry the procedure on like this

$$\boldsymbol{P}^-(\boldsymbol{P}^+)^n = 2(2\boldsymbol{n_d}+5)(\boldsymbol{P}^+)^{n-1} + \boldsymbol{P}^+\, 2(2\boldsymbol{n_d}+5)(\boldsymbol{P}^+)^{n-2} + (\boldsymbol{P}^+)^2\boldsymbol{P}^-(\boldsymbol{P}^+)^{n-2}.$$

Going on in this way we obtain

$$\mathbf{P}^-(\mathbf{P}^+)^n = \sum_{n'=0}^{n-1} (\mathbf{P}^+)^{n'} 2(2\mathbf{n_d} + 5)(\mathbf{P}^+)^{n-n'-1} + (\mathbf{P}^+)^n \mathbf{P}^- =$$

$$\sum_{n'=0}^{n-1} 4(\mathbf{P}^+)^{n'}\mathbf{n_d}\,(\mathbf{P}^+)^{n-n'-1} + 10n\,(\mathbf{P}^+)^{n-1} + (\mathbf{P}^+)^n\,\mathbf{P}^-. \qquad (10.9)$$

With the help of (10.6) one shows the relation

$$(\mathbf{P}^+)^{n'}\mathbf{n_d}\,(\mathbf{P}^+)^{n-n'-1} = (\mathbf{P}^+)^{n'+1}\mathbf{n_d}\,(\mathbf{P}^+)^{n-n'-2} + 2(\mathbf{P}^+)^{n-1} = \qquad (10.10)$$

$$(\mathbf{P}^+)^{n-1}\,\mathbf{n_d} + 2(n - 1 - n')\,(\mathbf{P}^+)^{n-1}$$ which we insert in (10.9).

$$\mathbf{P}^-(\mathbf{P}^+)^n = \sum_{n'=0}^{n-1} 4(\mathbf{P}^+)^{n-1}\mathbf{n_d} + \sum_{n'=0}^{n-1} 8(n - 1)(\mathbf{P}^+)^{n-1} - \sum_{n'=0}^{n-1} 8n'(\mathbf{P}^+)^{n-1} +$$

$$10n\,(\mathbf{P}^+)^{n-1} + (\mathbf{P}^+)^n\,\mathbf{P}^- =$$

$$4n\,(\mathbf{P}^+)^{n-1}\mathbf{n_d} + 8n(n - 1)(\mathbf{P}^+)^{n-1} - 8(1/2)(n - 1)n(\mathbf{P}^+)^{n-1} + 10n\,(\mathbf{P}^+)^{n-1} + (\mathbf{P}^+)^n\,\mathbf{P}^- =$$

$$(\mathbf{P}^+)^{n-1}(4\,\mathbf{n_d}n + 4n^2 + 6n) + (\mathbf{P}^+)^n\,\mathbf{P}^-$$ or

$$[\mathbf{P}^-, (\mathbf{P}^+)^n] = 2n(\mathbf{P}^+)^{n-1}\cdot(2(\mathbf{n_d} + n) + 3). \qquad (10.11)$$

Now we are able to look into the effect of $\mathbf{T}^2$ on a state of the spherical basis. Owing to (6.15) and (10.5a) we write

$$\mathbf{T}^2\,|\,N\,n_d\,t\;\;n_D\,J\,\rangle =$$

$$\mathbf{n_d}(\mathbf{n_d} + 3)\,|\,N\,n_d\,t\;\;n_D\,J\,\rangle - \mathbf{P}^+\mathbf{P}^-A_{n_d t}\,(\mathbf{P}^+)^{(n_d - t)/2}\,|\,N,\,n_d = t,\,t\;\;n_D\,J\,\rangle.$$

In the last ket state the number of *d*-bosons agrees with the seniority $t$. Note that no use has been made of the explicit form of $A_{n_d t}$. We put

$$(n_d - \tau)/2 = n_p\,, \qquad (10.12)$$

rebuild our result with the help of (10.11) and take the relation $\mathbf{P}^-\,|\,N,\,n_d = t,\,t\;n_D\,J\,\rangle = 0$ (6.31) into account as follows

$$\mathbf{T}^2\,|\,N\,n_d\,t\;\;n_D\,J\,\rangle = \mathbf{n_d}(\mathbf{n_d} + 3)\,|\,N\,n_d\,t\;\;n_D\,J\,\rangle\,-$$

$$A_{n_d t}\,\mathbf{P}^+([\mathbf{P}^-, (\mathbf{P}^+)^{n_p}] + (\mathbf{P}^+)^{n_p}\mathbf{P}^-)\,|\,N,\,n_d = t,\,t\;\;n_D\,J\,\rangle =$$

$$n_d(n_d + 3)\,|\,N\,n_d\,t\;\;n_D\,J\,\rangle - A_{n_d t}\,\mathbf{P}^+2n_p\,(\mathbf{P}^+)^{n_p-1}(2(t + n_p) + 3)\,)\,|\,N\;t\;\,t\;n_D\,J\rangle =$$

$$(n_d(n_d + 3) - 2n_p\,(2(t + n_p) + 3)\,|\,N\,n_d\,t\;\;n_D\,J\,\rangle = t\,(t + 3)\,|\,N\,n_d\,t\;\;n_D\,J\,\rangle$$

where (6.15) has been inserted again. Thus, we have

$$\mathbf{T}^2\,|\,N\,n_d\,t\;\;n_D\,J\,\rangle = t\,(t + 3)\,|\,N\,n_d\,t\;\;n_D\,J\,\rangle \qquad (10.13)$$

i. e. the states of the spherical basis are eigenfunctions of the seniority operator with the eigenvalues $t\,(t + 3)$.

From the commutation behaviour of $\boldsymbol{T}^{2}$, we can see also directly that eigenfunctions of $\boldsymbol{H}^{(\mathrm{I})}$ (10.1) are simultaneously eigenfunctions of $\boldsymbol{T}^{2}$. In chapters 7 and 8 it was shown that $\boldsymbol{N}$ and $\boldsymbol{J}^{2}$ commute with $\boldsymbol{H}$ and therefore with $\boldsymbol{H}^{(\mathrm{I})}$. Naturally, $\boldsymbol{n_d}$ and $\boldsymbol{H}^{(\mathrm{I})}$ commute with $\boldsymbol{H}^{(\mathrm{I})}$. According to (10.1) this operator consists essentially of the operators $\boldsymbol{N}$, $\boldsymbol{n_d}$, $\boldsymbol{J}^{2}$ and $\boldsymbol{T}^{2}$. Consequently, $\boldsymbol{T}^{2}$ commutes with $\boldsymbol{H}^{(\mathrm{I})}$. Moreover, $\boldsymbol{J}^{2}$ commutes with $\boldsymbol{T}^{2}$ which results in a common eigenfunction of the operators $\boldsymbol{H}^{(\mathrm{I})}$, $\boldsymbol{J}^{2}$ and $\boldsymbol{T}^{2}$.

Deriving the eigenvalue of $\boldsymbol{T}^{\mathbf{2}}$ (10.13), not all properties of the many-*d*-boson state are employed. From the eigenfunction, only the number of *d*-boson pairs with $J = 0$ or the number of unpaired *d*-bosons has to be known. On the other hand E. Chacon et al. (1976, 1977) succeeded in constructing the spherical basis starting from the eigenvalue equations of $\boldsymbol{T}^{2}$, $\boldsymbol{J}^{2}$ and $\boldsymbol{J_z}$ ( see also Frank (1994), p. 311).

At the end of this section, we treat the state

$$\boldsymbol{P}^{+} \mid N, n_d - 2, t \;\; n_D\, J \rangle,$$

of which we made use in section 6.2. By means of (10.6) and (10.7) we write

$$\boldsymbol{P}^{-}\boldsymbol{P}^{+} = [\boldsymbol{P}^{-}, \boldsymbol{P}^{+}] + \boldsymbol{P}^{+}\boldsymbol{P}^{-} = [\boldsymbol{P}^{-}, \boldsymbol{P}^{+}] + \boldsymbol{n_d}(\boldsymbol{n_d} + 3) - \boldsymbol{T}^{2} =$$

$$2(2\boldsymbol{n_d} + 5) + \boldsymbol{n_d}(\boldsymbol{n_d} + 3) - \boldsymbol{T}^{2} = \boldsymbol{n_d}^{2} + 7\boldsymbol{n_d} + 10 - \boldsymbol{T}^{2}.$$

Because of (10.13) we have

$$\boldsymbol{P}^{-}\boldsymbol{P}^{+} \mid N, n_d - 2, t \;\; n_D\, J \rangle =$$

$$((n_d - 2)^{2} + 7(n_d - 2) + 10 - t\,(t + 3)) \mid N, n_d - 2, t \;\; n_D\, J \rangle =$$

$$(n_d(n_d + 3) - \tau\,(\tau + 3)) \mid N, n_d - 2, t \;\; n_D\, J \rangle.$$

The corresponding diagonal matrix element reads

$$\langle N, n_d - 2, t \;\; n_D\, J \mid \boldsymbol{P}^{-}\boldsymbol{P}^{+} \mid N, n_d - 2, t \;\; n_D\, J \rangle = n_d(n_d + 3) - t\,(t + 3). \qquad (10.14)$$

The operator $\boldsymbol{P}^{-}$ is adjoint to $\boldsymbol{P}^{+}$ analogously to the operators $\boldsymbol{d}_{m}$ and $\boldsymbol{d}^{+}_{m}$ ( see section 5.2 ). Consequently $\langle N, n_d - 2, t \;\; n_D\, J \mid \boldsymbol{P}^{-}$ is conjugate complex with regard to the right hand side of the matrix element (10.14). Since $\boldsymbol{P}^{+}$ raises $n_d$ by 2 the expression

$$(n_d\,(n_d + 3) - t\,(t + 3))^{-½}\, \boldsymbol{P}^{+} \mid N, n_d - 2, t \;\; n_D\, J \rangle = \mid N\, n_d\, t \;\; n_D\, J \rangle \qquad (10.15)$$

is a normalised state with $n_d$ *d*-bosons. In an analogous way we obtain

$$\langle N\, n_d\, t \;\; n_D\, J \mid \boldsymbol{P}^{+}\boldsymbol{P}^{-} \mid N\, n_d \;\; t \;\; n_D\, J \rangle = n_d(n_d + 3) - t\,(t + 3) \qquad (10.16)$$

and $\quad (n_d(n_d + 3) - t\,(t + 3))^{-½}\, \boldsymbol{P}^{-} \mid N\, n_d\, t \;\; n_D\, J \rangle = \mid N, n_d - 2, t \;\; n_D\, J \rangle.$

### 10.3 Energy eigenvalues. Comparison with experimental data

If the Hamilton operator $\boldsymbol{H}^{(I)}$ (10.1) acts on a state of the spherical basis (10.3) according to (10.4) and (10.13) the following eigenenergy results

$$E^{(I)}_{N\,n_d\,t\,J} = e_n N + v_n N^2 + (e_d' + v_{nd} N) n_d + v_d n_d^2 + v_t\, t\,(t + 3) + v_j J(J + 1). \quad (10.17)$$

Independently of the complexity of the state, the eigenvalue has a simple, closed form. Neither the projection $M$ nor the ordering number $n_D$ influence the energy but the latter restricts the range of the angular momentum $J$ according to (10.5).

Comparing calculated with measured spectra of nuclear levels mostly one single nucleus is investigated neglecting the binding energy relative to other nuclei. The lowest level is characterised by $n_d = 0$ and therefore $t = J = 0$. If we put $e_n = v_n = 0$ the energy of this state is zero. Moreover at the beginning of this chapter $v_0 = v_2 = 0$ has been chosen and so we obtain $u_0 = e_s = 0$ (10.2) and the coefficients (10.17) read like this

$$e_d' = e + (3/10)c_0 - (2/7)c_2 - (18/35)c_4,$$

$$v_{nd} = (1/\sqrt{5})u_2,$$

$$v_d = (1/10)c_0 + (1/7)c_2 + (9/35)c_4 - (1/\sqrt{5})u_2, \quad (10.18)$$

$$v_t = -(1/10)c_0 + (1/7)c_2 - (3/70)c_4,$$

$$v_j = -(1/14)c_2 + (1/14)c_4.$$

Here we have written $e$ instead of $e_d$, as is usual. In most published comparisons with real nuclear spectra, the quantity $u_2$ is put equal to zero and the energies are given as follows

$$E^{(I)}_{n_d\,t\,J} = e_d' n_d + v_d n_d^2 + v_t\, t\,(t + 3) + v_j J(J + 1). \quad (10.19)$$

Fitting theoretical on experimental values one finds that for the most part $v_t$ and $v_j$ are positive and small relative to $e_d'$. The levels with greatest possible values for $t$ and $J$ for given $n_d$ i. e. with $t = n_d$ and $J = 2n_d$, constitute a conspicuous series showing the angular momenta $J =$ 0, 2, 4, 6 and so forth and slightly increasing steps in energy, which is named $Y$-series. Every level of this kind is the lowest one of the partial spectrum with the same angular momentum. Thus, levels of the $Y$-series show the following energies

$$E^{(Y)}_J \ \mathrm{o}\ E^{(I)}_{n_d=t=J/2} = (1/2)e_d' J + (1/4)v_d J^2 + (1/2)v_t J(J/2 + 3) + v_j J(J + 1) =$$

$$(1/8)[(4e - 2c_4)J + c_4 J^2] \quad (10.20)$$

where (10.18) and $u_2 = 0$ have been employed. Apart from the feeble quadratic term, the energy grows in (10.20) proportional to $J$ i. e. in the same way as the eigenenergies of the harmonic oscillator. This mathematical form appears also if the atomic nucleus is interpreted as a vibrating droplet for which reason the special case dealt with in this chapter is named “vibrational limit“. On the other hand, we will show that the Hamilton operator has features of the Lie algebra *u*(5), which is why the name “*u*(5) limit” is also used.

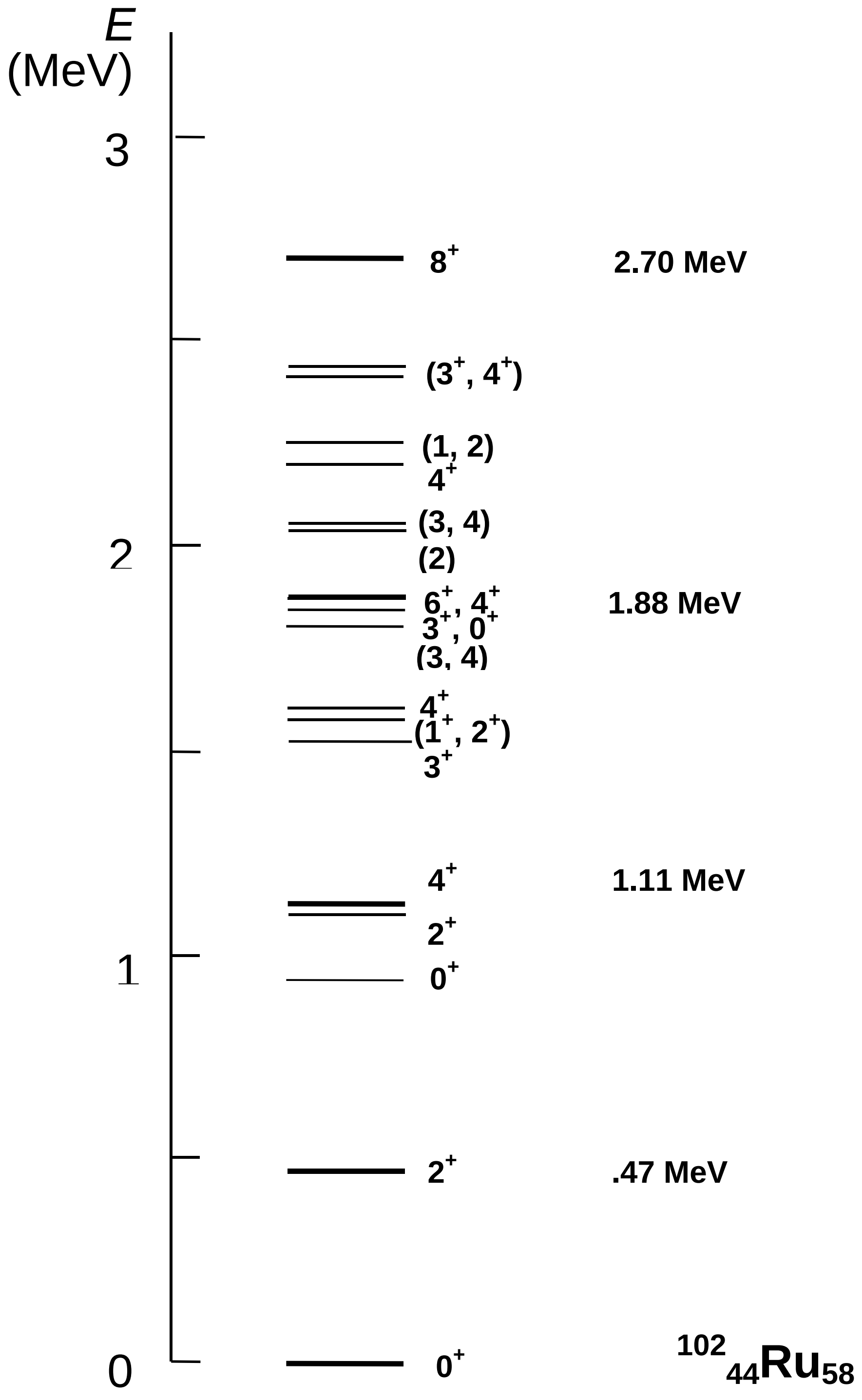

Figure 10.1. Measured spectrum of $^{102}$Ru. The energies of the *Y*-series are given (Arima and Iachello, 1976a, p. 273).

In figure 10.1 the levels of the *Y*-series of the nucleus $^{102}{}_{44}Ru_{58}$ are marked and the energies are given. The energy difference between two levels of the *Y*-series following each other reads according to (10.20)

$$DE^{(Y)}_J = E^{(Y)}_J - E^{(Y)}_{J-2} = e - c_4 + (1/2)c_4 J. \qquad (10.21)$$

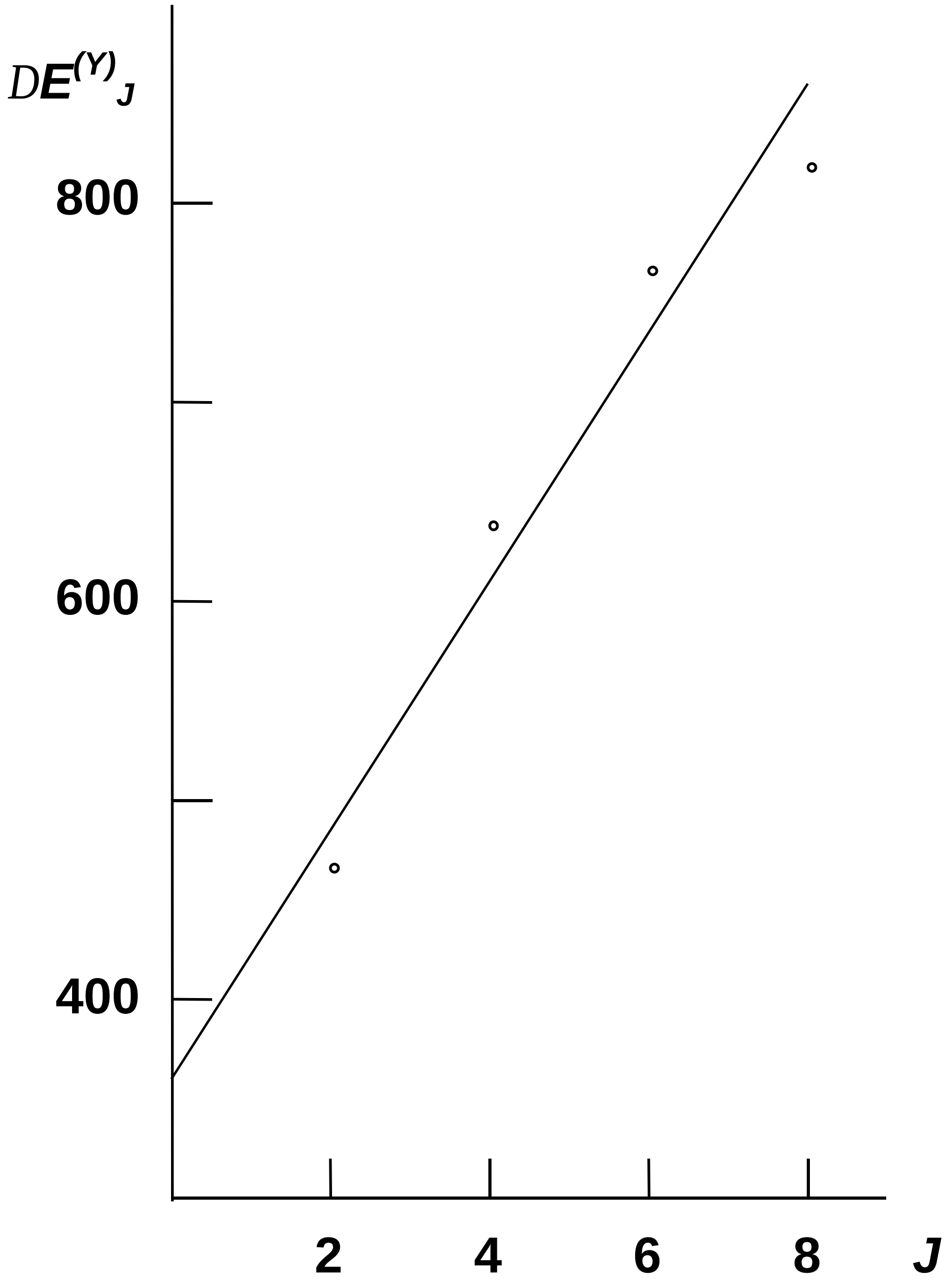

Figure 10.2. Energy differences $DE^{(Y)}_J = E^{(Y)}_J - E^{(Y)}_{J-2}$ between levels of the *Y*-series following each other of the nucleus $^{102}{}_{44}Ru_{58}$ according to figure 10.1. The straight line fitted in by eye yields the values $c_4$ = 125 keV and $e$ = 485 keV for the coefficients in (10.21).

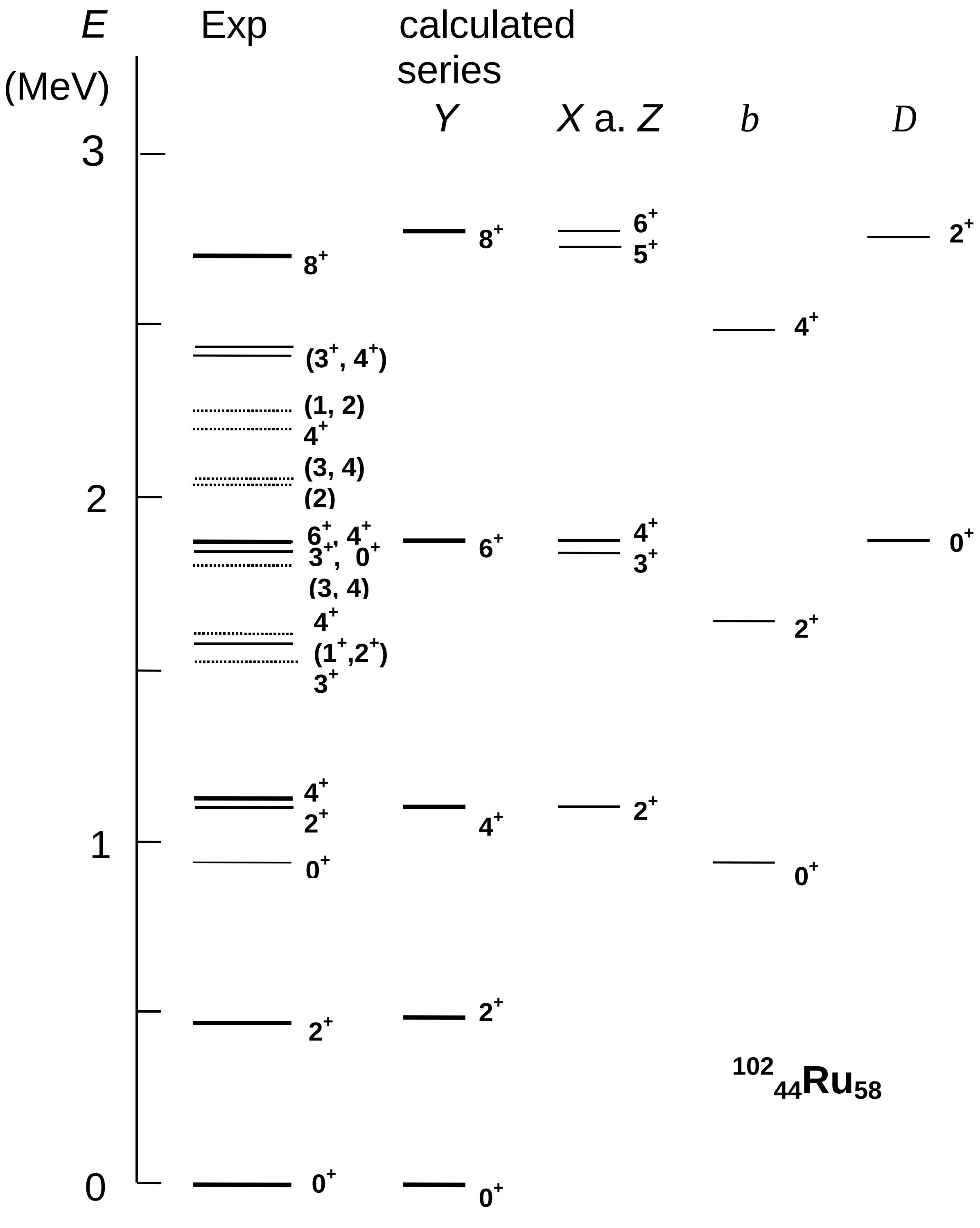

Figure 10.3. Comparison between the measured and the theoretical spectrum of $^{102}$Ru. The spectrum is calculated by means of (10.19) and (10.18) with the parameters $e$ = 481 keV, $c_0$ = -18 keV, $c_2$ = 141 keV, $c_4$ = 144 keV and $u_2$ = 0 (Arima and Iachello, 1976a, p. 273). The numbers above or below the lines indicate the numbers of $d$-bosons.

The measured values of $DE^{(Y)}_J$ of the same nucleus are plotted against $J$ in figure 10.2. Making use of (10.21) and of a graphical procedure the coefficients $c_4$ and $e$ are found. In figure 10.3 about a dozen levels are given which have been calculated by means of (10.19) and (10.18) adjusting the coefficients $c_0$, $c_2$, $c_4$ and $e$. The values of the last two quantities agree reasonably with those of figure 10.2. The levels are subdivided in the series $Y$, $X$, $Z$, $b$ and $D$ which have the following characteristics

| series | $t$ | $n_D$ | $J$ | |
|---|---|---|---|---|
| $Y$ | $n_d$ | 0 | $2n_d$ | |
| $X$ | $n_d$ | 0 | $2n_d - 2$ | |
| $Z$ | $n_d$ | 0 | $2n_d - 3$ | |
| others | $n_d$ | 0 | $2n_d - 4$ | (10.22) |
| | | | $2n_d - 5$ | |
| $b$ | $n_d - 2$ | 0 | $2t = 2n_d - 4$ | |
| $D$ | $n_d$ | 1 | $2n_d - 6$ | |

The rule (10.5) excludes a further series with $J = 2n_d - 1$ between the series $Y$ and $X$. As in other nuclear models here it is impossible to reproduce all measured levels. Every model is based on approximations, which restrict the field of applicability.

Figures 10.4 and 10.5 show the spectra of $^{110}_{46}Cd_{62}$ and $^{188}_{78}Pt_{110}$ with the corresponding levels calculated by means of (10.19). Comparing these three measured ( or calculated ) spectra reveals a clear agreement of the energy structures, which is expected because (10.19) does not contain the boson number $N$.

Figure 10.6 depicts the series $Y$ and $X$ for $^{100}_{46}Pt_{54}$ ( $N = 4$ ). Apparently, here the boson number has been extrapolated over the value 4 otherwise $J £ 8$ would hold.

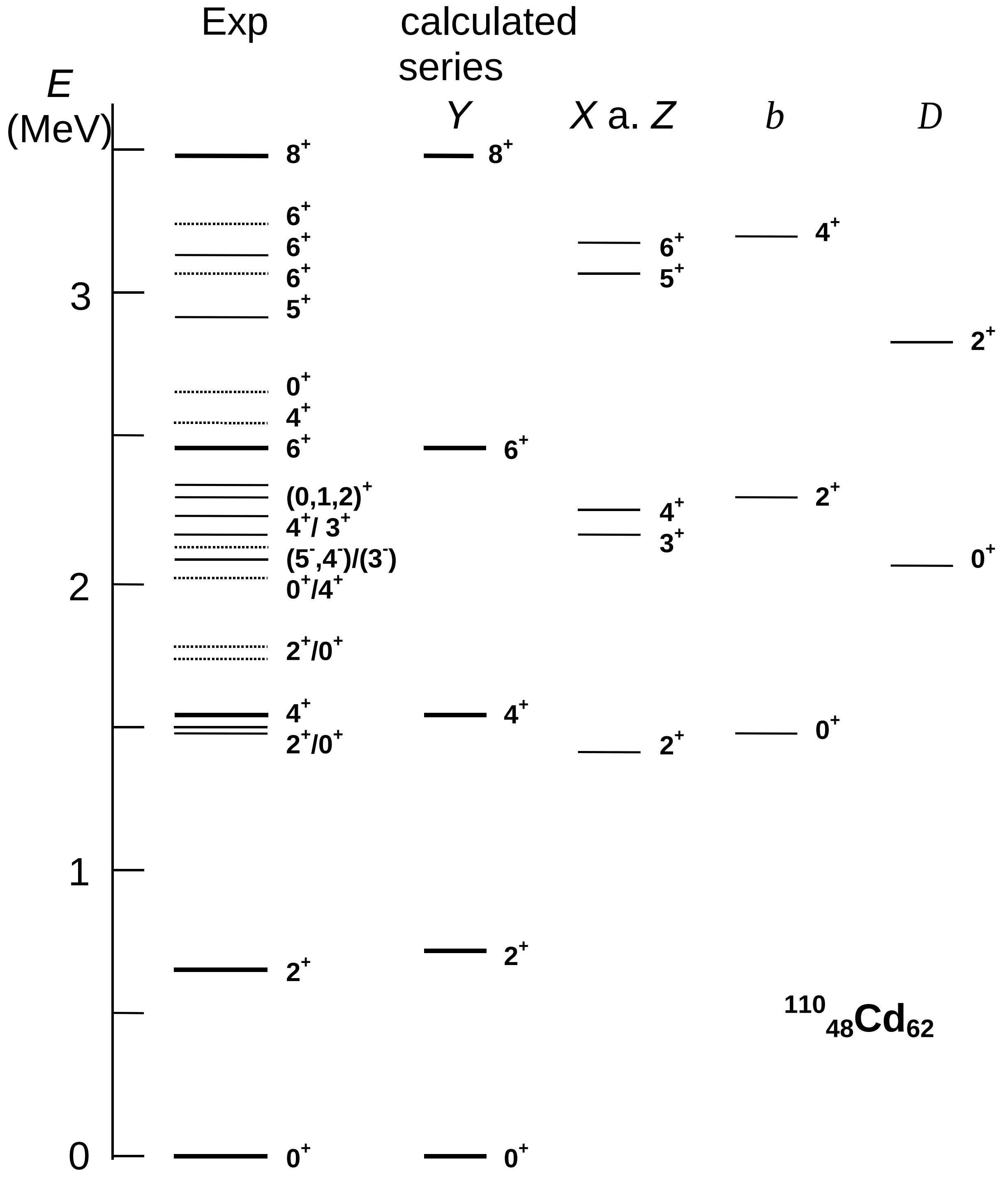

Figure 10.4. Comparison between the measured and the theoretical spectrum of $^{110}$Cd. The spectrum is calculated by means of (10.19) and (10.18) with the parameters $e$ = 722 keV, $c_0$ = 29 keV, $c_2$ = - 42 keV, $c_4$ = 98 keV and $u_2 = 0$ (Arima and Iachello, 1976a, p. 274).

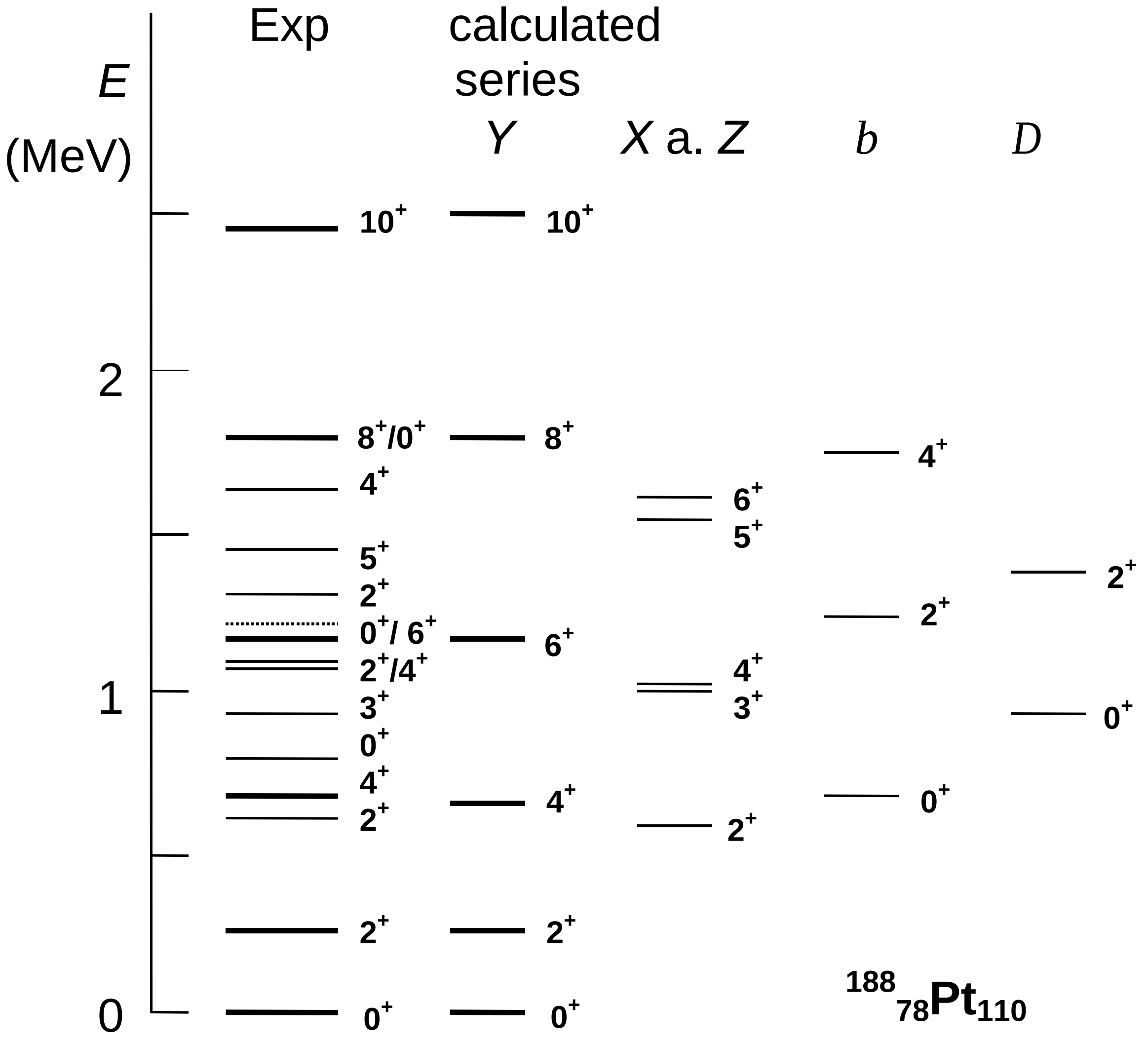

Figure 10.5. Comparison between the measured and the theoretical spectrum of $^{188}$Pt. The spectrum is calculated by means of (10.19) and (10.18) with the parameters $e$ = 281 keV, $c_0$ = 148 keV, $c_2$ = 30 keV, $c_4$ = 110 keV and $u_2$ = 0 (Arima and Iachello, 1976a, p. 274).

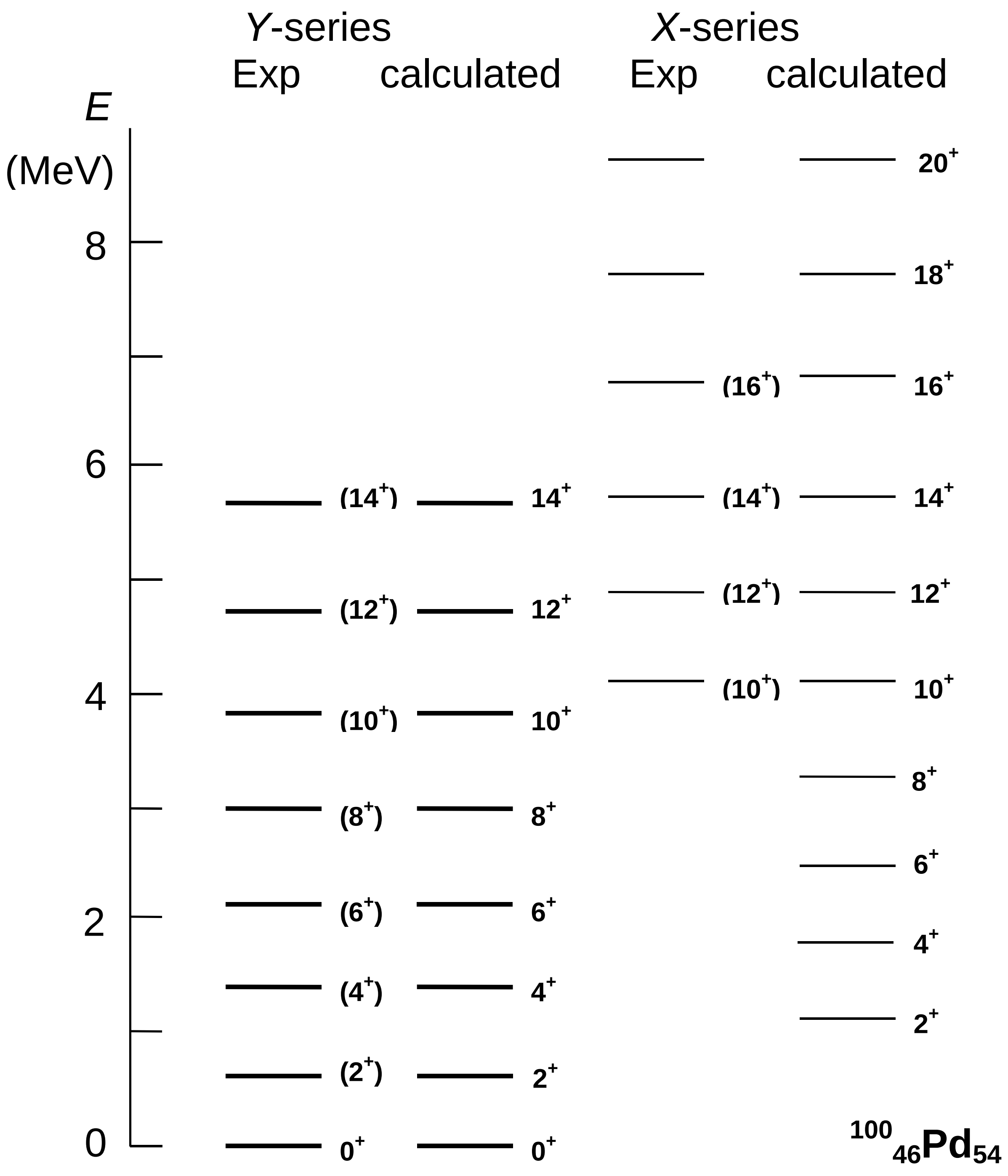

Figure 10.6. Comparison between the measured and the calculated energies of the *Y*-series and the *X*-series in $^{100}$Pd. The parameters employed are $e$ = 680 keV, $c_2$ = -160 keV und $c_4$ = 45 keV (Arima and Iachello, 1976a, p. 275).

According to (7.16) the quantities $c_0$, $c_2$, $c_4$ and $e$ stand for matrix elements of interaction operators and they should be individually equal for every nucleus contrary to the reality. The discrepancy can be explained in the following way

(i) the interaction between *s*- and *d*-bosons is neglected in this limit,
(ii) the effective boson-boson interaction can depend on the total number of bosons,
(iii) no difference is made between bosons attributed to protons and those belonging to neutrons.

The limitation (i) will be dropped later for the IBM1 and the approximation (iii) is abolishes in the model IBM2.

Assigning measured to calculated levels is uncertain if no other criterion than the energy is considered. The rates of electromagnetic transitions from level to level are well-known quantities, which can help to improve the identification of states. We will deal with them in the next chapter.

Only a restricted number of nuclei behave according to the vibrational limit. Systematic investigations showed that nuclei with neutron- and/or proton numbers near the magic numbers 50, 82 and 126 belong to this limit ( see figure 10.7 ).

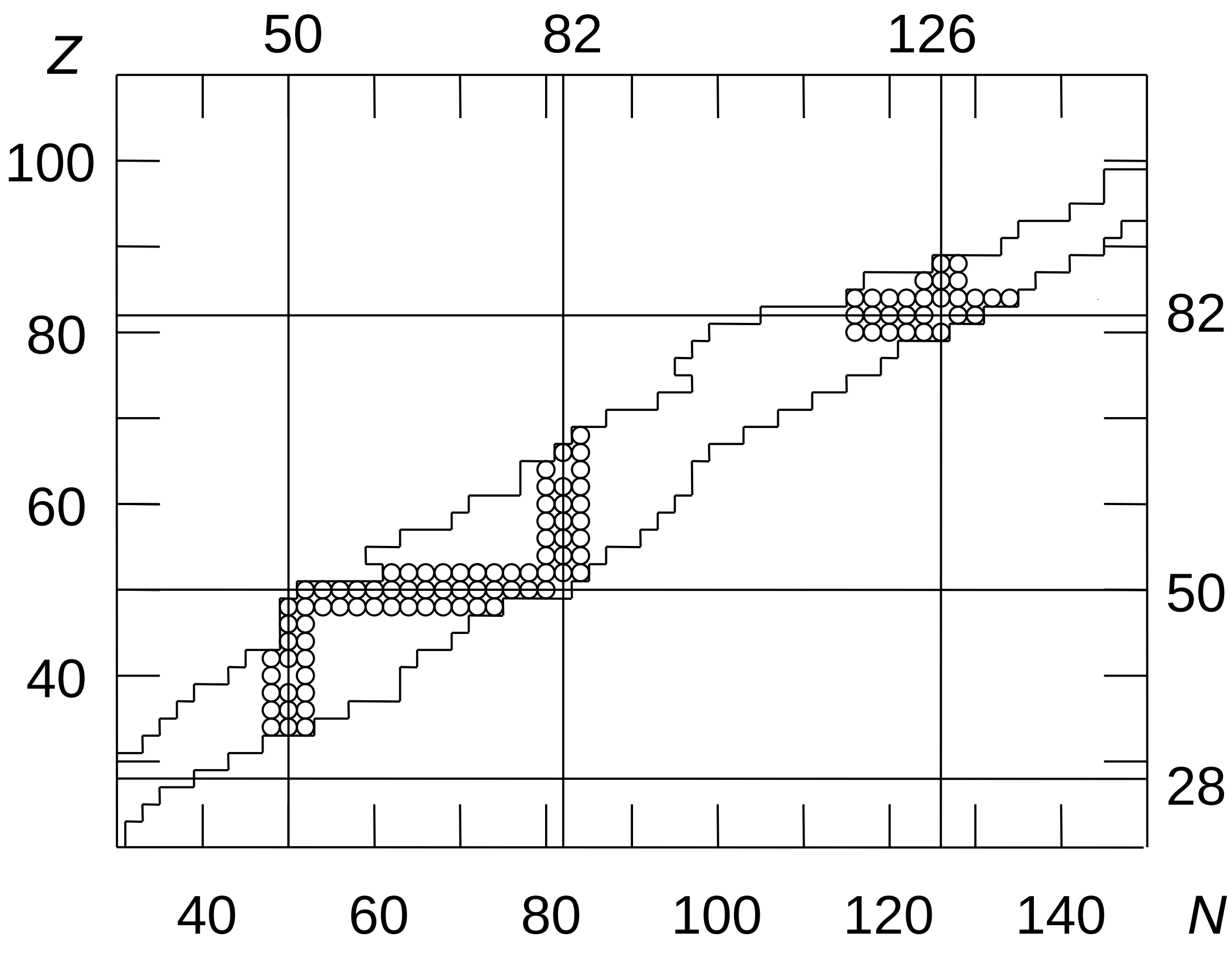

Figure 10.7. Region of the periodic table of the even-even nuclei. *Z* = number of protons, *N* = number of neutrons. The circles denote nuclei of the *u*(5)-limit (Iachello and Arima, 1987, p. 42).

# 11 Electromagnetic transitions in the *u*(5)-limit

Excited nuclei can reach lower energy levels by emitting electromagnetic radiation. In this chapter mainly the electric quadrupole radiation will be treated and the transition probabilities will be formulated for the IBM. First, a short introduction will be given and in the second section the definition of the interaction operator and its representation by creation- and annihilation operators for bosons follows. In the third section the connection between the matrix elements of the interaction operator and the transition probabilities will be described. The derivation of closed expressions for these matrix elements ( section 11.4 ) and a comparison with experimental data follow. In the section 11.6 and 11.7 special transitions and levels are treated and finally we look into the quadrupole moment.

## 11.1 Multipole radiation

The electromagnetic field of a radiating nucleus is represented by a vector potential $\underline{A}(\underline{r})$. It is characterised by a spherical harmonic $Y_{LM}(\underline{r})$ with $\underline{r} = \underline{r}/r$ (Brussaard and Glaudemans, 1977, p. 179). It describes the multipole radiation of the order $2^L$ possessing the angular momentum $L$. When a nucleus emits a photon, this angular momentum is carried away from the initial nuclear state having spin $J_i$ which results in a final state with spin $J_f$ . Thus, the triangle condition

$$J_i + L \geq J_f \geq | J_i - L | \qquad (11.1)$$

is satisfied. There are two types of multipole radiation, the electric and the magnetic. They differ not only in their spatial shape but also in the parity of the field. For example, positive parity means that two points of the field function lying diametrically with respect to the origin show the same function value. For negative parity, these values differ only in sign.

Electric multipole radiation has positive parity for even $L$-values and negative parity for odd ones. The magnetic multipole radiation shows just the opposite behaviour. If two fields are coupled which possess the same parity the combination shows positive parity and negative parity results if the components have different parity.

In the interacting boson model 1 and 2, single boson states ( actually material waves ) are coupled which have all positive parity ( even angular momenta $L$ = 0, 2 result in positive parity, which is connected with the behaviour of the spherical harmonics ). Thus, all collective states of these models have positive parity. Because the emission of a photon leads from a state of positive parity to another with the same parity all types of radiation must show positive parity. Consequently, for multipole radiation the following $L$-values are possible electric : $L$ = 2, 4, 6, ... (11.2)
magnetic : $L$ = 1, 3, 5, ... .

In order to formulate the probability of a radiative transition to a lower level we will looke into the energy experienced by the excited nucleus in the field of radiation i. e. into the interaction energy of the radiation field with the charges or magnetic moments of the nucleus.

## 11.2 The operator of the electromagnetic interaction

In quantum mechanics, the classical expression of the interaction energy is replaced by an operator. For the electric multipole radiation, this operator of the electromagnetic interaction can be written employing the long-wavelength approximation ( Brussaard and Glaudemans, 1977, p. 182 ) like this

$$\boldsymbol{O}(\mathrm{E}, L, M) = \sum_{k=1}^{A} e(k)\, r(k)^L\, Y_{LM}(\underline{r}(k)). \qquad (11.3)$$

The sum covers all nucleons of the nucleus and $e(k)$ is the charge of the $k$th nucleon. Obviously $\boldsymbol{O}(\mathrm{E}, L, M)$ is a single nucleon operator and it corresponds formally to the single-boson operator (5.1). The operator for magnetic multipole radiation is of this type as well. Due to the spherical harmonics, both kinds of interaction operators are tensor operators ( section 6.3 ). We transfer both properties of this operator - it is a tensor and it represents the single particle type - to the corresponding interaction operator of the interacting boson model, which we write analogously to (11.3) as follows

$$\boldsymbol{O}(\mathrm{E}, L, M) = \sum_{i=1}^{N} \boldsymbol{o}(\mathrm{E}, L, M, \underline{x}_i). \qquad (11.4)$$

The sum extend over all active bosons and $\boldsymbol{o}(\mathrm{E}, L, M, \underline{x}_i)$ is a tensor operator of the rank $L$. Provisionally we are attributing to the $i$th boson the set of variables $\underline{x}_i$. For the magnetic radiation an analogous relation is true. The expression (11.4) cannot be deduced completely or be described more in detail. In fact, it belongs to the basic postulates of the IBM. According to (5.25) and (5.14) the operator $\boldsymbol{O}(\mathrm{E}, L, M)$ can be written in the second quantisation as

$$\boldsymbol{O}(\mathrm{E}, L, M) = \sum_{l_p, l_q, m_p, m_q} \langle p, l_p, m_p \,|\, \boldsymbol{o}(\mathrm{E}, L, M) \,|\, q, l_q, m_q \rangle\, \boldsymbol{b}^+_{l_p\, m_p} \boldsymbol{b}_{l_q\, m_q}\,. \qquad (11.5)$$

Matrix elements of tensor operators can be split in two factors with the aid of the Wigner - Eckart theorem (A5.8) as follows

$$\langle p, l_p, m_p \,|\, \boldsymbol{o}(\mathrm{E}, L, M) \,|q, l_q, m_q \rangle =$$

$$(l_q\, m_q\, L\, M \,|\, l_p\, m_p)\, (2l_p + 1)^{-1/2} \langle p, l_p \,||\, \boldsymbol{o}(\mathrm{E},L) \,||q, l_q \rangle. \qquad (11.6)$$

The right hand side of (11.6) contains a Clebsch-Gordan coefficient and the reduced matrix element, which is independent of the projection quantum numbers $m_p$, $m_q$ and $M$. According to (6.19) and (A1.14) we write

$$\boldsymbol{b}^+_{l_p, m_p} \boldsymbol{b}_{l_q, m_q} = (-1)^{-m_q}\, \boldsymbol{b}^+_{l_p\, m_p} \boldsymbol{b}^{\sim}_{l_q\, m_q} =$$

$$(-1)^{-m_q} \sum_{L' M'} (l_p\, m_p\, l_q, -m_q \,|\, L'\, M')\, [\boldsymbol{b}^+_{l_p} \times \boldsymbol{b}^{\sim}_{l_q}]^{(L')}{}_{M'}\,. \qquad (11.7)$$

We insert (11.6) and (11.7) in (11.5) and obtain

$$\boldsymbol{O}(\mathrm{E}, L, M) = \sum_{L' M' l_p l_q} \sum_{m_p m_q} (-1)^{-m_q} (l_q\, m_q\, L\, M \mid l_p\, m_p)\, (l_p\, m_p\, l_q, -m_q \mid L'\, M') \cdot$$

$$(2l_p + 1)^{-1/2} \langle p, l_p \,||\, \boldsymbol{o}(\mathrm{E},L) \,||\, q, l_q \rangle\ [\boldsymbol{b}^{+}_{l_p} \times \boldsymbol{b}^{\sim}_{l_q}]^{(L')}_{M'}. \qquad (11.8)$$

Due to (A1.4) up to (A1.6) the relation

$$(-1)^{-m_q} (l_q\, m_q\, L\, M \mid l_p\, m_p) = (-1)^{l_q} ((2l_p + 1)/(2L + 1))^{1/2} (l_q\, m_q\, l_p, -m_p \mid L, -M)$$

$$= (-1)^{l_q} ((2l_p + 1)/(2L + 1))^{1/2} (l_p\, m_p\, l_q\, {-m_q} \mid L\, M) \qquad (11.9)$$

holds, which we insert in (11.8). Since $l_q$ is even and because of

$\sum_{m_p m_q} (l_p\, m_p\, l_q, -m_q \mid L\, M)\, (l_p\, m_p\, l_q, -m_q \mid L'\, M') = d_{LL'}\, d_{MM'}$ (A1.10) we have

$$\boldsymbol{O}(\mathrm{E}, L, M) = \sum_{L' M' l_p l_q} d_{LL'}\, d_{MM'}\, (2L + 1)^{-1/2} \langle p, l_p \,||\, \boldsymbol{o}(\mathrm{E},L) \,||\, q, l_q \rangle\ [\boldsymbol{b}^{+}_{l_p} \times \boldsymbol{b}^{\sim}_{l_q}]^{(L')}_{M'}$$

$$= \sum_{l_p l_q} (2L + 1)^{-1/2} \langle p, l_p \,||\, \boldsymbol{o}(\mathrm{E},L) \,||\, q, l_q \rangle\ [\boldsymbol{b}^{+}_{l_p} \times \boldsymbol{b}^{\sim}_{l_q}]^{(L')}_{M'}. \qquad (11.10)$$

Among the radiative transitions the electric quadrupole transition is most suitable for demonstrating the nature of a collective state (Talmi, 1993, p. 746 ). In this case ( $L$ = 2, $l_p$ and $l_q$ = 0, 2 ) the expression (11.10) reduces to three terms

$$\boldsymbol{O}(\mathrm{E}, 2, m) = a_2\, \boldsymbol{d}^{+}_{m}\, \boldsymbol{s} + a_2'\, \boldsymbol{s}^{+}\, \boldsymbol{d}^{\sim}_{m} + b_2\, [\boldsymbol{d}^{+} \times \boldsymbol{d}^{\sim}]^{(2)}_{m}$$

with $a_2 = \langle d \,||\, \boldsymbol{o}(\mathrm{E},2) \,||\, s \rangle/\sqrt{5}$ ,

$a_2' = \langle s \,||\, \boldsymbol{o}(\mathrm{E},2) \,||\, d \rangle/\sqrt{5}$ ,

$b_2 = \langle d \,||\, \boldsymbol{o}(\mathrm{E},2) \,||\, d \rangle/\sqrt{5}$ . (11.11)

Because $\boldsymbol{o}$(E,2) is a tensor operator one can show that $a_2 = a_2'^{*}$ holds (Talmi, 1993, p. 112). It's usual to assume that $a_2$ is real i. e. $a_2 = a_2'$ and we can write

$$\boldsymbol{O}(\mathrm{E}, 2, m) = a_2\, (\boldsymbol{d}^{+}_{m}\, \boldsymbol{s} + \boldsymbol{s}^{+}\, \boldsymbol{d}^{\sim}_{m}) + b_2\, [\boldsymbol{d}^{+} \times \boldsymbol{d}^{\sim}]^{(2)}_{m}. \qquad (11.12)$$

This is the operator of the electric quadrupole transitions of the IBM. It is not limited to a special case ( limit ).

We now carry out a rough estimate of the quotient $b_2 / a_2$ . As a model we consider a closed core with a single boson moving like a particle in the potential of the harmonic oscillator. It is represented by the function $R_l(r)\, Y_{l\,m}(\underline{r})$ with $\underline{r}$ = $\underline{r}/r$, which consists of the spherical harmonics $Y_{l\,m}(\underline{r})$ and of the $r$-dependent part $R_l(r)$ ( see Lawson, 1980, p. 7 ). We denote the single particle operator of the quadrupole radiation according to (11.3) by $\boldsymbol{O}(\mathrm{E}, 2, M) = \boldsymbol{o}(\mathrm{E}, 2, M) = e\, r^2\, Y_{2M}(\underline{r})$. Thus, we write

$$\langle d \,||\, \boldsymbol{o}(\mathrm{E},2) \,||\, s \rangle = \mathrm{e} \langle Y_2 \,||\, Y_2 \,||\, Y_0 \rangle \int R_2(r)\, R_0(r)\, r^4\, dr \text{ and}$$

$$\langle d \,||\, \boldsymbol{o}(\mathrm{E},2) \,||\, d \rangle = \mathrm{e} \langle Y_2 \,||\, Y_2 \,||\, Y_2 \rangle \int (R_2(r))^2\, r^4\, dr. \qquad (11.13)$$

From Brussaard and Glaudemans (1977, p. 419) we take

$$\langle Y_2 \| Y_2 \| Y_0 \rangle = (25/(4\pi))^{1/2} \begin{pmatrix} 2 & 2 & 0 \\ 0 & 0 & 0 \end{pmatrix} = (5/(4\pi))^{1/2} \text{ and}$$

$$\langle Y_2 \| Y_2 \| Y_2 \rangle = (25/(4\pi))^{1/2} \sqrt{5} \begin{pmatrix} 2 & 2 & 2 \\ 0 & 0 & 0 \end{pmatrix} = - 5/(14\pi)^{1/2}. \qquad (11.14)$$

Lawson (1980, p.7) gives the following relations

$$\int R_{2,\, n=0}(\alpha\, r)\, R_{0,\, n=0}(\alpha\, r)\, r^4\, dr = 3\sqrt{15}/(2\alpha^2) \text{ and}$$

$$\int (R_{2,\, n=0}(\alpha\, r))^2\, r^4\, dr = 7/(2\alpha^2). \qquad (11.15)$$

From (11.11) up to (11.15) follows

$$\beta_2 / \alpha_2 = \langle d \| \boldsymbol{o}(\mathrm{E},2) \| d \rangle / \langle d \| \boldsymbol{o}(\mathrm{E},2) \| s \rangle = - \sqrt{(14/27)} = - 0.72. \qquad (11.16)$$

For “physical solutions“ O. Scholten (1991, p. 99) formulates the condition $0 > \beta_2 / \alpha_2 > - \sqrt{7}/2 = - 1.323$, which is satisfied by (11.16). In practice the coefficients $\alpha_2$ and $\beta_2$ serve as parameters, with which measured decay rates are fitted.

### 11.3 Transition probabilities

The theory of the electromagnetic radiation connected with quantum mechanics yields the following formula for the probability per time unit of a radiative transition from the initial state *i* with spin $J_i$ to the final state *f* with $J_f$ ( Brussaard and Glaudemans, 1977, p. 189)

$$T(L, J_i \rightarrow J_f) = 8\pi(L + 1)L^{-1}((2L + 1)!!)^{-2} (q^{2L+1}/\hbar)\, B(L, J_i \rightarrow J_f). \qquad (11.17)$$

*L* is the angular momentum of the radiation field and (2*L* + 1)!! stands for (2*L* + 1)(2*L* - 1)(2*L* - 3)· ... ·3·1. The quantity *q* depends directly on the energy difference $\Delta E = E_i - E_f$ of the states involved in the transition like this

$$q = \Delta E / (\hbar c). \qquad (11.18)$$

Equation (11.17) holds both for electric and magnetic multipole radiation. A multitude of nuclei being in the state *i* decays exponentially with the mean lifetime $\tau_m$ which is connected with $T(L, J_i \rightarrow J_f)$ as follows

$$\tau_m(L, J_i \rightarrow J_f) = 1/\, T(L, J_i \rightarrow J_f). \qquad (11.19)$$

Using the uncertainty relation $\Delta E \cdot \Delta t \approx \hbar$ one can define the width of a level

$$\Gamma = \hbar/\tau_m = \hbar T \qquad . \qquad (11.20)$$

The quantity $B(L, J_i \rightarrow J_f)$ in (11.17) is the reduced transition probability. According to (11.17) up to (11.20) it is related in a simple way to the mean lifetime or the width of the initial level and therefore we can regard it as a measurable quantity.

From quantum mechanics we learn that $B(L, J_i \rightarrow J_f)$ is connected with the reduced matrix element of the operator $\boldsymbol{O}(L, M)$ of the interaction between the multipole radiation field and the nucleus as follows

$$B(L, J_i \rightarrow J_f) = \langle J_f \,||\, \boldsymbol{O}(L) \,||\, J_i \rangle^2/(2J_i + 1). \qquad (11.21)$$

This relation holds for electric and magnetic transitions. The reduced matrix element $\langle J_f \,||\, \boldsymbol{O}(L) \,||\, J_i \rangle$ is related to the normal matrix element due to the Wigner - Eckart theorem ( appendix 5 ) like this

$$\langle J_f M_f | \boldsymbol{O}(L\, M) | J_i M_i \rangle = (J_i M_i L\, M \,|\, J_f M_f)\, (2J_f + 1)^{-1/2} \langle J_f \,||\, \boldsymbol{O}(L) \,||\, J_i \rangle. \qquad (11.22)$$

For electric radiative transitions with $L = 2$ we insert the expression (11.12) in the normal matrix element of (11.22) and obtain

$$\langle J_f M_f | \boldsymbol{O}(\mathrm{E}, 2, M) | J_i M_f \rangle = \qquad (11.23)$$

$$\langle n_{sf} n_{df} t_f n_{Df} J_f M_f |\; a_2 (\boldsymbol{d}^+_m \boldsymbol{s} + \boldsymbol{s}^+ \tilde{\boldsymbol{d}}_m) + b_2 [\boldsymbol{d}^+ \times \tilde{\boldsymbol{d}}]^{(2)}_m \,|\, n_{si} n_{di} t_i n_{Di} J_i M_i \rangle.$$

This matrix element of the electric quadrupole transition contains three terms. The first becomes active ( is different from zero ) if the final state *f* has one *d*-boson more than the initial one ( *i* ), the second is attributed to the inverse situation and the third is important, if both states contain the same number of *d*-bosons. In (11.23) states of the spherical basis i. e. of the vibrational limit are chosen which have a definite number $n_d$ of *d*-bosons. In this limit radiative quadrupole transitions occur only between states which differ in the number of *d*-bosons by $\Delta n_d = 0$ , + 1 or – 1.

## 11.4 Reduced matrix elements for $|\Delta n_d| = 1$

Now we will work on the first term of the matrix element (11.23) and formulate just the reduced matrix element of the electric quadrupole radiation making use of (11.22) and dropping $n_D$. We have

$$\langle n_{sf} = n_{si} - 1, n_{df} = n_{di} + 1, t_f J_f \,||\, \boldsymbol{d}^+ \boldsymbol{s} \,||\, n_{si} n_{di} t_i J_i \rangle =$$

$$(2J_f + 1)^{1/2} (J_i M_i 2\, m \,|\, J_f M_f)^{-1} \cdot$$

$$\langle n_{sf} = n_{si} - 1, n_{df} = n_{di} + 1, t_f J_f M_f | \boldsymbol{d}^+_m \boldsymbol{s} \,|\, n_{si} n_{di} t_i J_i M_i \rangle. \qquad (11.24)$$

Due to (5.17), $\boldsymbol{s}\, |n_{si} n_{di} t_i J_i M_i \rangle = \sqrt{n_{si}}\, |n_{si} - 1, n_{di} t_i J_i M_i \rangle$ holds. Because the states are factorised in a part with *s*- and a part with *d*-bosons (6.18) the “integration“ over the *s*-part in (11.24) yields only the factor $\sqrt{n_s}$. Therefore, we have

$$\langle n_{sf} = n_{si} - 1, n_{df} = n_{di} + 1, t_f J_f \,||\, \boldsymbol{d}^+ \boldsymbol{s} \,||\, n_{si} n_{di} t_i J_i \rangle =$$

$$\sqrt{n_{si}} \langle n_{df} = n_{di} + 1, t_f J_f \,||\, \boldsymbol{d}^+ \,||\, n_{di} t_i J_i \rangle. \qquad (11.25)$$

Temporarily we are concentrating on reduced matrix elements and leave $n_s$ out according to the element on the right hand side of (11.25). We increase $n_d$ by steps of 1 and begin with $n_{di}$ = 0. In order to obtain $\langle\, n_{df} = 1, J_f = 2 \,||\, \boldsymbol{d}^+ \,||\, \rangle$ we form

$$\langle\, n_{df} = 1, J_f = 2, M_f \,|\, \boldsymbol{d}^+_m \,|\, n_{di} = 0, J_i = 0, 0 \,\rangle = d_{M_f m} \;,$$

which results from the fact that the operator $\boldsymbol{d}^+_m$ generates the state $|\, n_d = 1, J = 2, m \,\rangle$ on the right hand side. Owing to (11.22), the relations

$$1 = \langle\, n_{df} = 1, J_f = 2, m \,|\, \boldsymbol{d}^+_m \,|\, n_{di} = 0, J_i = 0, 0 \,\rangle =$$

$$(0\;0\;2\;m \,|\, 2\;m)\; 5^{-1/2} \,\langle\, n_{df} = 1,\; J_f = 2 \,||\, \boldsymbol{d}^+ \,||\, n_{di} = 0 \,\rangle, \text{ i. e.}$$

$$\langle\, n_{df} = 1,\; J_f = 2 \,||\, \boldsymbol{d}^+ \,||\, n_{di} = 0 \,\rangle = \sqrt{5} \qquad (11.26)$$

hold because the Clebsch-Gordan coefficient is 1.

We now turn to $n_{di}$ = 1 and $n_{df}$ = 2. First we calculate the normal matrix element of $\boldsymbol{d}^+$ like this

$$\langle\, n_{df} = 2, J_f\, M_f \,|\, \boldsymbol{d}^+_m \,|\, n_{di} = 1, J_i = 2, M_i \,\rangle = \langle\, n_{df} = 2, J_f\, M_f \,|\, \boldsymbol{d}^+_m \boldsymbol{d}^+_{M_i} \,|\rangle. \qquad (11.27)$$

Using (A1.14) and (6.5), we write

$$\langle\, n_{df} = 2, J_f\, M_f \,|\, \boldsymbol{d}^+_m \,|\, n_{di} = 1, J_i = 2, M_i \,\rangle =$$

$$\textstyle\sum_{JM} (2\; m\; 2\; M_i \,|\, J\, M)\, \langle\, n_{df} = 2, J_f\, M_f \,|[\, \boldsymbol{d}^+ \times \boldsymbol{d}^+ \,]^{(J)}_M \,|\rangle =$$

$$\textstyle\sum_{JM} (2\; m\; 2\; M_i \,|\, J\, M)\, \sqrt{2}\, \langle\, n_{df} = 2, J_f\, M_f \,|\, n_d = 2, J\, M \,\rangle.$$

According to the orthonormality of the $n_d$=2-states and of (A1.4) we obtain

$$\langle\, n_{df} = 2, J_f\; M_f \,|\, \boldsymbol{d}^+_m \,|\, n_{di} = 1, J_i = 2, M_i \,\rangle =$$

$$\sqrt{2}\; (2\; M_i\, 2\; m \,|\, J_f M_f), \quad J_f \text{ even.} \qquad (11.29)$$

Moreover, according to (11.22) we have

$$\langle\, n_{df} = 2, J_f\; M_f \,|\, \boldsymbol{d}^+_m \,|\, n_{di} = 1, J_i = 2, M_i \,\rangle =$$

$(2\; M_i\; 2\; m \,|\, J_f\, M_f\,)(2J_f + 1)^{-1/2} \langle\, n_{df} = 2,\; J_f \;||\, \boldsymbol{d}^+ \,||\, n_{di} = 1,\; J_i = 2 \,\rangle$, which yields

$$\langle\, n_{df} = 2,\; J_f \;||\, \boldsymbol{d}^+ \,||\, n_{di} = 1,\; J_i = 2 \,\rangle = \sqrt{2}\, \sqrt{(2J_f + 1)}. \qquad (11.30)$$

We now take $n_{df}$ = 3 and $n_{di}$ = 2 i. e. we will calculate $\langle\, n_{df} = 3,\; J_f \;||\, \boldsymbol{d}^+ \,||\, n_{di} = 2,\; J_i \,\rangle$. We restrict ourselves to final states with the highest seniority $t_f = n_{df}$ = 3. According to (3.9) or (10.5) this means that we are interested in values $J_f$ = 3, 4 and 6. First, we work on the final state employing (6.9) and (3.7)

$$|\, n_{df} = 3, J'\, J_f\, M_f \,\rangle = A\, \sqrt{(3/2)}\; (-1)^{J_f}\, [\boldsymbol{d}^+ \times [\boldsymbol{d}^+ \times \boldsymbol{d}^+]^{(J')}\,]^{(J_f)}_{M_f} \,|\rangle =$$

$$[1 + 2(2J' + 1)\{^{2}_{2}\ ^{2}_{J_f}\ ^{J'}_{J'}\}]^{-1/2} \sqrt{(1/2)}\ (-1)^{J_f}\ [\boldsymbol{d}^+ \times [\boldsymbol{d}^+ \times \boldsymbol{d}^+]^{(J')}]^{(J_f)}{}_{M_f} |\rangle. \qquad (11.31)$$

Making use of (A3.6) and of the symmetries of the 6-$j$ symbol, we write

$$| n_{df} = 3, J' J_f M_f \rangle =$$

$$[1 + 2(2J' + 1)(-1)^{J_f}\ W(J' J_f\ 2\ J' ; 2\ 2)]^{-1/2} \sqrt{(1/2)}\ (-1)^{J_f}\ [\boldsymbol{d}^+ \times [\boldsymbol{d}^+ \times \boldsymbol{d}^+]^{(J')}]^{(J_f)}{}_{M_f} |\rangle.$$

We restrict ourselves to $J' = 4$, because with this choice all values for $J_f$ mentioned above can be reached. Moreover, we employ (A1.1) and (6.5) like this

$$| n_{df} = 3, J' = 4, J_f M_f \rangle =$$

$$[1 + 2{\cdot}9\ (-1)^{J_f}\ W(4\ J_f\ 2\ 4 ; 2\ 2)]^{-1/2}\ (-1)^{J_f} \cdot \sum_{m'} (2\ m', J' = 4, M_f - m' \,|\, J_f M_f) \cdot$$

$$\boldsymbol{d}^+_{m'} |n_d = 2, J' = 4, M_f - m' \rangle. \qquad (11.32)$$

Now we calculate the normal ( not reduced ) matrix element and represent the bra part with the help of (11.32) employing the hermitian adjoint operator $\boldsymbol{d}_{m'}$ this way

$$\langle n_{df} = 3, J' = 4, J_f M_f \,|\, \boldsymbol{d}^+_m \,|\, n_{di} = 2, J_i M_i \rangle =$$

$$[1 + 2{\cdot}9\ (-1)^{J_f}\ W(4\ J_f\ 2\ 4 ; 2\ 2)]^{-1/2}\ (-1)^{J_f} \cdot \sum_{m'} (2\ m', J' = 4, M_f - m' \,|\, J_f M_f) \cdot$$

$$\langle n_d = 2, J' = 4, J_f, M_f - m' \,|\, \boldsymbol{d}_{m'} \boldsymbol{d}^+_m |\, n_{di} = 2, J_i, M_i \rangle \qquad (11.33)$$

Making use of $\boldsymbol{d}_{m'} \boldsymbol{d}^+_m = d_{m'\,m} + \boldsymbol{d}^+_m \boldsymbol{d}_{m'}$ and of the orthonormality of the states $| n_d = 2, J, M \rangle$ we obtain

$$\langle n_{df} = 3, J' = 4, J_f M_f \,|\, \boldsymbol{d}^+_m \,|\, n_{di} = 2, J_i M_i \rangle =$$

$$[1 + 2{\cdot}9\ (-1)^{J_f}\ W(4\ J_f\ 2\ 4 ; 2\ 2)]^{-1/2}\ (-1)^{J_f} \cdot$$

$$[(2\ m, 4, M_f - m \,|\, J_f M_f)\ d_{J_i,4} + \sum_{m'} (2\ m', 4, M_f - m' \,|\, J_f M_f) \cdot$$

$$\langle n_d = 2, J' = 4, J_f, M_f - m' \,|\, \boldsymbol{d}^+_m \boldsymbol{d}_{m'} \,|\, n_{di} = 2, J_i, M_i \rangle] \qquad (11.34)$$

We look into the right hand side of the last matrix element making use of (6.5) and (A1.1) as follows

$$\boldsymbol{d}_{m'} \,|\, n_{di} = 2, J_i, M_i\rangle =$$

$$2^{-1/2} \sum_{m''} (2\ m''\ 2, M_i - m'' \,|\, J_i M_i)\ \boldsymbol{d}_{m'} \boldsymbol{d}^+_{m''} \boldsymbol{d}^+_{M_i - m''} |\rangle. \qquad (11.35)$$

Because of $\boldsymbol{d}_{m'} \boldsymbol{d}^+_{m''} \boldsymbol{d}^+_{M_i - m''} = d_{m' m''} \boldsymbol{d}^+_{M_i - m''} + \boldsymbol{d}^+_{m''} d_{m', M_i - m''} + \boldsymbol{d}^+_{m''} \boldsymbol{d}^+_{M_i - m''} \boldsymbol{d}_{m'}$ we have

$$\boldsymbol{d}_{m'} \,|\, n_{di} = 2, J_i, M_i\rangle = 2^{-1/2} (2\ m'\ 2, M_i - m' \,|\, J_i M_i)\ |\, d_{M_i - m'} \rangle +$$

$$2^{-1/2}\,(2, M_i - m', 2\; m' \mid J_i\, M_i)\; \mid d_{M_i - m'} \rangle + 0. \qquad (11.36)$$

The summand zero results from the annihilation operator acting on the vacuum state. Because $J_i$ is even, we can write

$$\boldsymbol{d}_{m'} \mid n_{di} = 2, J_i, M_i\rangle = 2^{1/2}\,(2\; m'\, 2, M_i - m' \mid J_i\, M_i)\; \mid d_{M_i - m'} \rangle. \qquad (11.37)$$

Now we insert (11.37) in (11.34) and obtain

$$\langle\, n_{df} = 3, J' = 4, J_f\, M_f \mid \boldsymbol{d}^+_m \mid n_{di} = 2, J_i\, M_i \,\rangle =$$

$$[1 + 2{\cdot}9\,(-1)^{J_f}\,W(4\; J_f\; 2\; 4\; ;\; 2\; 2)]^{-1/2}\,(-1)^{J_f}\cdot$$

$$[(2\; m, 4, M_f - m \mid J_f\, M_f)\; d_{J_i,4} + \textstyle\sum_{m'}(2\; m', 4, M_f - m' \mid J_f\, M_f)\cdot$$

$$\sqrt{2}\,(2\; m'\, 2, M_i - m' \mid J_i\, M_i)\,\langle\, n_d = 2, J' = 4, J_f, M_f - m' \mid \boldsymbol{d}^+_m \mid d_{M_i - m'} \,\rangle \qquad (11.38)$$

For the last matrix element, we take the expression (11.29). Moreover, we make use of the Wigner-Eckart theorem (A5.8) this way

$$\langle\, n_{df} = 3, J' = 4, J_f\, M_f \mid \boldsymbol{d}^+_m \mid n_{di} = 2, J_i\, M_i \,\rangle =$$

$$[1 + 2{\cdot}9\,(-1)^{J_f}\,W(4\; J_f\; 2\; 4\; ;\; 2\; 2)]^{-1/2}\,(-1)^{J_f}\cdot$$

$$[(2\; m, 4, M_f - m \mid J_f\, M_f)\; d_{J_i,4} + \textstyle\sum_{m'}(2\; m', 4, M_f - m' \mid J_f\, M_f)\cdot$$

$$\sqrt{2}\,(2\; m'\, 2, M_i - m' \mid J_i\, M_i)\,\sqrt{2}\,(2, M_i - m', 2\; \mu \mid J' = 4, M_f - m')] =$$

$$(J_i\, M_i\; 2\; m \mid J_f\, M_f)\,(2J_f + 1)^{-1/2}\,\langle\, n_{df} = 3, J_f\; \|\, \boldsymbol{d}^+ \,\|\; n_{di} = 2, J_i \,\rangle. \qquad (11.39)$$

We multiply both sides of the last equation with $(J_i\; M_i\; 2\; m \mid J_f\; M_f)$ and sum over $m$ and $M_i$. Making use of (A1.10), (A1.4) and (A3.4) we obtain.

$$\langle\, n_{df} = 3, J_f\, \|\, \boldsymbol{d}^+ \,\|\, n_{di} = 2, J_i \,\rangle = [1 + (-1)^{J_f}{\cdot}2{\cdot}9{\cdot}W(4\; J_f\; 2\; 4\; ;\; 2\; 2)]^{-1/2}\cdot$$

$$(-1)^{J_f}(2J_f + 1)^{1/2}\,[(-1)^{J_f}\,d_{J_i,\,4} + 6\sqrt{(2J_i + 1)}\; W(J_i\; 2\; J_f\; 4;\; 2\; 2)], \quad (J_i \text{ even}). \qquad (11.40)$$

Thus, we have represented the reduced matrix element of $\boldsymbol{d}^+$ on the basis of the vibrational limit with maximal seniority up to $n_{di} = 2$ ( $n_{df} = 3$ ). Arima and Iachello (1976a, p. 269) have carried on the procedure up to $n_{di} = 3$ and have elaborated a closed algebraic expression depending from $n_{di}$ . For $n_{di} = 0$, 1 or 2 it agrees with our relations (11.26), (11.30) and (11.40).

The reduced matrix element of the second operator of the electric quadrupole transition (11.23) can be written analogously to (11.25) as

$$\langle\, n_{sf} = n_{si} + 1, n_{df} = n_{di} - 1, J_f\, \|\, \boldsymbol{s}^+ \boldsymbol{d}^{\sim} \,\|\, n_{si}\, n_{di}\, J_i \,\rangle =$$

$$\sqrt{(n_{si} + 1)}\,\langle\, n_{df} = n_{di} - 1,\; J_f\, \|\, \boldsymbol{d}^{\sim} \,\|\, n_{di}\, J_i \,\rangle. \qquad (11.41)$$

With the help of (A5.8) and (A1.4) up to (A1.6) one obtains

$$\langle\, n_d - 1, J'\, \|\, \boldsymbol{d}^{\sim} \,\|\, n_d\; J \,\rangle =$$

$$\sqrt{(2J' + 1)}\ (J\ M\ 2\ m\ |\ J'\ M')^{-1} \langle\ n_d - 1,\ J'\ M'\ |\ \boldsymbol{d}^{\sim}_m\ |\ n_d\ J\ M\ \rangle =$$

$$\sqrt{(2J' + 1)}\ (J\ M\ 2\ m\ |\ J'\ M')^{-1} (-1)^m \langle\ n_d\ ,\ J\ M\ |\ \boldsymbol{d}^{+}_{-m}\ |\ n_d - 1,\ J'\ M'\ \rangle =$$

$$\sqrt{(2J + 1)}\ (J'\ M'\ 2\ -m\ |\ J\ M\ )^{-1} (-1)^{J - J'} \langle\ n_d\ ,\ J\ M\ |\ \boldsymbol{d}^{+}_{-m}\ |\ n_d - 1,\ J'\ M'\ \rangle =$$

$$(-1)^{J - J'} \langle\ n_d\ ,\ J\ ||\ \boldsymbol{d}^{+}\ ||\ n_d -1,\ J'\ \rangle. \qquad (11.42)$$

In the equations (11.41) and (11.42), $n_d$ is not limited. They are true as well if the seniority $t$ does not agree with $n_d$ .

If the radiative transition of the type $n_{di}$ ® $n_{df} = n_{di} + 1$, which appears in (11.24) up to (11.40), is energetically forbidden the inverse process $n_{di}$ ® $n_{df} = n_{di} - 1$ may be possible which is described in (11.42) down to (11.40).

The reduced probability for transitions between states of the $Y$-series in the vibrational limit ( section 10.3 ) can be written now for an arbitrary boson number $n_d$. Emitting E2-radiation the state $|\ n_s\ n_d,\ J_i = 2n_d = M_i\ \rangle$ turns into $|\ n_s + 1,\ n_d - 1,\ J_f = 2(n_d - 1) = M_f\ \rangle$. The reduced matrix element reads according to (11.41), (11.42) and (A5.8) as follows

$$\langle\ n_{sf} = n_{si} + 1,\ n_{df} = n_{di} - 1,\ J_f\ ||\ \boldsymbol{s}^{+} \boldsymbol{d}^{\sim}\ ||\ n_{si}\ n_{di}\ J_i\ \rangle =$$

$$\sqrt{(n_{si} + 1)}\ \sqrt{(2J_f + 1)}\ (J_i\ J_i\ 2, -2\ |\ J_f\ J_f\ )^{-1} \cdot$$

$$\langle\ n_{di} - 1,\ J_f,\ M_f = J_f\ \ |\ \boldsymbol{d}_2\ |\ n_{di}\ J_i\ M_i = J_i\ \rangle\ (-1)^{-2}.$$

Because of $J_f = J_i - 2$ we have

$$(J_i\ J_i\ 2, -2\ |\ J_f\ J_f\ ) = \sqrt{(2J_f + 1)}\ (2\ 2\ J_f\ J_f\ |\ J_i\ J_i) / \sqrt{(2J_i + 1)} =$$

$$\sqrt{(2J_f + 1)} / \sqrt{(2J_i + 1)}.$$

The initial state is “stretched“ and we have set it in the $z$-direction and so it consists only on $d_2$-single states. Therefore the operator $\boldsymbol{d}_2$**,** which appears above, transforms the initial state directly in the final one and adds the factor $\sqrt{n_d}$, from which follows

$$\langle\ n_{si} + 1,\ n_{di} - 1,\ J_f\ ||\ \boldsymbol{s}^{+} \boldsymbol{d}^{\sim}\ ||\ n_{si}\ n_{di}\ J_i\ \rangle = \sqrt{(n_{si} + 1)}\ \sqrt{(2J_i + 1)}\ \sqrt{n_{di}}\ . \qquad (11.43)$$

Owing to (11.21) and (11.23) we can write

$$B(\text{E2},\ J_i\ ®\ J_f\ ) = (a_2)^2 \langle\ n_{si} + 1,\ n_{di} - 1,\ J_f\ ||\ \boldsymbol{s}^{+} \boldsymbol{d}^{\sim}\ ||\ n_{si}\ n_{di}\ J_i\ \rangle / (2J_i + 1) =$$

$$(a_2)^2\ (n_{si} + 1)\ n_{di\,.} \qquad (11.44)$$

Especially for the lowest transition in the $Y$-series the following relation holds

$$B(\text{E2},\ 2^{+}_1\ ®\ 0^{+}_1\ ) = (a_2)^2 (N - 1 + 1) \cdot 1 = (a_2)^2 N. \qquad (11.45)$$

## 11.5 Comparison with experimental data of electric quadrupole transitions

In figures 10.3 up to 10.6 the values for $n_d$ are 0, 1, 2, .. . On the right hand side of each *Y*-level there are levels with about the same energy which agree in number $n_d$ with the corresponding *Y*-level ( see also 10.22 ). Consequently for energy reasons transitions from $n_d$ to $n_d$ – 1 prevail. Measurements reveal that transitions with $\Delta n_d = 0$ are much weaker and an ascent from $n_d$ to $n_d + 1$ is practically excluded. For transitions in the *Y*-series the expressions, (11.43) and (11.44) can be applied. According to (11.41) and (11.42) generally holds

$$\langle\, n_{sf} = n_{si} + 1,\ n_{df} = n_{di} - 1,\ J_f \,\|\, \boldsymbol{s}^{+} \tilde{\boldsymbol{d}} \,\|\, n_{si}\, n_{di}\, J_i \,\rangle =$$

$$\sqrt{(n_{si} + 1)}\ (-1)^{J_i - J_f} \langle\, n_{di}\ \ J_i \,\|\, \boldsymbol{d}^{+} \,\|\, n_{di} - 1,\ J_f \,\rangle$$

where the last reduced matrix element is calculated by means of (11.26), (11.30) or (11.40) according to the number $n_{di}$ . Doing so the indices *i* and *f* have to be interchanged. From (11.21), one obtains the reduced transition probability.

In table 11.1 transitions of the type $n_d \rightarrow n_d - 1$ are listed. The reduced matrix elements *ME* and the transition probabilities *B* are given. They are compared with measured values of $^{110}{}_{48}Cd_{62}$ . The model nucleus contains 7 active bosons. In view of the sensitivity of the quantity *B* ( quadratic expression ) these results corroborate the theory and the interpretations of the involved levels. Measured reduced probabilities *B* for transitions with $|\, n_d \,| > 1$ are two orders of magnitude smaller than the values of table 11.1. In our model these transitions are inadmissible. For the transition E2, $2_1 \rightarrow 0_1$ Milner (1969) gives the measured value $B(E2,\ 2_1 \rightarrow 0_1) = 934\ (+/- 38)\ e^2\ fm^4$ which yields using (11.45)

$$\alpha_2 = 11{,}6\ e\ fm^2. \qquad (11.46)$$

Table 11.1. Comparison between calculated and measured *B*(E2) values for the nucleus $^{110}$Cd .

| Initial state spin* | $n_d$ | series | final state spin* | $n_d$ | series | $ME^2$ | $B/(\alpha_2)^2$ | index | quotients of reduced transition probabilities | calcul. | exper.** |
|---|---|---|---|---|---|---|---|---|---|---|---|
| $2_1^+$ | 1 | *Y* | $0_1^+$ | 0 | *Y* | 5 | 7 | a | | | |
| $4_1^+$ | 2 | *Y* | $2_1^+$ | 1 | *Y* | 18 | 12 | b | *B*(b)/*B*(a) | 1.71 | 1.53 (.19) |
| $2_2^+$ | 2 | *X* | $2_1^+$ | 1 | *Y* | 10 | 12 | c | *B*(c)/*B*(a) | 1.71 | 1.08 (.29) |
| $3_1^+$ | 3 | *Z* | $2_2^+$ | 2 | *X* | 15 | $5\cdot 15/7$ | d | | | |
| $3_1^+$ | 3 | *Z* | $4_1^+$ | 2 | *Y* | 6 | $5\cdot 6/7$ | e | *B*(e)/*B*(d) | 0.4 | 0.47 (.2) |
| $4_2^+$ | 3 | *X* | $2_2^+$ | 2 | *X* | 11 | $5\cdot 11/7$ | f | | | |
| $4_2^+$ | 3 | *X* | $4_1^+$ | 2 | *Y* | 10 | $5\cdot 10/7$ | g | *B*(g)/*B*(f) | 0.91 | 0.23 (.3) |

The series and the levels correspond to figure 10.4. The quantity *ME* is the reduced matrix element of the *d*-boson states according to (11.42).
* The indices beside the spin value serves as spectroscopic identification.
** The estimated errors ( Arima and Iachello, 1976a, p. 287 ) are put in parentheses. The data a up to c stem from Milner (1969) and d up to g are taken from Krane and Steffen (1970).

Table 11.2 shows further quotients of reduced transition probabilities compared with measurements.

Table 11.2. Reduced transition probabilities for quadrupole radiation of Xe-isotopes (Arima and Iachello, 1976a, p. 276).

| Quotients of the *B*'s | calculated | measured values | | | |
|---|---|---|---|---|---|
| | | $^{132}_{54}Xe_{78}$ | $^{130}_{54}Xe_{76}$ | $^{126}_{54}Xe_{72}$ | $^{124}_{54}Xe_{74}$ |
| $B(3_1^+ \rightarrow 4_1^+) / B(3_1^+ \rightarrow 2_2^+)$ | 0.4 | 0.51 | 0.24-0.25 | 0.46-0.72 | 0.16 |
| $B(4_2^+ \rightarrow 4_1^+) / B(4_2^+ \rightarrow 2_2^+)$ | 0.91 | | 0.95-1.05 | 0.94 | 0.90 |

The measured values of transition probabilities which are forbidden in the IBM1 are two orders of magnitude smaller.

### 11.6 Transitions with $|\Delta n_d| = 0$

In our treatment of the quadrupole radiation the transitions with unchanged numbers of *d*-bosons ( $\Delta n_d = 0$, i. e. $n_{di} = n_{df}$ ) were left over. In order to make them up, first we look into the normal matrix element of the part of the transition operator in question in (11.23) and write with the aid of (11.22)

$$(J_i\, M_i\, 2\, m \,|\, J_f\, M_f\,)\,(2J_f + 1)^{-1/2} \langle\, n_d\, \alpha_f\, J_f \,||\, [\boldsymbol{d}^+ \times \tilde{\boldsymbol{d}}\,]^{(2)} \,||\, n_d\, \alpha_i\, J_i \,\rangle =$$

$$\langle\, n_d\, \alpha_f\, J_f\, M_f \,|\, [\boldsymbol{d}^+ \times \tilde{\boldsymbol{d}}\,]^{(2)}{}_m \,|\, n_d\, \alpha_i\, J_i\, M_i \,\rangle =$$

$$\Sigma_{m\,m'} (2\ m\ 2\ m' \,|\, 2\ m) \langle n_d\ a_f\ J_f\ M_f \,|\, \boldsymbol{d}^+_m \boldsymbol{d}^\sim_{m'} \,|\, n_d\ a_i\ J_i\ M_i \rangle. \quad (11.47)$$

The specification $a$ comprises $n_D$ and the seniority $t$. We now use the concept of the completeness of state functions of a quantum mechanical system. For spatial functions $j_b(\underline{r})$ of this kind the following equation holds

$$\Sigma_b\, j_b(\underline{r})\, j_b{}^*(\underline{r}') = d(\underline{r} - \underline{r}').$$

Consequently the functions $F_i$ and $F_f$ which belong to the same system obey the relation

$$\int F_f{}^*(\underline{r})\, F_i(\underline{r})\, d\underline{r} \ ^{o}\ \iint F_f{}^*(\underline{r})\, d(\underline{r} - \underline{r}') F_i(\underline{r}')\, d\underline{r}\, d\underline{r}' =$$

$$\Sigma_b \int F_f{}^*(\underline{r})\, j_b(\underline{r})\, d\underline{r} \cdot \int j_b{}^*(\underline{r}')\, F_i(\underline{r}')\, d\underline{r}'.$$

Thus, we can write down the last matrix element of (11.47) this way

$$\langle n_d\ a_f\ J_f\ M_f \,|\, \boldsymbol{d}^+_m \boldsymbol{d}^\sim_{m'} \,|\, n_d\ a_i\ J_i\ M_i \rangle =$$

$$\Sigma_{a\,J\,M} \langle n_d\ a_f\ J_f\ M_f \,|\, \boldsymbol{d}^+_m \,|\, n_d - 1,\ a\ J\ M \rangle \langle n_d - 1,\ a\ J\ M \,|\, \boldsymbol{d}^\sim_{m'} \,|\, n_d\ a_i\ J_i\ M_i \rangle =$$

$$\Sigma_{a\,J\,M} (J\ M\ 2\ m \,|\, J_f\ M_f)\, (2J_f + 1)^{-1/2}\, (J_i\ M_i\ 2\ m' \,|\, J\ M)\, (2J + 1)^{-1/2} \cdot$$

$$\langle n_d\ a_f\ J_f \,||\, \boldsymbol{d}^+ \,||\, n_d - 1,\ a\ J \rangle \langle n_d - 1,\ a\ J \,||\, \boldsymbol{d}^\sim \,||\, n_d\ a_i\ J_i \rangle. \quad (11.48)$$

In the last line the Wigner-Eckart theorem (A5.8) or (11.22) has been applied twice. We now insert (11.48) in (11.47), multiply by $(J_i\ M_i\ 2\ m \,|\, J_f\ M_f)$, sum over $m$ ( and $M_i$ ) and obtain by means of (A1.10), (A3.4) and (A3.6) the relation

$$\langle n_d\ a_f\ J_f \,||\, [\boldsymbol{d}^+ \times \boldsymbol{d}^\sim]^{(2)} \,||\, n_d\ a_i\ J_i \rangle =$$

$$\Sigma_{m\,m\,m'} (J_i\ M_i\ 2\ m \,|\, J_f\ M_f)\, (2\ m\ 2\ m' \,|\, 2\ m) \cdot$$

$$\Sigma_{a\,J\,M} (J\ M\ 2\ m \,|\, J_f\ M_f)\, (J_i\ M_i\ 2\ m' \,|\, J\ M)\, (2J + 1)^{-1/2} \cdot \quad (11.49)$$

$$\langle n_d\ a_f\ J_f \,||\, \boldsymbol{d}^+ \,||\, n_d - 1,\ a\ J \rangle \langle n_d - 1,\ a\ J \,||\, \boldsymbol{d}^\sim \,||\, n_d\ a_i\ J_i \rangle =$$

$$(-1)^{J_i - J_f} \sqrt{5}\, \Sigma_{a\,J} \{{}^{2}_{J_i}\ {}^{2}_{J}\ {}^{2}_{J_f}\} \langle n_d\ a_f\ J_f \,||\, \boldsymbol{d}^+ \,||\, n_d - 1,\ a\ J \rangle \langle n_d - 1,\ a\ J \,||\, \boldsymbol{d}^\sim \,||\, n_d\ a_i\ J_i \rangle.$$

As an example, we calculate the matrix element of the E2-transition from the level $4_1$ to $2_2$. According to figures 10.3 up to 10.5 these states are the lowest one with $n_d = 2$ and maximal seniority. Since there is only one level with $n_d = 1$ ($2_1$-level ) the expression (11.49) contains here merely one summand as follows

$$\langle n_d = 2,\ J_f = 2 \,||\, [\boldsymbol{d}^+ \times \boldsymbol{d}^\sim]^{(2)} \,||\, n_d\ J_i = 4 \rangle =$$

$$\sqrt{5}\, \{{}^{2}_{4}\ {}^{2}_{2}\ {}^{2}_{2}\} \langle n_d = 2,\ J_f = 2 \,||\, \boldsymbol{d}^+ \,||\, n_d = 1,\ J = 2 \rangle \cdot$$

$$\langle n_d = 1,\ J = 2 \,||\, \boldsymbol{d}^\sim \,||\, n_d = 2,\ J_i = 4 \rangle = \sqrt{5}\, (2/35)\, \sqrt{10}\, \sqrt{18} = 12/7. \quad (11.50)$$

The values of the last matrix elements are taken from table 11.1 and the 6-$j$ symbol comes from (A3.14).

The sum in (11.49) extends over all states with $n_d$ - 1 including possibly higher levels than ( $n_d$, $a_i$, $J_i$ ) or lower ones than ( $n_d$, $a_f$, $J_f$ ). Therefore, the sum can comprise levels not having maximal seniority, which means that the prerequisite of (11.40) is not met.

## 11.7 Configurations with $n_d$ > $t$

With that, we turn to quadrupole transitions whose states have not maximal seniority and belong therefore to the series $b$ or other according to (10.22). Such transitions have relatively low probabilities in nuclei of the vibrational limit but anyway we will need the affiliated matrix element for the treatment of the general Hamilton operator of the IBM1 ( chapter 12 ).

I. Talmi (1993, p. 766 - 775) has shown in an easily readable fashion that matrix elements of the kind

$$\langle\, n_d',\ t' \text{£}\ n_d',\ n_D',\ J' \,||\, \boldsymbol{d}^{+} \,||\, n_d = n_d' - 1,\ t \text{ £ } n_d,\ n_D,\ J \,\rangle \qquad (11.51)$$

are connected by simple factors with matrix elements having maximal seniority in both included states ( the effect of $\boldsymbol{d}^{+}$ on the state on the right hand side increases $t$ by 1 or lowers it by 1 because an $J$=0-pair can arise ). The corresponding matrix element of the operator $\boldsymbol{d}^{\sim}$ is formulated by means of (11.42).

In an analogous manner, I.Talmi has treated the matrix element

$$\langle\, n_d,\ t' \text{ £ } n_d,\ n_D',\ J' \,||\, [\boldsymbol{d}^{+} \times \boldsymbol{d}^{\sim}]^{(2)} \,||\, n_d,\ t \text{ £ } n_d,\ n_D,\ J \,\rangle$$

where $t = t' - 2$, $t'$ or $t' + 2$ holds.

## 11.8 Quadrupole moments

Electric multiple moments are measures of the charge distribution of nuclear states and especially of their deviation from spherical symmetry. The quadrupole moment $Q_z$ conveys how the charge is arranged along the $z$-axis. Classically $Q_z$ is defined like this

$$Q_z = \Sigma_{k=1}{}^{A}\, e(k)\, (3z(k)^2 - r(k)^2). \qquad (11.52)$$

$A$ is the number of nucleons and $e(k)$ is the charge of the $k$th nucleon. The spherical harmonics of the second order read

$$r^2\, Y_{2\,0}(\underline{r}) = \sqrt{}(5/16\pi)\, (3z^2 - r^2).$$

We insert them in (11.52) and obtain

$$Q_z = \sqrt{}(16\pi/5)\, \Sigma_{k=1}{}^{A}\, e(k)\, r(k)^2\, Y_{2\,0}(\underline{r}(k)). \qquad (11.53)$$

In quantum mechanics this function act as an operator, which we name $\boldsymbol{Q}^{(2)}$ ($\circ Q_z$). From (11.3) we see that $\boldsymbol{Q}^{(2)}$ agrees largely with the operator $\boldsymbol{O}$(E, 2, 0) i. e.

$$\boldsymbol{Q}^{(2)} = \sqrt{(16\pi/5)}\ \boldsymbol{O}(\mathrm{E}, 2, 0). \qquad (11.54)$$

According to (11.4) this operator results in a single-boson operator. Analogously to (11.12) it is written by means of creation- and annihilation operators like this

$$\boldsymbol{Q}^{(2)} = \sqrt{(16\pi/5)}\ (\ a_2\,(\boldsymbol{d}^+_{\boldsymbol{0}}\,\boldsymbol{s} + \boldsymbol{s}^+\,\boldsymbol{d}^{\sim}_{\boldsymbol{0}}) + b_2\,[\boldsymbol{d}^+ \times \boldsymbol{d}^{\sim}]^{(2)}{}_0\,). \qquad (11.55)$$

In the vibration limit the operator $\boldsymbol{d}^+_{\boldsymbol{0}}\ \boldsymbol{s} + \boldsymbol{s}^+\ \boldsymbol{d}^{\sim}_{\boldsymbol{0}}$ does not contribute to the expectation value of $\boldsymbol{Q}^{(2)}$ because this operator combines only boson states with different numbers $n_d$ ( $\Delta n_d$ = + 1 or - 1 ). The quadrupole moment $Q^{(2)}$ of the nuclear state $|\ n_d\ a\ J\ M\ \rangle$ is defined as the expectation value of $\boldsymbol{Q}^{(2)}$ where the nuclear spin is put in the $z$-axis i. e. $M = J$. Therefore this moment reads

$$Q^{(2)} = \sqrt{(16\pi/5)}\ b_2\,\langle\ n_d\ a\ J,\ M = J\ |\ [\boldsymbol{d}^+ \times \boldsymbol{d}^{\sim}]^{(2)}{}_0\ |\ n_d\ a\ J,\ M = J\ \rangle. \qquad (11.56)$$

The specification $a$ is the same as in (11.47). In order to obtain consistent expressions we generalise the concept of the reduced probability of the electric quadrupole radiation (11.23, 11.21). Starting from (11.21), (11.12) and (11.22), we define

$$(B(\mathrm{E2}, (n_d\ a\ J) \to (n_d\ a\ J)))^{1/2} = (2J + 1)^{-1/2}\ b_2\,\langle\ n_d\ a\ J\ ||[\boldsymbol{d}^+ \times \boldsymbol{d}^{\sim}]^{(2)}||\ n_d\ a\ J\ \rangle =$$

$$(J\ J\ 2\ 0\ |\ J\ J)^{-1}\ b_2\,\langle\ n_d\ a\ J,\ M = J\ |\ [\boldsymbol{d}^+ \times \boldsymbol{d}^{\sim}]^{(2)}{}_0\ |\ n_d\ a\ J,\ M = J\ \rangle. \qquad (11.57)$$

Because of $(J\ J\ 2\ 0\ |\ J\ J) = \sqrt{J}\,(2J - 1)^{1/2}(J + 1)^{-1/2}\,(2J + 3)^{-1/2}$ and from (11.56) and (11.57) we have the following relation

$$Q^{(2)} = \qquad (11.58)$$

$$\sqrt{(16\pi/5)}\ \sqrt{J}\,(2J - 1)^{1/2}(J + 1)^{-1/2}\,(2J + 3)^{-1/2}\,(B(\mathrm{E2}, (n_d\ a\ J) \to (n_d\ a\ J)))^{1/2}.$$

This expression shows that calculating $Q^{(2)}$, one can use the formalism developed for *B.*

Now we calculate quantitatively the quadrupole moment of the level $2_1$ in the vibrational limit of the IBM1. According to (11.58), (11.57) and (11.21) we write

$$Q^{(2)}(2_1) = \sqrt{(16\pi/5)}\ \sqrt{(2/7)}\ \sqrt{(1/5)}\,\langle\ n_d = 1,\ J = 2\ ||[\ \boldsymbol{d}^+ \times \boldsymbol{d}^{\sim}]^{(2)}||\ n_d = 1,\ J = 2\ \rangle.$$

With the help of (11.49), (11.26), (11.42) and (A3.14) we continue like this

$$Q^{(2)}(2_1) = \sqrt{(16\pi/5)}\ \sqrt{(2/7)}\ \sqrt{(1/5)}\ b_2\,\sqrt{5}\,\{^2{}_2\ {}^2{}_0\ {}^2{}_2\} \cdot$$

$$\langle\ n_d = 1,\ J = 2\ ||\ \boldsymbol{d}^+\ ||\ \rangle\,\langle\ ||\ \boldsymbol{d}^{\sim}\ ||\ n_d = 1,\ J = 2\ \rangle =$$

$$\sqrt{(16\pi/5)}\ \sqrt{(2/7)}\ b_2 = 1.695\ b_2. \qquad (11.59)$$

One obtains the same result from (11.56) using the relation

$$[\,\boldsymbol{d}^{+} \times \tilde{\boldsymbol{d}}\,]^{(2)}_{0} = \sum_{m} (2\ m\ 2\ {-m}\ |\ 2\ 0)\ \boldsymbol{d}^{+}_{m}\,\tilde{\boldsymbol{d}}_{-m} = \sum_{m} (m^{2} - 2)/\sqrt{14}\ \boldsymbol{d}^{+}_{m}\,\boldsymbol{d}_{m}\,. \quad (11.60)$$

# 12 The treatment of the complete Hamiltonian of the IBM1

In chapter 10 the Hamiltonian has been reduced to the form of the vibrational limit. In this chapter we go back to the complete Hamiltonian of the IBM1 (7.17 or 9.39). In the next section the eigenstates of this Hamilton operator will be developed on the basis of the eigenfunctions of the vibrational limit i. e. on the spherical basis. The well-known diagonalisation procedure is outlined. To this end, the matrix elements of the Hamiltonian have to be formed on the spherical basis, which is brought up in the second section. In the following section the electric quadrupole radiation is treated in this model. Finally comparisons with measured nuclear states and a hint for a roughly simplified model will be given.

## 12.1 Eigenstates

According to (9.39) and (10.1) the Hamilton operator of the IBM1 can be written as

$$\boldsymbol{H} = \boldsymbol{H}^{(\mathbf{I})} + v_r \boldsymbol{R}^{\mathbf{2}} + v_q \boldsymbol{Q}^2 \text{ with}$$

$$\boldsymbol{H}^{(\mathbf{I})} = e_n N + v_n N^2 + (e_d' + v_{nd} N)\boldsymbol{n_d} + v_d \boldsymbol{n_d}^2 + v_t \boldsymbol{T}^2 + v_j \boldsymbol{J}^2. \qquad (12.1)$$

In section 10.2 we have shown that the states of the spherical basis, $|\, N\, n_d\, t\, n_D\, J\, M\,\rangle$, are eigenfunctions of the operator $\boldsymbol{H}^{(\mathbf{I})}$ with the eigenvalues $E^{(\mathrm{I})}{}_{N\, n_d\, t\, J}$ (10.17). The orthogonality and the completeness of this basis, which have been used in section 11.6, make it possible to represent functions satisfying the same boundary conditions by expanding them in series of basis functions. Therefore, the eigenfunctions of $\boldsymbol{H}$, which are numbered by the index $l$ ( small letter of $L$ ), can be written like this

$$|\, l\, N\, J\, M\,\rangle\rangle = \sum_{i=1}{}^{Z} c^{(l)}{}_{N\,J}(i)\, |\, i\, N\, J\, M\,\rangle. \qquad (12.2)$$

The index $i$ represents the triple of the ordering numbers $n_{d},\ t$ and $n_D$ of the basis states. The sum comprises $Z$ states of this kind. The coefficients $c^{(l)}{}_{N\,J}\,(i)$ normalise the state $|\, l\, N\, J\, M\,\rangle\rangle$ to 1.

The eigenenergy of the $l$ th state (12.2) is characterized by $E^{(l)}{}_{N\,J}$ and satisfies the relation

$$\boldsymbol{H}\, |\, l\, N\, J\, M\,\rangle\rangle = E^{(l)}{}_{N\,J}\, |\, l\, N\, J\, M\,\rangle\rangle\ . \qquad (12.3)$$

Here we insert (12.2) and have

$$\sum_{i=1}{}^{Z} c^{(l)}{}_{N\,J}(i)\, \boldsymbol{H}\, |\, i\, N\, J\, M\,\rangle = E^{(l)}{}_{N\,J} \sum_{i=1}{}^{Z} c^{(l)}{}_{N\,J}(i)\, |\, i\, N\, J\, M\,\rangle. \qquad (12.4)$$

By multiplying with $\langle\, k\, N\, J\, M\, |$ and by integrating we obtain

$$\sum_{i=1}^{Z} c^{(l)}_{NJ}(i)\,\langle\, k\,N\,J\,M \mid \boldsymbol{H} \mid i\,N\,J\,M \,\rangle = E^{(l)}_{NJ}\; c^{(l)}_{NJ}(k). \qquad (12.5)$$

We name the matrix element above $H_{N\,J\,M}\,(k\;i\,)$. It will be treated in section 12.2. The system of $Z$ linear, homogeneous equations

$$\sum_{i=1}^{Z} H_{NJM}\,(k\;i\,)\,c^{(l)}_{NJ}(i\,) = E^{(l)}_{NJ}\; c^{(l)}_{NJ}(k), \qquad k = 1, \ldots ,Z \qquad (12.6)$$

is soluble only if the determinant of the matrix of this system vanishes i. e.

$$|\,\mathit{H}_{NJM} - \mathbf{1}\cdot E^{(l)}_{NJ}\,| = 0. \qquad (12.7)$$

The matrix $\mathit{H}_{N\,J\,M}$ contains the elements $H_{N\,J\,M}\,(k\;i\,)$ and the matrix $\mathbf{1}$ is the unit matrix. The secular equation (12.7) yields $Z$ roots $E^{(l)}_{N\,J}$ ( $l$ = 1, ... , $Z$), which are the energy eigenvalues of the complete Hamilton operator for $N$ bosons with the total angular momentum $J$. By inserting $E^{(l)}_{N\,J}$ in (12.6) one obtains the affiliated eigenvector ($c^{(l)}_{N\,J}\,(1)$, ... , $c^{(l)}_{N\,J}\,(Z)$), which produces the state $|\; l\;N\;J\;M\,\rangle\rangle$ according to (12.2). It can be shown that the matrix $\mathit{H}_{N\,J\,M}$ of the Hamilton operator is diagonalised by the matrix of the eigenvectors, for which reason this method is named diagonalisation procedure.

## 12.2 Matrix elements of the Hamiltonian

Now we shall calculate the matrix elements of (12.5), which are constituted by states of the spherical basis. Remember that the indices $i$ and $k$ denote triples of quantum numbers. Consequently, $i$ (or $k$ ) can be written as a function $i(n_d,\ t,\ n_D)$. According to (12.5), (12.1) and (10.17) we write

$$\langle\, k\,N\,J\,M \mid \boldsymbol{H} \mid i\,N\,J\,M \,\rangle = \langle\, k\,N\,J\,M \mid \boldsymbol{H}^{(\mathbf{I})} + v_r\boldsymbol{R}^{\mathbf{2}} + v_q\boldsymbol{Q}^{2} \mid i\,N\,J\,M \,\rangle =$$

$$[e_nN + v_nN^{2} + (e_d{'} + v_{nd}\,N)n_d + v_dn_d^{2} + v_t\;t\;(t + 3) + v_j\,J(J + 1)]\;d_{i\,k} +$$

$$v_r\langle\, k\,N\,J\,M \mid \boldsymbol{R}^{\mathbf{2}} \mid i\,N\,J\,M \,\rangle \;+ v_q\,\langle\, k\,N\,J\,M \mid \boldsymbol{Q}^{2} \mid i\,N\,J\,M \,\rangle. \qquad (12.8)$$

We turn to the last term in (12.8) but one, i. e. to the matrix element of the operator $\boldsymbol{R}^{\mathbf{2}}$ (9.3), which reads

$$\boldsymbol{R}^{2} = \boldsymbol{N}(\boldsymbol{N} + 4) - (\surd 5\cdot[\boldsymbol{d}^{+} \times \boldsymbol{d}^{+}]^{(0)} - \boldsymbol{s}^{+}\boldsymbol{s}^{+})(\surd 5[\boldsymbol{d}^{\sim} \times \boldsymbol{d}^{\sim}]^{(0)} - \boldsymbol{ss}).$$

Making use of (9.2) we can write

$$\boldsymbol{R}^{\mathbf{2}} = \qquad (12.9)$$

$$N(N + 4) + \boldsymbol{T}^{\mathbf{2}} - \boldsymbol{n_d}(\boldsymbol{n_d} + 3) - \surd 5[\boldsymbol{d}^{+} \times \boldsymbol{d}^{+}]^{(0)}\boldsymbol{ss} - \surd 5\;\boldsymbol{s}^{+}\boldsymbol{s}^{+}[\boldsymbol{d}^{\sim} \times \boldsymbol{d}^{\sim}]^{(0)} - \boldsymbol{s}^{+}\boldsymbol{s}^{+}\boldsymbol{ss}.$$

With the aid of the commutation rules for the $s$-bosons, of the number operator $\boldsymbol{n_s} = \boldsymbol{s}^{+}\boldsymbol{s}$ and (10.13) we obtain

$$\langle\, k(n_d{'}\;t{'}\;n_D{'})\;N\,J\,M \mid \boldsymbol{R}^{\mathbf{2}} \mid i(n_d\;t\;n_D)\;N\,J\,M \,\rangle =$$

$$[N(n + 4) + t(t + 3) - n_d(n_d + 3) - \boldsymbol{n_s}^{\mathbf{2}} + \boldsymbol{n_s}\,]\;d_{i\,k} -$$

$\sqrt{5}[\langle k\, N\, J\, M\, |\, [\boldsymbol{d}^+ \times \boldsymbol{d}^+]^{(0)} |\, i\, N\, J\, M \rangle \sqrt{n_s}\sqrt{(n_s - 1)}\; d_{n_d', n_d + 2}$ +

$\langle k\, N\, J\, M\, |\, [\boldsymbol{d}^{\sim} \times \boldsymbol{d}^{\sim}]^{(0} \,|\, i\, N\, J\, M \rangle \sqrt{(n_s + 1)}\sqrt{(n_s + 2)}\; d_{n_d', n_d - 2}].$ (12.10)

We put the relation $\boldsymbol{n_s} = \boldsymbol{N} - \boldsymbol{n_d}$ and $\sqrt{5}\, [\boldsymbol{d}^+ \times \boldsymbol{d}^+]^{(0)} = \sum_m (-1)^m\, \boldsymbol{d}^+_m\, \boldsymbol{d}^+_{-m}$ in (12.10), which yields

$\langle k(n_d'\; t'\; n_D')\; N\, J\, M\, |\, \boldsymbol{R}^2\, |\, i(n_d\; t\; n_D)\; N\, J\, M \rangle =$

$[(N - n_d)\,(2n_d + 5) + n_d + t(t + 3)]\; d_{ik}$ - (12.11)

$\sum_m (-1)^m [\,\langle k(n_d'\; t'\; n_D')\; N\, J\, M\, |\, \boldsymbol{d}^+_m \boldsymbol{d}^+_{-m}\, |\, i(n_d\; t\; n_D)\; N\, J\, M \rangle \sqrt{n_s}\sqrt{(n_s - 1)}\; d_{n_d', n_d + 2}$ +

$\langle k(n_d'\; t'\; n_D')\; N\, J\, M\, |\, \boldsymbol{d}^{\sim}_m \boldsymbol{d}^{\sim}_{-m} |\, i(n_d\; t\; n_D)\; N\, J\, M \rangle \sqrt{(n_s + 1)}\sqrt{(n_s + 2)}\; d_{n_d', n_d - 2}].$

Matrix elements like $\langle\, k\; N\; J\; M\, |\, \boldsymbol{d}^+_m\, \boldsymbol{d}^+_{-m}\, \;|\; i\; N\; J\; M \,\rangle$ in principle have been treated in (11.48). Correspondingly the relation

$\langle k\, N\, J\, M\, |\, \boldsymbol{d}^+_m \boldsymbol{d}^+_{-m}\; |\; i\, N\, J\, M \rangle =$

$\sum_{i''\, J''\, M''} \langle k\, N\, J\, M\, |\, \boldsymbol{d}^+_m |\; i''\, N\, J''\, M'' \rangle \langle i''\, N\, J''\, M''\, |\; \boldsymbol{d}^+_{-m}\; |\; i\, N\, J\, M \rangle$ (12.12)

holds. For the operator $\boldsymbol{d}^{\sim}_m\, \boldsymbol{d}^{\sim}_{-m}$ in (12.11) an analogous expression can be written down. Thus, we have formulated completely the matrix elements of the operator $\boldsymbol{R}^2$ in (12.8).

We now turn to the last matrix element in (12.8), which contains the operator $\boldsymbol{Q}^2$. According to (9.4) it reads

$\boldsymbol{Q}^2 = \sum_m (-1)^m\, \boldsymbol{Q}_m \boldsymbol{Q}_{-m}$

with $\boldsymbol{Q}_m = \boldsymbol{d}^+_m \boldsymbol{s} + \boldsymbol{s}^+ \boldsymbol{d}^{\sim}_m - (\sqrt{7}/2)\, [\boldsymbol{d}^+ \times \boldsymbol{d}^{\sim}]^{(2)}_m$ . We form

$$\boldsymbol{Q}_m \boldsymbol{Q}_{-m} = \boldsymbol{d}^+_m \boldsymbol{s}\, \boldsymbol{d}^+_{-m} \boldsymbol{s} + \boldsymbol{d}^+_m \boldsymbol{s} \boldsymbol{s}^+ \boldsymbol{d}^{\sim}_{-m} + \boldsymbol{s}^+\, \boldsymbol{d}^{\sim}_m\, \boldsymbol{d}^+_{-m} \boldsymbol{s} + \boldsymbol{s}^+\, \boldsymbol{d}^{\sim}_m\, \boldsymbol{s}^+\, \boldsymbol{d}^{\sim}_{-m} -$$
$$(\sqrt{7}/2)\, (\boldsymbol{d}^+_m \boldsymbol{s}\, [\boldsymbol{d}^+ \times \boldsymbol{d}^{\sim}]^{(2)}_{-m} + \boldsymbol{s}^+\, \boldsymbol{d}^{\sim}_m [\boldsymbol{d}^+ \times \boldsymbol{d}^{\sim}]^{(2)}_{-m} +$$
$$[\boldsymbol{d}^+ \times \boldsymbol{d}^{\sim}]^{(2)}_m\, \boldsymbol{d}^+_{-m} \boldsymbol{s} + [\boldsymbol{d}^+ \times \boldsymbol{d}^{\sim}]^{(2)}_m\, \boldsymbol{s}^+ \boldsymbol{d}^{\sim}_{-m}\,) +$$
$$(7/4)\, [\boldsymbol{d}^+ \times \boldsymbol{d}^{\sim}]^{(2)}_m\, [\boldsymbol{d}^+ \times \boldsymbol{d}^{\sim}]^{(2)}_{-m}\, . \quad (12.13)$$

Making use of

$\sum_m (-1)^m\, (\boldsymbol{d}^+_m \boldsymbol{s} \boldsymbol{s}^+ \boldsymbol{d}^{\sim}_{-m} + \boldsymbol{s}^+\, \boldsymbol{d}^{\sim}_m\, \boldsymbol{d}^+_{-m} \boldsymbol{s}\,) = \boldsymbol{n_d}\, (\boldsymbol{n_s} + 1) + (5 + \boldsymbol{n_d})\boldsymbol{n_s} =$ (12.14)

$\boldsymbol{n_d} + 2\, \boldsymbol{n_d}\, \boldsymbol{n_s} + 5\, \boldsymbol{n_s}$ we write

$\langle k\, N\, J\, M\, |\, \boldsymbol{Q}^2\, |\, i\, N\, J\, M \rangle =$

$\sum_m (-1)^m \langle k(n_d'\; t'\; n_D')\; N\, J\, M\, |\, \boldsymbol{Q}_m \boldsymbol{Q}_{-m} |\, i(n_d\; t\; n_D)\; N\, J\, M \rangle =$

$(5N + 2Nn_d - 4n_d - 2n_d^2\,)d_{ik}$ +

$$\sum_m (-1)^m \langle k(n_d'\, t'\, n_D')\, N\, J\, M \,|\, [\boldsymbol{d}^+_m \boldsymbol{d}^+_{-m} \sqrt{n_s}\sqrt{(n_s - 1)}\, d_{n_d',n_d+2} + \qquad (12.15)$$

$$\boldsymbol{d}^{\sim}_m \boldsymbol{d}^{\sim}_{-m} \sqrt{(n_s + 1)}\, \sqrt{(n_s + 2)}\, d_{n_d',n_d-2} -$$

$$(\sqrt{7}/2)\sqrt{n_s}\, (\boldsymbol{d}^+_m [\boldsymbol{d}^+ \times \boldsymbol{d}^{\sim}]^{(2)}_{-m} + [\boldsymbol{d}^+ \times \boldsymbol{d}^{\sim}]^{(2)}_m \boldsymbol{d}^+_{-m})\, d_{n_d',n_d+1} -$$

$$(\sqrt{7}/2)\sqrt{(n_s + 1)}\, (\boldsymbol{d}^{\sim}_m [\boldsymbol{d}^+ \times \boldsymbol{d}^{\sim}]^{(2)}_{-m} + [\boldsymbol{d}^+ \times \boldsymbol{d}^{\sim}]^{(2)}_m \boldsymbol{d}^{\sim}_{-m}) d_{n_d',n_d-1} +$$

$$(7/4)\, [\boldsymbol{d}^+ \times \boldsymbol{d}^{\sim}]^{(2)}_m [\boldsymbol{d}^+ \times \boldsymbol{d}^{\sim}]^{(2)}_{-m}\, d_{n_d',n_d}\ ]\,|\, i(n_d\, t\, n_D)\, N\, J\, M \rangle\,.$$

The first two terms in the matrix element on the right hand side of (12.15) have to be treated by means of (12.12). The next two terms contain matrix elements of the type

$$\langle k\, N\, J\, M \,|\, \boldsymbol{d}^+_m [\boldsymbol{d}^+ \times \boldsymbol{d}^{\sim}]^{(2)}_{-m} \,|\, i\, N\, J\, M \rangle = \qquad (12.16)$$

$$\sum_{i''\, J''\, M''} \langle k\, N\, J\, M \,|\, \boldsymbol{d}^+_m \,|\, i''\, N\, J''\, M'' \rangle \langle i''\, N\, J''\, M'' \,|\, [\boldsymbol{d}^+ \times \boldsymbol{d}^{\sim}]^{(2)}_{-m} \,|\, i\, N\, J\, M \rangle.$$

Similarly, the last term on the right hand side of (12.15) is written as

$$\langle k\, N\, J\, M \,|[\boldsymbol{d}^+ \times \boldsymbol{d}^{\sim}]^{(2)}_m [\boldsymbol{d}^+ \times \boldsymbol{d}^{\sim}]^{(2)}_{-m}] \,|\, i\, NJM \rangle = \qquad (12.17)$$

$$\sum_{i''\, J''\, M''} \langle k\, N\, J\, M \,|[\boldsymbol{d}^+ \times \boldsymbol{d}^{\sim}]^{(2)}_m \,|\, i''\, N\, J''\, M'' \rangle \langle i''\, N\, J''\, M'' \,|\, [\boldsymbol{d}^+ \times \boldsymbol{d}^{\sim}]^{(2)}_{-m} \,|\, i\, N\, J\, M \rangle.$$

Making use of $[\boldsymbol{d}^+ \times \boldsymbol{d}^{\sim}]^{(2)}_m = \sum_{m} (2\ m\ 2,\ m\ -\ m \,|\, 2\ m)\ \boldsymbol{d}^+_m \boldsymbol{d}^{\sim}_{m-m}$ the remaining terms in (12.16) and (12.17) can be calculated by means of (11.48). The matrix elements of the operators $\boldsymbol{d}^+_m$ and $\boldsymbol{d}^{\sim}_m$ are reviewed in section 11.4. Matrix elements of complex *d*-boson configurations are calculated with the help of the coefficients of fractional parentage (Talmi, 1993, S. 763, 766 and Bayman, 1966), which have been mentioned shortly in section 6.2. Evidently, the numerical calculation of matrix elements and the diagonalisation can become laborious. The computer program package PHINT (appendix A7) coded by O. Scholten (1991) performs this work and calculates eigenenergies, eigenstates and reduced transition probabilities.

The method applied in this chapter is named configuration mixing because pure states of the spherical basis are linearly combined. The resulting states don't show any group theoretical symmetries in contrast to the states $|\, N\, n_d\ t\ n_D\, J\, M \rangle$. For this reason, there is talk of symmetry breaking.

## 12.3 Electric quadrupole radiation

In order to calculate the reduced transition probability (11.21) we employ initial and final states of the type (12.2) and form the reduced matrix element of the transition operator (11.12) like this $\langle\langle\, l_f\, N\, J_f \,||\, \boldsymbol{O}(\mathrm{E}, 2) \,||\, l_i\ N\, J_i \rangle\rangle =$

$$\langle\langle\, l_f\, N\, J_f \,||\, a_2 (\boldsymbol{d}^+ \boldsymbol{s} + \boldsymbol{s}^+ \boldsymbol{d}^{\sim}) + b_2 [\boldsymbol{d}^+ \times \boldsymbol{d}^{\sim}]^{(2)} || \, l_i\ N\, J_i \rangle\rangle = \qquad (12.18)$$

$$\sum_{k\, n} c^{(l_f)}_{N\, J_f}(k)\, c^{(l_i)}_{N\, J_i}(n) \cdot \langle k\, N\, J_f \,||\, a_2 (\boldsymbol{d}^+ \boldsymbol{s} + \boldsymbol{s}^+ \boldsymbol{d}^{\sim}) + b_2 [\boldsymbol{d}^+ \times \boldsymbol{d}^{\sim}]^{(2)} ||\, n\, N\, J_i \rangle.$$

The last matrix elements have been treated in the normal form in section 11.4. They are transformed to the reduced form with the help of (11.6). By means of (11.17) and (11.21) the expression (12.18) is connected with the transition probability per time unit $T$ or with the mean lifetime $t_m$ according to (11.19).

## 12.4 Comparison with experimental data

Figures 12.1 and 12.2 show comparisons between measured and calculated level energies of the nuclei $^{110}$Cd and $^{102}$Ru, for which the complete Hamilton operator of the IBM1 has been employed. The parameters of the simplified form (7.31) of the Hamiltonian $e$, $c_0$, $c_2$, $c_4$, $v_2$ and $v_o$ are given. The calculated level schemes of both nuclei don't differ much from the schemes of figures 10.3 and 10.4 because it is about a numerical adaptation to nuclei which belong clearly to the vibrational limit. For $^{102}$Ru the calculated and measured reduced transition probabilities agree quite well ( Kaup, 1983, p. 16).

Han, Chuu and Hsieh (1990) have calculated about 260 level energies of Sm-, Gd- and Dy-isotopes and about 140 transitions probabilities $B$(E,2), which they compared with measured values. By adjusting the boson number $N$ as a free parameter, an essential improvement could be achieved compared with earlier publications.

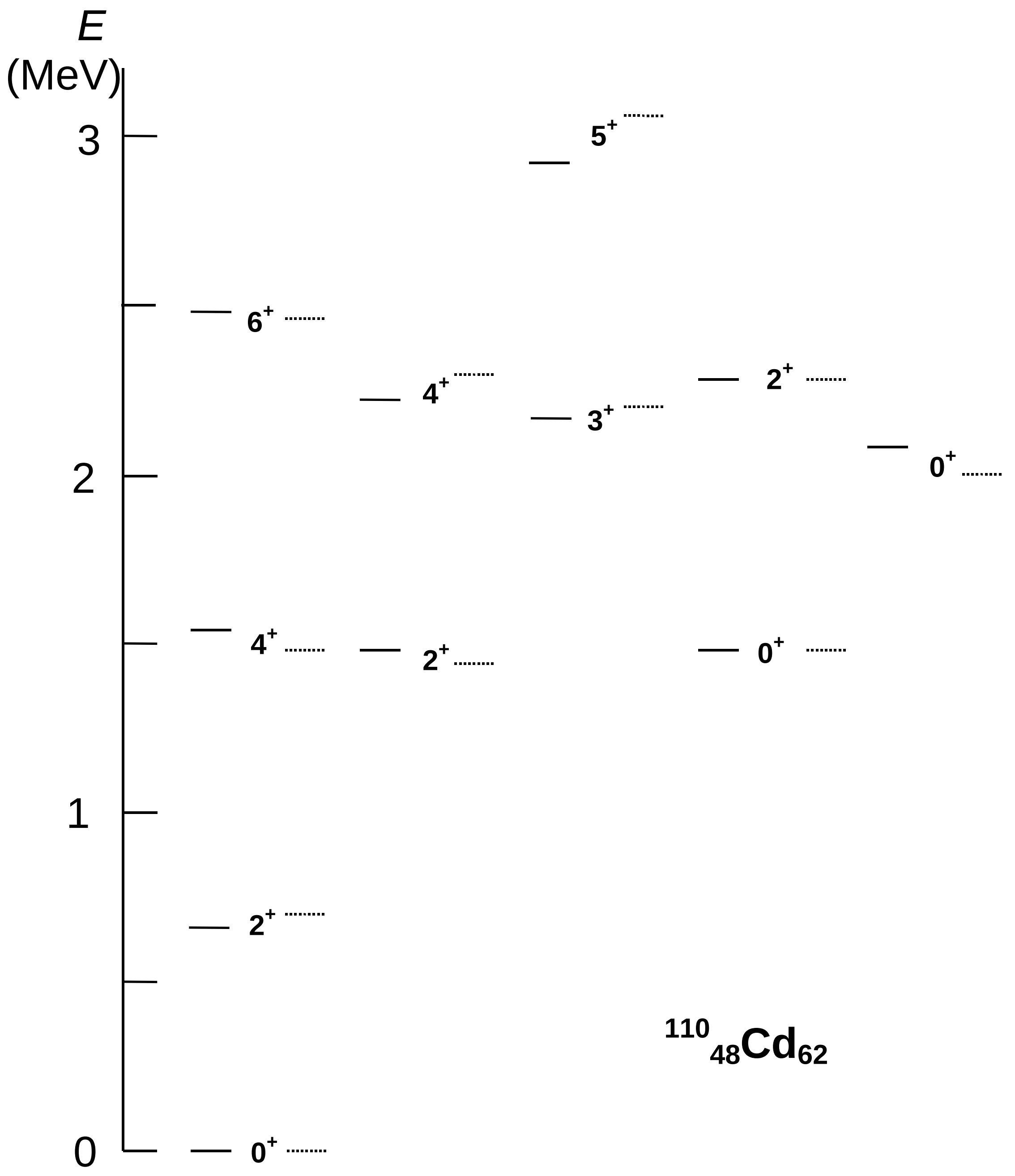

Figure 12.1. Comparison between the measured and the theoretical spectrum of $^{110}$Cd. It is calculated with the parameters $e$ = 740 keV, $c_0$ = 30 keV, $c_2$ = -120 keV, $c_4$ = 100 keV, $v_0$ = 71 keV and $v_2$ = -133 keV (Arima and Iachello, 1976a, p. 288). The broken lines denote calculated energies. The order of the levels corresponds to the scheme (10.22) and figure 10.4 respectively.

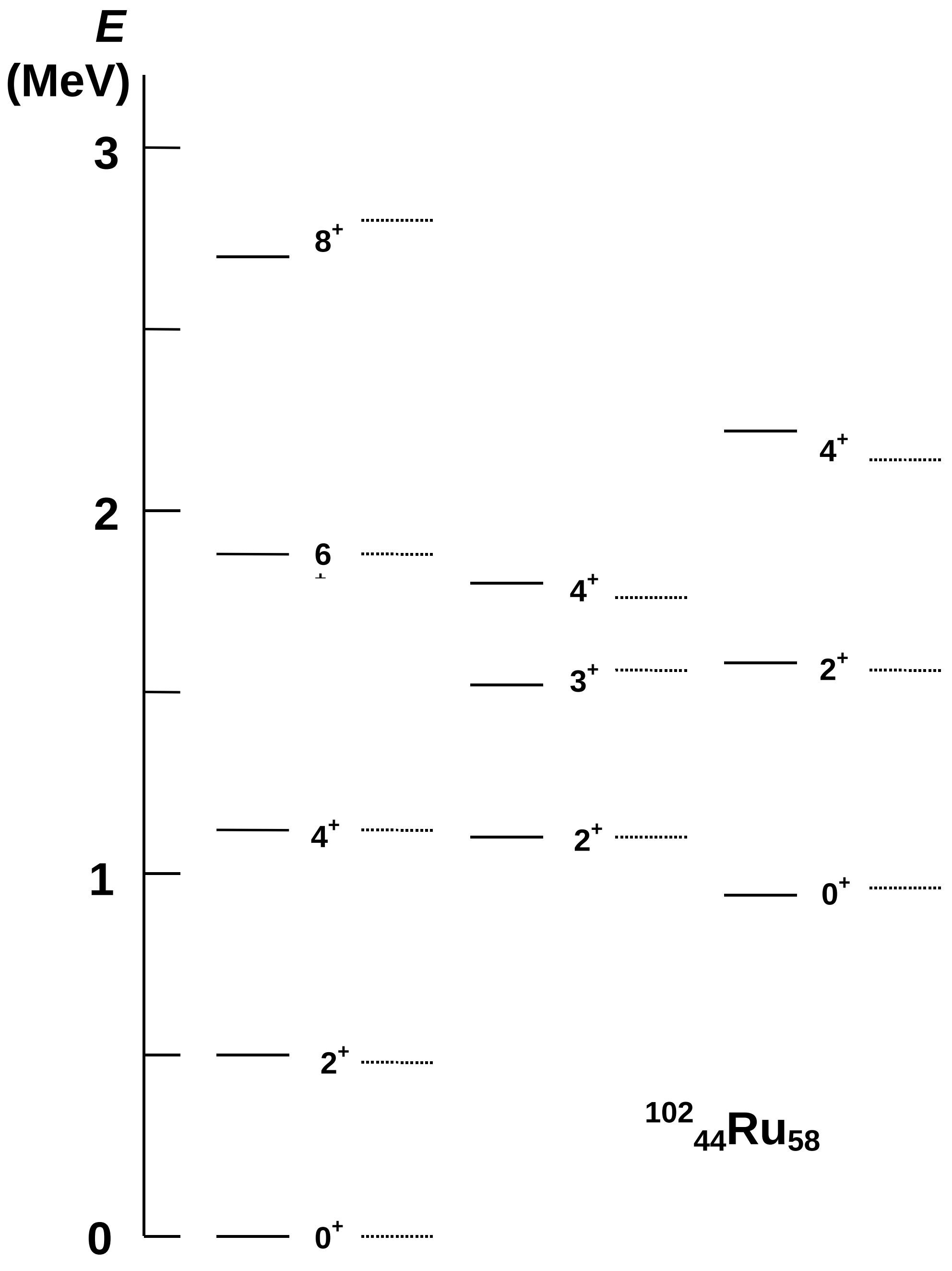

Figure 12.2. Comparison between the measured and the theoretical spectrum of $^{102}$Ru. It is calculated with the parameters $e$ = 561 keV, $c_0$ = -177 keV, $c_2$ = -174 keV, $c_4$ = 81 keV, $v_2/\sqrt{2}$ = 78 keV, $v_0/2$ = - 86 keV (Kaup 1983, p. 16).The order of the levels corresponds to the scheme (10.22) and figure 10.3 respectively.

### 12.5 An empirical Hamilton operator

The authors Casten and Warner ( 1988, p. 440 ) describe a Hamilton operator with drastic simplifications compared with (12.1). They set $v_r = 0$ and $\boldsymbol{H}^{(l)} = e\,\boldsymbol{n_d}$ i. e.

$$\boldsymbol{H} = e\,\boldsymbol{n_d} + v_q\,\boldsymbol{Q}^2 \quad \text{with}$$

$$\boldsymbol{Q}_m = \boldsymbol{d}^+{}_m\,\boldsymbol{s} + \boldsymbol{s}^+\tilde{\boldsymbol{d}}_m + c\,[\boldsymbol{d}^+ \times \tilde{\boldsymbol{d}}]^{(2)}{}_m\,. \qquad (12.19)$$

The expression $\boldsymbol{Q}_m$ in (12.19) differs from (9.4) in the parameter $c$. Similar simplifications as in (12.19) are usual in the interacting boson model 2 ( chapter 15 ). Following the ideas of the IBM2, the expression (12.19) is made dependent on the number of proton-bosons, $N_p$ , and on the number of neutron-bosons, $N_n$. The conspicuous fact that nuclear properties depend largely on the product $N_p \times N_n$ (Casten, 1990, p. 235) is taken into account by writing the energy $e$ of a single $d$-boson like this

$$e = e_0 \cdot \mathrm{e}^{-Q\,(N_p \times N_n - N_0)}\,. \qquad (12.20)$$

Properties of about 100 nuclei in three different regions of the nuclear card could be reproduced by varying weakly the parameters $e_0$ , $Q$, $N_0$, $v_q$ and $c$

# 13 Lie algebras

In chapter 9 the importance of the Lie algebras for interpreting and evaluating the IBM-Hamiltonian has been stressed. In the present chapter the essential characteristics of these algebras will be outlined starting from well-known examples. In the first section the rules are described which hold for the elements of a Lie algebra. Looking into special matrices in sections 2 and 3 three classical, real Lie algebras are identified and their dimensions are put together in section 4. In the last section above all Casimir operators ( chapter 9 and 14 ) will be treated.

## 13.1 Definition

A Lie algebra $L$ comprises infinitely many elements *a, b, c, ...* which show common characteristics. For instance Lie algebras can be represented by matrices with certain symmetry properties or by operators. In particular the elements of a Lie algebra must be able to form Lie products or commutators [*a* , *b*]. For quadratic matrices *a* and *b* the well-known relation

$$[a , b] = ab - ba \qquad (13.1)$$

holds, which results again in a matrix of the same type. If the elements of a Lie algebra are operators instead of matrices the commutator [*a* , *b*] is still defined but the right hand side of (13.1) need not be satisfied.

The definition of a Lie algebra demands the following properties from its elements *a, b, c*

a) the commutator of two elements is again an element of the algebra

[*a* , *b*] î $L$ for all *a, b* î $L$, (13.2)

b) a linear combination $a$ *a* + $b$ *b* of two elements with the numbers $a$ and $b$ is again an element of the algebra and the relation

[$a$ *a* + $b$ *b* , *c*] = $a$ [*a* , *c*] + $b$ [*b* , *c*] (13.3)

holds ( therefore the element 0 ( zero ) belongs also to the algebra),

c) interchanging both elements of a commutator results in

$$[a , b] = - [b , a], \qquad (13.4)$$

d) finally “Iacobi’s identity“ has to be satisfied as follows

$$[a , [b , c]] + [b , [c , a]] + [c , [a , b]] = 0. \qquad (13.5)$$

Anyhow, other kinds of coupling the elements are possible in principle. As an example the Casimir operators (13.32) can be mentioned, which result from multiplicative couplings of elements.

Furthermore, the definition demands that a Lie algebra has a finite dimension $n$ i. e. it comprises a set of $n$ elements $a_1, a_2, \dots, a_n$ which act as a basis, by which every element $x$ of the algebra can be represented like this

$$x = \sum_{i=1}^{n} x_i a_i . \quad (13.6)$$

In other words the algebra constitutes a $n$-dimensional vector space. The structure of the algebra can be formulated in terms of the basis. Because the commutator of two basis elements $a_p$ and $a_q$ belongs also to the algebra, according to (13.6) the relation

$$[a_p , a_q] = \sum_{r=1}^{n} c^{(r)}_{p\,q}\, a_r \quad (13.7)$$

holds. The numbers $c^{(r)}_{p\,q}$ are named structure constants. One formulates the commutator of the arbitrary elements $x$ (see 13.6) and $y$ ( = $\sum_j z_j a_j$ ) in terms of the structure constants like this

$$[x , y] = \sum_{i,j=1}^{n} x_i z_j [a_i , a_j] = \sum_{i,j,r=1}^{n} x_i z_j c^{(r)}_{i\,j}\, a_r . \quad (13.8)$$

If an algebra corresponds element by element to an other both are isomomorphic and we mark them with the same symbol. In the following we will restrict ourselves to real algebras which are characterised by real coefficients $x_i$ in the linear combinations (13.6). With regard to the interacting boson model we will deal with the following types of real algebras: $u(N)$, $su(N)$ and $so(N)$.

## 13.2 The *u*(*N*)- and the *su*(*N*)-algebra

We maintain that quadratic and antihermitian matrices of the rank $N$ ( $N\times N$-matrices ) form a Lie algebra. Antihermiticity means that the adjoint of the matrix $a$ i. e. the conjugate complex and transposed ( reflected on the diagonal ) matrix $a^\dagger$ is identical with $-a$ that is

$$a^\dagger = -a. \quad (13.9)$$

Providing real coefficients $\alpha$ and $\beta$, for two antihermitian matrices $a$ and $b$ with the same rank naturally the following relation holds

$$(\alpha\, a + \beta\, b)^\dagger = \alpha\, a^\dagger + \beta\, b^\dagger = -\alpha\, a - \beta\, b = -(\alpha\, a + \beta\, b). \quad (13.10)$$

Thus linear combinations of antihermitian matrices are again of this type. We make use of the well-known relation

$$(ab)^\dagger = b^\dagger a^\dagger \quad (13.11)$$

in order to look into the commutator of antihermitian matrices as follows

$$[a , b]^\dagger = (ab)^\dagger - (ba)^\dagger = b^\dagger a^\dagger - a^\dagger b^\dagger = ba - ab = -[a , b]. \quad (13.12)$$

Thus the commutator of two antihermitian matrices is again antihermitian. Now it is not difficult to show that the conditions (13.3) up to (13.5) are met. This means that the antihermitian matrices of the rank *N* constitute a Lie algebra. It is named *u*(*N*) algebra because it is closely related to the Lie group of the <u>u</u>nitary matrices.

A matrix is antihermitian if every element of the matrix equals the negative, conjugate complex value of the element which is in the transposed position. Any *N*×*N*-matrix of this kind can be composed linearly by the following quadratic matrices

$$\begin{pmatrix} i & 0 & \cdot & \\ 0 & 0 & \cdot & \\ \cdot & \cdot & \cdot & \\ & & & \cdot \end{pmatrix}, \quad \begin{pmatrix} 0 & 0 & \cdot & \\ 0 & i & 0 & \\ \cdot & 0 & 0 & \\ & & & \cdot \end{pmatrix}, \ldots, \quad \begin{pmatrix} 0 & i & 0 & \\ i & 0 & \cdot & \\ 0 & \cdot & \cdot & \\ & & & \cdot \end{pmatrix}, \quad \begin{pmatrix} 0 & 0 & i & 0 \\ 0 & 0 & 0 & \\ i & 0 & \cdot & \\ 0 & & & \cdot \end{pmatrix},$$

$$\ldots, \quad \begin{pmatrix} 0 & 1 & 0 & \cdot \\ -1 & 0 & & \\ 0 & & \cdot & \\ & & & \cdot \end{pmatrix}, \quad \begin{pmatrix} 0 & 0 & 1 & 0 \\ 0 & 0 & 0 & \\ -1 & 0 & \cdot & \\ 0 & & & \cdot \end{pmatrix}, \ldots \quad . \qquad (13.13)$$

Simple counting reveals that there are $N^2$ matrices in (13.13). As just mentioned they represent the basis of the Lie algebra *u*(*N*). The number of basis elements is named dimension *n*. Thus for the Lie algebra *u*(*N*) we have

$$n = N^2 \, . \qquad (13.14)$$

Now we will investigate antihermitian matrices of the rank *N* with vanishing traces ( sum of the diagonal elements ). We expect that they constitute a Lie algebra as well. Obviously a linear combination of these matrices again has a vanishing trace. Furthermore, for quadratic matrices *a* and *b* the relation

$$trace\,(ab) = trace\,(ba). \qquad (13.15)$$

holds. Thus the trace of a commutator vanishes always. This means that the conditions (13.2) up to (13.5) are satisfied and that we have a Lie algebra.

The trace of the basis matrices is always zero if in (13.13) all *N* diagonal matrices are replaced by (*N* – 1) matrices of the following form

$$\begin{pmatrix} i & 0 & . & & \\ 0 & -i & 0 & & \\ & 0 & . & & \\ & & & . & \\ & & & & . \end{pmatrix}, \begin{pmatrix} 0 & 0 & . & & \\ 0 & i & 0 & & \\ . & 0 & -i & 0 & \\ & & 0 & . & \\ & & & & . \end{pmatrix}, \begin{pmatrix} 0 & 0 & 0 & & \\ 0 & 0 & . & & \\ 0 & . & i & & \\ & & & -i & \\ & & & & . \end{pmatrix}, \ldots \ . \qquad (13.16)$$

Every diagonal, antihermitian matrix with trace 0 can be built by linearly combining matrices from (13.16). The elements (13.16) and the non-diagonal elements of (13.13) constitute the basis of the new Lie algebra. It comprises one matrix less than the one of the $u(N)$ algebra (13.13). Hence its dimension amounts to

$$n = N^2 - 1. \qquad (13.17)$$

This algebra represents a special kind of the $u(N)$ algebra and is therefore named $su(N)$.

## 13.3 The *so*(*N*) algebra

The third Lie algebra, which we are investigating here, is constituted by real, antisymmetric and quadratic matrices of the rank $N$. The transposed $a^{tr}$ of such a matrix $a$ equals to $-a$

$$a^{tr} = -a\ .$$

Obviously each linear combination of these matrices is still antisymmetric. With the help of

$$(a\, b)^{tr} = b^{tr}\, a^{tr} \qquad (13.18)$$

one shows in the same way as in the previous section that real, antisymmetric $N{\times}N$ matrices constitute a Lie algebra. As it is closely related to the Lie group of the orthogonal matrices it is named $o(N)$. Each diagonal element of an antisymmetric matrix is zero, i.e. its trace vanishes. Therefore its Lie algebra is named $so(N)$ as well in analogy to $su(N)$.

The basis elements of this algebra show the following structure

$$\begin{pmatrix} 0 & 1 & 0 & & \\ -1 & 0 & & & \\ 0 & 0 & . & & \\ & & & . & \\ & & & & . \end{pmatrix}, \begin{pmatrix} 0 & 0 & 1 & & \\ 0 & 0 & 0 & & \\ -1 & 0 & . & & \\ & & & . & \\ & & & & . \end{pmatrix}, \begin{pmatrix} 0 & 0 & 0 & & \\ 0 & 0 & 1 & & \\ 0 & -1 & 0 & & \\ & & & . & \\ & & & & . \end{pmatrix}, \ldots \ . \qquad (13.19)$$

and their number is $N(N-1)/2$. The dimension $n$ of the Lie algebra $so(N)$ is thus

$$n = N(N-1)/2. \qquad (13.20)$$

Specially for $N = 3$ the $so(3)$ basis elements read

$$a_1 = \begin{pmatrix} 0 & 1 & 0 \\ -1 & 0 & 0 \\ 0 & 0 & 0 \end{pmatrix}, \quad a_2 = \begin{pmatrix} 0 & 0 & 1 \\ 0 & 0 & 0 \\ -1 & 0 & 0 \end{pmatrix}, \quad a_3 = \begin{pmatrix} 0 & 0 & 0 \\ 0 & 0 & 1 \\ 0 & -1 & 0 \end{pmatrix}. \tag{13.21}$$

By calculating one obtains the following commutator relation

$$[a_1 , a_2] = - a_3 \text{ with cyclic permutations.} \tag{13.22}$$

The structure constants (13.7) of this algebra are therefore 1, 0 or -1. The minus sign in (13.22) disappears if one changes the sign of every basis matrix. The substitution

$$e_1 = ( a_1 + ia_2)/\sqrt{2},\ \ e_{-1} = ( a_1 - ia_2)/\sqrt{2},\ \ e_0 = - ia_3 \quad \text{yields} \tag{13.23}$$

$$[e_{-1} , e_1] = e_0,\ \ [e_{-1}, e_0] = e_{-1},\ \ [e_0 , e_1] = e_1 . \tag{13.24}$$

Obviously the elements $e_1$, $e_{-1}$ and $e_0$ constitute a Lie algebra with the dimension 3. It is *so*(3) as can be shown. The commutation relations (13.24) agree completely with the rules (8.7) of the angular momentum operators $J_m$ . Thus they constitute the Lie algebra *so*(3).

## 13.4 Dimensions of three classical algebras

In table 13.1 the lowest dimensions *n* of the three Lie algebras mentioned above are put together. Group theory shows that there is only a small number of simple and real Lie algebras. Therefore, a simple algebra can be identified directly by means of the dimension. For the IBM1 the *n*-values of table 13.1 are sufficient. They facilitate to project a given Lie algebra on one of the algebras *u*(*N*), *su*(*N*) or *so*(*N*). The corresponding matrix algebra is a representation of the given Lie algebra. It is possible to transform the basis of one algebra (see 13.23 and 13.24) in order to obtain a corresponding algebra.

Table 13.1. The lowest dimensions *n* of the Lie algebras *u*(*N*), *su*(*N*) and *so*(*N*) according to (13.14), (13.17) and (13.20).

| Algebra : | *u*(*N*) | *su*(*N*) | *so*(*N*) | dimension *n* |
|---|---|---|---|---|
| Rank : | *N* | *N* | *N* | |
| | | 2 | 3 | 3 |
| | 2 | | | 4 |
| | | | 4 | 6 |
| | | 3 | | 8 |
| | 3 | | | 9 |
| | | | 5 | 10 |
| | | 4 | 6 | 15 |
| | 4 | | | 16 |
| | | | 7 | 21 |
| | | 5 | | 24 |
| | 5 | | | 25 |
| | | | 8 | 28 |
| | | 6 | | 35 |
| | 6 | | 9 | 36 |

## 13.5 Operators constituting Lie algebras, their basis functions and their Casimir operators

In the IBM Lie algebras constituted by operators prevail. A very important property of the operators is the existence of sets of basis functions $y^{(j)}{}_k$ which are affiliated to the operators. For example such basis functions are frequently represented by eigenstates of a physical system. They constitute a basis vector, the dimension of which can be chosen. Its elements can be combined linearly and generate a multidimensional vector space. They are characterised by ordering numbers.

In order to clarify the situation we consider the angular momentum operators $\boldsymbol{J_1}$, $\boldsymbol{J_{-1}}$ and $\boldsymbol{J_0}$ (8.6). They obey the commutation rules (8.7)

$$[\, \boldsymbol{J_{-1}}, \boldsymbol{J_0} \,] = \boldsymbol{J_{-1}}\,, \qquad [\, \boldsymbol{J_{-1}}, \boldsymbol{J_1} \,] = \boldsymbol{J_0}\,, \qquad [\, \boldsymbol{J_0}, \boldsymbol{J_1} \,] = \boldsymbol{J_1}\,. \tag{13.25}$$

In section 13.3 we have seen that they constitute the Lie algebra *so*(3). On the other hand, from quantum mechanics we know that the operators $\boldsymbol{J_1}$, $\boldsymbol{J_{-1}}$ and $\boldsymbol{J_0}$ generate functions or quantum states $j_{\,jm}$, which are characterised by the quantity $j$ ( an integer for bosons ). For a given $j$ there are $2j + 1$ functions, which are labelled by $m$ ( projection of the spin or of the angular momentum ). If an operator $\boldsymbol{J_i}$ acts on $j_{\,jm}$ the result is not an entirely new function but as we know from quantum mechanics the relations

$$J_1 j_{jm} = -2^{-1/2}[j(j+1) - m(m+1)]^{1/2} j_{j,m+1},$$

$$J_{-1} j_{jm} = 2^{-1/2}[j(j+1) - m(m-1)]^{1/2} j_{j,m-1},$$

$$J_0 j_{jm} = m j_{jm} \qquad (13.26)$$

hold. Thus when acting on a function $j_{jm}$ the operators $J_1, J_{-1}$ or $J_0$ generate again a function of this kind with the same *j*. In general an operator of a Lie algebra creates a linear combination of basis functions when it acts on such a function.

We turn to the operator

$$J^2 = J_x^2 + J_y^2 + J_z^2 = -J_1 J_{-1} - J_{-1} J_1 + J_0^2 \qquad (13.27)$$

defined in (8.8). With the aid of (13.25) one shows that the following relation holds

$$[J^2, J_i] = 0 \text{ for } i = -1, 0, 1. \qquad (13.28)$$

We see that the operator $J^2$, which does not belong to the Lie algebra ( $J_1, J_{-1}, J_0$), commutes with the basis elements and therefore with all elements of the algebra. Operators which behave like $J^2$ are named **Casimir operatores**. In physics they are important because they generate eigenvalues. In the case of $J^2$ we have according to (8.5)

$$J^2 j_{jm} = j(j+1)\, j_{jm} . \qquad (13.29)$$

Thus, when the Casimir operator $J^2$ acts on the basis function $j_{jm}$ , this is reproduced and supplied with a factor, which depends only on *j*.

Now we can put together the properties of the Lie algebras, which we need for the IBM. We refer to the situation of the angular momentum and abstain from group theoretical proofs.

## 13.6 Properties of operators

We start with basis operators $F^{(i)}$ with $i = 1, .. ,n$ which are linearly independent, constitute a Lie algebra according to (13.2) up to (13.5) and are connected by the structure constants $c^{(r)}_{p\,q}$ (13.7). There exist sets of basis functions $y^{(j)}_k$ ( *j* stands for the set ) on which the operators $F^{(i)}$ act like this

$$F^{(i)} y^{(j)}_k = \sum_{l=1}^{n} G^{(i,j)}_{k\,l}\, y^{(j)}_l \qquad (13.30)$$

analogously to (13.26). The Casimir operators $X$, which commute with all operators $F^{(i)}$ according to (13.28) as follows

$$[X , F^{(i)}] = 0 , \qquad (13.31)$$

are formed in terms of the elements $F^{(i)}$. We need the quadratic type $X_2$ of that operator. The last has the following form

$$X_2 = \sum_{i,j}^{n} X_{ij} F^{(i)} F^{(j)}. \tag{13.32}$$

The coefficients $X_{ij}$ of the quadratic Casimir operator depend from the structure constants $c^{(r)}{}_{pq}$ (13.7). For example the coefficients of the Casimir operator $\boldsymbol{J}^2$ read according to (13.27) : $X_{1,-1} = X_{-1,1} = -1$, $X_{00} = 1$ and all other vanish.

The sets of functions $y^{(j)}{}_k$, which have so-called irreducible matrices $G^{(i,j)}{}_{kl}$ (13.30), generate eigenvalues $C^{(j)}{}_2$ under the influence of the Casimir operator $X_2$ like this

$$X_2\, y^{(j)}{}_k = C^{(j)}{}_2\, y^{(j)}{}_k. \tag{13.33}$$

According to (13.29) the eigenvalue of the quadratic Casimir operator $\boldsymbol{J}^2$ of the Lie algebra *so*(3) has the form $j(j + 1)$.

Using creation and annihilation operators for bosons, Frank and Van Isacker (1994, p. 310) show that the eigenvalues of the Casimir operators of the *so*(*N*)-algebras have the following form

$$C^{(r)}{}_{2,\, so(N)} = r\,(r + N - 2). \tag{13.34}$$

The quantity $N$ is the matrix dimension of the *so*-algebra and $r$ is an integer which is limited by a function of the boson number. For instance for $N = 3$ (13.34) agrees with (13.29), provided that $r$ is replaced by the angular momentum $J$ of the state. We know that in the *u*(5)-limit of the IBM the quantity $J$ has the values $2n_d$, $2n_d - 2$, $2n_d - 3$, ... , 1, 0. In the next chapter we will meet further applications of formula (13.34).

For the Lie algebra *su*(3) Cornwell (1990, p. 596) derives the following eigenvalue of the quadratic Casimir operator ( apart from a factor 1/9 )

$$C^{(n,m)}{}_{2,su(3)} = n^2 + m^2 + nm + 3n + 3m. \tag{13.35}$$

The integers $n$ and $m$ characterise the irreducible matrices ($G^{(i,nm)}{}_{kl}$) (13.30) of the algebra and the affiliated functions $y^{(nm)}{}_l$ .The expression (13.35) will arise in the next chapter in connection with the second special case of the IBM1.

The basis operators of the IBM ( chapter 14 ) not only constitute a Lie algebra in all but they contain subsets which are subalgebras. This means for instance that commutators of operators of a subalgebra are linear combinations of elements of this subset (13.2). The Casimir operator of the higher algebra commutes with the Casimir operator of the subalgebra, which follows from $X F^{(i)} = F^{(i)} X$ (13.31) and from (13.32). The Casimir operators of the subalgebras, which we will meet in the next chapter, are summands in the Casimir operators of the higher algebra.

Furthermore, it is true that the basis function $y^{(j)}{}_k$ (13.30) of the higher algebra is also the basis function of the subalgebra. Therefore this function is labelled both with the index of the Casimir eigenvalue $C_2$ of the higher algebra and with the corresponding index of the subalgebra. For example in the basis function $j_{jm}$ (13.29) the quantity $m$ is the characteristic parameter of the subalgebra *so*(2) ( consisting only of the operator $\boldsymbol{J_z} = \boldsymbol{J_0}$ ) and $j$ is the index of the higher algebra *so*(3), which comprises $\boldsymbol{J_{-1}}$ , $\boldsymbol{J_1}$ and $\boldsymbol{J_0}$ .

The whole publication including 16 chapters plus appendix and index can be downloaded from www.walterpfeifer.ch.